\documentclass[a4paper,oneside,11pt]{scrartcl}

\def\RPlus{\ensuremath{\mathbin{\rule[.13em]{.66em}{.22em}\hspace{-.44em}\rule[-.08em]{.22em}{.66em}\,}}} 
\usepackage[english]{babel}
\usepackage{graphicx}
\usepackage{dsfont}
\usepackage{bm}
\usepackage{amssymb}
\usepackage{amsthm}
\usepackage{amsmath}
\usepackage{amsfonts}
\usepackage{booktabs}
\usepackage{amsbsy}
\usepackage{fontenc}
\usepackage[utf8]{inputenc}
\usepackage{fancybox}
\usepackage[hypcap=false]{caption}
\usepackage{float}
\usepackage{subfigure} 
\usepackage{lmodern}
\usepackage[numbib,nottoc]{tocbibind}
\usepackage[pdfstartview={FitH},linkbordercolor={0 1 0}, colorlinks]{hyperref}
\usepackage{esint}
\usepackage{fancyhdr}
\usepackage{pdfpages}
\usepackage{trfsigns}
\usepackage{wasysym} 
\usepackage{siunitx} 
\usepackage{mathtools} 
\usepackage{relsize} 
\usepackage{scrhack} 
\usepackage{mleftright} 
\usepackage{tikz} 
\usepackage{stmaryrd} 
\usepackage[many]{tcolorbox} 
\tcbuselibrary{theorems} 

\usetikzlibrary{cd}

\makeatletter
\DeclareFontEncoding{LS1}{}{}
\DeclareFontSubstitution{LS1}{stix}{m}{n}
\DeclareMathAlphabet{\mathKel}{LS1}{stixscr}{m}{n}
\DeclareMathAlphabet{\mathcal}{LS1}{stixscr}{m}{n}
\makeatother

\allowdisplaybreaks 

\def\be{\begin{equation}}
\def\ee{\end{equation}}
\def\bs{\begin{subequations}}
\def\es{\end{subequations}}
\def\ba#1\ea{\begin{align}#1\end{align}}
\def\bes{\begin{equation*}}
\def\ees{\end{equation*}}
\def\bas#1\eas{\begin{align*}#1\end{align*}}

\theoremstyle{plain}
\newtheorem{theorem}{Theorem}[section]

\newtcbtheorem
  [use counter*=theorem,number within=section]
  {lemmata}
  {Lemma}
  {%
		breakable,
		enhanced,
    colback=gray!25,
    colframe=gray!0!black,
    fonttitle=\bfseries,
  }
  {lem}

\newtcbtheorem
  [use counter*=theorem,number within=section]
  {propositions}
  {Proposition}
  {%
		breakable,
		enhanced,
    colback=gray!25,
    colframe=gray!0!black,
    fonttitle=\bfseries,
  }
  {prop}
\newtcbtheorem
  [use counter*=theorem,number within=section]
  {theorems}
  {Theorem}
  {%
		breakable,
		enhanced,
    colback=gray!25,
    colframe=gray!0!black,
    fonttitle=\bfseries,
  }
  {thm}
\newtcbtheorem
  [use counter*=theorem,number within=section]
  {corollaries}
  {Corollary}
  {%
		breakable,
		enhanced,
    colback=gray!25,
    colframe=gray!0!black,
    fonttitle=\bfseries,
  }
  {cor}

\theoremstyle{remark}
\newtcbtheorem
  [use counter*=theorem,number within=section]
  {remarks}
  {Remark}
  {%
		breakable,
		enhanced,
    colback=gray!5,
    colframe=gray!50!black,
    fonttitle=\bfseries,
  }
  {rem}

\newtheorem{remark}[theorem]{Remarks}

\theoremstyle{definition}

\newtcbtheorem
  [use counter*=theorem,number within=section]
  {definitions}
  {Definition}
  {%
		breakable,
		enhanced,
    colback=gray!25,
    colframe=gray!0!black,
    fonttitle=\bfseries,
  }
  {def}
\newtcbtheorem
  [use counter*=theorem,number within=section]
  {examples}
  {Example}
  {
		breakable,
		enhanced,
    colback=gray!5,
    colframe=gray!50!black,
    fonttitle=\bfseries,
  }
  {ex}
\newtcbtheorem
  [use counter*=theorem,number within=section]
  {situations}
  {Situation}
  {
		breakable,
		enhanced,
    colback=gray!5,
    colframe=gray!50!black,
    fonttitle=\bfseries,
  }
  {sit}
\newtcbtheorem
  [use counter*=theorem,number within=section]
  {fieldredefinitions}
  {Theorem}
  {
		breakable,
		enhanced,
    colback=gray!25,
    colframe=gray!0!black,
    fonttitle=\bfseries,
  }
  {fieldredef}

\DeclareFontFamily{U}{mathx}{\hyphenchar\font45}
\DeclareFontShape{U}{mathx}{m}{n}{
      <5> <6> <7> <8> <9> <10>
      <10.95> <12> <14.4> <17.28> <20.74> <24.88>
      mathx10
      }{}
\DeclareSymbolFont{mathx}{U}{mathx}{m}{n}
\DeclareFontSubstitution{U}{mathx}{m}{n}
\DeclareMathAccent{\widecheck}{0}{mathx}{"71}
\DeclareMathAccent{\wideparen}{0}{mathx}{"75}

\begin{document}
\pagenumbering{Roman}
\renewcommand{\thefootnote}{\fnsymbol{footnote}}
\begin{titlepage}

\author{Simon-Raphael Fischer\footnote{Email: \href{mailto:sfischer@ncts.tw}{sfischer@ncts.tw}} }
\title{Infinitesimal gauge transformations induced by Lie algebroid connections, in the context of Yang-Mills-Higgs gauge theory} 
\date{\today} 
\maketitle
\thispagestyle{empty}

\begin{center}
National Center for Theoretical Sciences, Mathematics Division, National Taiwan University\\
No. 1, Sec. 4, Roosevelt Rd., Taipei City 106, Taiwan Room 503, Cosmology Building, Taiwan
\ \\
\ \\
\ \\
\textbf{Abstract}
\begin{abstract}
  \small{There is the notion of action Lie algebroids, containing information about Lie algebras and their actions, which is why it is natural to generalise gauge theories to a formulation using Lie algebroids; these allow structure functions in general. This is for example done in the formulation of curved Yang-Mills-Higgs gauge theory, formulated by Alexei Kotov and Thomas Strobl. We will discuss how to formulate infinitesimal gauge transformations using Lie algebroids in such a way that these close as algebra. For this we are going to generalise infinitesimal gauge transformations of Yang-Mills-Higgs gauge theories to derivations on vector bundle $V$-valued functionals. In the context of gauge theory, we will motivate that those vector bundles $V$ will be the pullback of another bundle $W$, and the gauge transformations as derivations will be induced by a Lie algebroid connection on $W$, using a more general notion of pullback connections. This also supports the usage of arbitrary types of connections on $W$ in the definition of the infinitesimal gauge transformation, not just canonical flat ones as in the classical formulation. As usual, these derivations are parametrised and their parameters have to be generalised to functionals, especially the parameters themselves can have a non-trivial gauge transformation. We then discuss under which conditions this algebra of derivations gives a closed algebra, also by providing a natural Lie bracket on the space of parameters, and we are going to see that closure is strongly related to the vanishing of a tensor known as the basic curvature.
	%
}
 \end{abstract}
\end{center}

\end{titlepage}


\tableofcontents

\renewcommand{\thefootnote}{\arabic{footnote}}

\setlength{\parindent}{12 pt}


\pagenumbering{arabic}

\section{Introduction}

Alexei Kotov and Thomas Strobl formulated an infinitesimal gauge theory, using Lie algebroids to describe the structural data, called curved Yang-Mills-Higgs gauge theories: Essentially, the structural Lie algebra equipped with an action on a manifold $N$ of values of physical fields like the Higgs field is replaced by a general Lie algebroid $E \to N$. We introduce Lie algebroids later, but one possible difference is that now we can have structure functions instead of constants as usually in gauge theory and particle physics.

We will neither describe nor introduce such gauge theories here in full details, see for example \cite{CurvedYMH} and the references therein for a summary or my Ph.D.~thesis \cite{MyThesis} for a thorough introduction and rediscovery of curved Yang-Mills-Higgs gauge theories, where the following results were also presented. However, we want to provide a generalized formulation of gauge transformations, independently of curved Yang-Mills-Higgs gauge theories, although highly motivated by my own studies of such theories. This definition serves as a coordinate-free generalisation of infinitesimal gauge transformations, useful for theories like curved Yang-Mills-Higgs gauge theories and may even motivate the formulation of such theories.

Section \ref{BasicDefinitions} provides some basic definitions and notions, while Section \ref{ClassicalGaugeTrafoDiscussion} is generally about how one can reformulate infinitesimal gauge transformations, using Lie algebroid connections, especially in the context of action Lie algebroids, also known as transformation Lie algebroids. The Higgs field will be a smooth map $M \to N$, where $M$ plays the role of the spacetime, and we will motivate that it is natural to think of an action Lie algebroid defined over $N$, thus, the general description uses a Lie algebroid $E$ over $N$. In order to define infinitesimal gauge transformations as derivations on functionals depending on $M$, we are going to make a pullback of an $E$-connection; which is why we discuss in the previous Section \ref{PullbacksOfEConnections} under which conditions the pullback of Lie algbebroid connections is possible. Furthermore, while Subsection \ref{ClassicalBasicSetup} is repeating the typical definition of infinitesimal gauge transformations, we will see in Subsection \ref{ClassicalWithLieAlgebroidConnections} that the definition of the infinitesimal gauge transformation of the Higgs field has a certain 1:1 correspondence to the existence of the pullback of Lie algebroid connections.

Subsection \ref{ClassicalWithLieAlgebroidConnections} will conlude that the discussed pullback of Lie algebroid connections naturally generalises the typical definition of classical gauge transformations; we will generalise this definition in Section \ref{GeneralisedGaugeTrafoStuffYeah}, supporting formalisms like curved Yang-Mills-Higgs gauge theories. In Subsection \ref{InfinitesimalGaugeTransformation} we will introduce and motivate the definitions of the infinitesimal gauge transformations in a coordinate-free manner with respect to this general setting. In order to define the infinitesimal gauge transformation on functionals which can also have values in certain vector bundles, we will use Lie algebroid connections; essentially a generalised construction of what we will have seen in Subsection \ref{ClassicalWithLieAlgebroidConnections}. Finally, in Subsection \ref{CommutatorOfGaugeTrafos} we will discuss and provide a discussion about under which conditions the algebra of derivations describing the infinitesimal gauge transformations closes with respect to their commutator. For this the vanishing of a tensor known as the basic curvature will be crucial, a condition which is reasonable and expected in the context of gauge theory. Recall that infinitesimal gauge transformations are parametrised, and we will generalise the parameters to functionals in those discussions, such that the parameters itself may have non-trivial gauge transformations; in fact, it will not be avoidable to assume that there are gauge transformations for the parameters. Therefore we have to discuss how to construct a natural Lie bracket on these generalised parameters, in order to study whether or not the algebra of infinitesimal gauge transformations closes.

\section{Basic definitions}\label{BasicDefinitions}

In the following, we denote with $V^*$ the dual of a vector bundle $V \to N$ over a smooth manifold $N$, and $\Phi^*V$ denotes the pull-back of $V$ by $\Phi: M \to N$, a smooth map from a smooth manifold $M$ to $N$. We have a similar notation for the pull-back of sections, especially we will have sections $F$ as an element of $\Gamma\left( \left(\bigotimes_{m=1}^{l} E_m^*\right) \otimes E_{l+1} \right)$, where $E_1, \dots, E_{l+1} \to N$ ($l \in \mathbb{N}$) are real vector bundles of finite rank over a smooth manifold $N$, and $\Gamma(\cdot)$ denotes the space of smooth sections. Then we view the pull-back $\Phi^*F$ as an element of $\Gamma\left( \mleft(\bigotimes_{m=1}^{l} \mleft(\Phi^*E_m\mright)^*\mright) \otimes \Phi^*E_{l+1} \right)$, and it is essentially given by
\bas
	(\Phi^*F)(\Phi^*\nu_1, \dotsc , \Phi^*\nu_l)
	&=
	\Phi^*\mleft( F\mleft( \nu_1, \dotsc, \nu_l \mright) \mright)
\eas
for all $\nu_1 \in \Gamma(E_1), \dotsc, \nu_l \in \Gamma(E_l)$. In general we also make use of that sections of $\Phi^*E$ can be viewed as sections of $E$ along $\Phi$, where $E \stackrel{\pi}{\to} N$ is any vector bundle over $N$. Let $\mu \in \Gamma(\Phi^*E)$, then it has the form $\mu_p = (p, u_p)$ for all $p \in M$, where $u_p \in E_{\Phi(p)}$, the fibre of $E$ at $\Phi(p)$; and a section $\nu$ of $E$ along $\Phi$ is a smooth map $M \to E$ such that $\pi \circ \nu = \Phi$. Then on one hand $\mathrm{pr}_2 \circ \mu$ is a section along $\Phi$, where $\mathrm{pr}_2$ is the projection onto the second component, and on the other hand $M \ni p \mapsto (p, \nu_p)$ defines an element of $\Gamma(\Phi^*E)$. With that one can show that there is a 1:1 correspondence of $\Gamma(\Phi^*E)$ with sections along $\Phi$. Similarly, vector bundle morphisms $L: G \to E$ over $\Phi$ have 1:1 correspondences to base-preserving vector bundle morpishms $G \to \Phi^*E$, where $G \to M$ is a vector bundle over $M$. We do not necessarily mention it when we make use of such trivial identifications, it should be clear by the context. For example $\mathrm{D}\Phi$ denotes the total differential of $\Phi$ (also called tangent map). It can be viewed as a vector bundle morphism $\mathrm{T}M \to \mathrm{T}N$ over $\Phi$, and we often view it as an element of $\Omega^1(M; \Phi^*\mathrm{T}N)$ by $\mathfrak{X}(M) \ni Y \mapsto \mathrm{D}\Phi(Y)$, where $\mathrm{D}\Phi(Y) \in \Gamma(\Phi^*\mathrm{T}N), M \ni p \mapsto \mathrm{D}_p\Phi(Y_p)$.

Additionally, with $\Omega^k(N; E)$ ($k \in \mathbb{N}_0$) we denote $k$-forms on $N$ with values in a vector bundle $E \to N$, and we always use the Einstein's sum convention. If one has a connection $\nabla$ on a vector bundle $V \to N$, then one has the notion of the exterior covariant derivative on $\Omega^p(M;E)$, denoted by $\mathrm{d}^\nabla$. In the case of a trivial vector bundle $V=N \times W \to N$, where $W$ is some vector space, we will often use the \textbf{canonical flat connection} for $\nabla$, defined by $\nabla \nu = 0$, where $\nu$ is a constant section of $N \times W$, see \textit{e.g.}~\cite[Example 5.1.7; page 260f.]{hamilton} for a geometric interpretation as horizontal distribution. The canonical flat connection is clearly uniquely defined (if a trivialization is given) because constant sections generate all sections and due to the Leibniz rule and linearity of $\nabla$. Let $\mleft( e_a \mright)_a$ be a constant global frame of $N \times W$, thence,
\bas
\mathrm{d}^\nabla \omega
&=
\mathrm{d} \omega^a \otimes e_a
\eas
for all $\omega \in \Omega^p(M; W)$, where we write $\omega= \omega^a \otimes e_a$. Hence, we define
\ba
\mathrm{d}\omega
&\coloneqq
\mathrm{d}^\nabla \omega,
\ea
when $\nabla$ is the canonical flat connection. $\mathrm{d}$ is clearly a differential.

As usual, there will be definitions of certain objects depending on other elements, and for keeping notations simple we will not always explicitly denote all dependencies. It will be clear by context on which it is based on, that is, when we define an object $A$ using the notion of Lie algebra actions $\gamma$ and we write "Let $A$ be [as defined before]", then it will be clear by context which Lie algebra action is going to be used, for example given in a previous sentence writing "Let $\gamma$ be a Lie algebra action".

We also need the following definitions.

\begin{definitions}{Graded extension of products, \newline \cite[generalization of Definition 5.5.3; page 275]{hamilton}}{GradingOfProducts}
Let $l \in \mathbb{N}$ and $E_1, \dots E_{l+1} \to N$ be vector bundles over a smooth manifold $N$, and $F \in \Gamma\left( \left(\bigotimes_{m=1}^{l} E_m^*\right) \otimes E_{l+1} \right)$. Then we define the \textbf{graded extension of $F$} as
	\bas
\Omega^{k_1}(N; E_1) \times \dots \times \Omega^{k_l}(N; E_l)
&\to \Omega^{k}(N; E_{l+1}), \\
(A_1, \dots, A_l)
&\mapsto
F\mleft(A_1\stackrel{\wedge}{,} \dotsc \stackrel{\wedge}{,} A_l\mright),
\eas
where $k := k_1+\dots k_l$ and $k_i \in \mathbb{N}_0$ for all $i\in \{1, \dots, l\}$. $F\mleft(A_1\stackrel{\wedge}{,} \dotsc \stackrel{\wedge}{,} A_l\mright)$ is defined as an element of $\Omega^{k}(N; E_{l+1})$ by
\bas
&F\mleft(A_1\stackrel{\wedge}{,} \dotsc \stackrel{\wedge}{,} A_l\mright)\mleft(Y_1, \dots, Y_{k}\mright)
\coloneqq \\
&\frac{1}{k_1! \cdot \dots \cdot k_l!} \sum_{\sigma \in S_{k}} \mathrm{sgn}(\sigma) ~ F\left( A_1\left( Y_{\sigma(1)}, \dots, Y_{\sigma(k_1)} \right), \dots, A_l\left( Y_{\sigma(k-k_l+1)}, \dots, Y_{\sigma(k)} \right) \right)
\eas
for all $Y_1, \dots, Y_{k} \in \mathfrak{X}(N)$, where $S_{k}$ is the group of permutations of $\{1, \dots, k\}$ and $\mathrm{sgn}(\sigma)$ the signature of a given permutation $\sigma$. 

$\stackrel{\wedge}{,}$ may be written just as a comma when a zero-form is involved.

Locally, with respect to given frames $\mleft( e^{(i)}_{a_i} \mright)_{a_i}$ of $E_i$, this definition has the form
\ba\label{CoordExprOfGradedExtension}
F\mleft(A_1\stackrel{\wedge}{,} \dotsc \stackrel{\wedge}{,} A_l\mright)
&=
F\mleft(e^{(1)}_{a_1}, \dotsc, e^{(l)}_{a_l}\mright) \otimes A_1^{a_1} \wedge \dotsc \wedge A_l^{a_l}
\ea
for all $A_i = A_i^{a_i} \otimes e^{(i)}_{a_i}$, where $A_i^{a_i}$ are $k_i$-forms on $N$.
\end{definitions}

\begin{remark}
\leavevmode\newline
Assume $F \in \Gamma\left( \mleft(\bigwedge_{m=1}^{l} \mathrm{T}^*N \mright) \otimes E \right) \cong \Omega^l(N; E)$ for some vector bundle $E$, \textit{i.e.}~$F$ is an $l$-form on $N$ with values in $E$. The pull-back $\Phi^*F$ by $\Phi$ can be then viewed as an element of $\Gamma\left( \bigwedge_{m=1}^{l} \mleft(\Phi^*\mathrm{T}N\mright)^* \otimes \Phi^*E \right)$.

Do not confuse this pull-back with the pull-back of forms, here denoted by $\Phi^!F$, which is an element of $\Gamma\left( \mleft(\bigwedge_{m=1}^{l} \mathrm{T}^*M \mright) \otimes \Phi^*E \right) \cong \Omega^l(M; \Phi^*E)$ defined by
\ba
\mleft.\mleft(\Phi^!F\mright)(Y_1, \dots, Y_l)\mright|_p
&\coloneqq
F_{\Phi(p)}\mleft(\mathrm{D}_p\Phi\mleft(\mleft.Y_1\mright|_p\mright), \dots, \mathrm{D}_p\Phi\mleft(\mleft.Y_l\mright|_p\mright)\mright)
\ea
for all $p \in M$ and $Y_1, \dots, Y_l \in \mathfrak{X}(M)$. Then
\ba\label{EqPullBackFormelFuerVerschiedeneDefinitionen}
\Phi^!F 
&=
\frac{1}{l!}~
\mleft(\Phi^*F\mright) ( \underbrace{\mathrm{D}\Phi \stackrel{\wedge}{,} \dotsc \stackrel{\wedge}{,} \mathrm{D}\Phi}_{l \text{ times}} )
\ea
by using the anti-symmetry of $F$ and Def.~\ref{def:GradingOfProducts}, \textit{i.e.}
\bas
&\mleft.\frac{1}{l!}~
\Big(\mleft(\Phi^*F\mright) ( \mathrm{D}\Phi \stackrel{\wedge}{,} \dotsc \stackrel{\wedge}{,} \mathrm{D}\Phi ) \Big) (Y_1, \dots, Y_l)\mright|_p \\
&\hspace{1cm}
=
\frac{1}{l!}~
\sum_{\sigma \in S_{l}} \mathrm{sgn}(\sigma) ~ \underbrace{(\Phi^*F)\mleft(\mathrm{D}\Phi\mleft(Y_{\sigma(1)}\mright), \dots, \mathrm{D}\Phi\mleft(Y_{\sigma(l)}\mright)\mright)}_{\mathclap{= \mathrm{sgn}(\sigma) ~ (\Phi^*F)\mleft(\mathrm{D}\Phi\mleft(Y_{1}\mright), \dots, \mathrm{D}\Phi\mleft(Y_{l}\mright)\mright)}}\Big|_p \\
&\hspace{1cm}
=
\frac{1}{l!}~ \underbrace{\mleft( \sum_{\sigma \in S_{l}} 1 \mright)}_{= l!} ~
F_{\Phi(p)}\mleft(\mathrm{D}_p\Phi\mleft(\mleft.Y_{1}\mright|_p\mright), \dots, \mathrm{D}_p\Phi\mleft(\mleft.Y_{l}\mright|_{p}\mright)\mright) \\
&\hspace{1cm}
= \mleft.\mleft(\Phi^!F\mright)(Y_1, \dots, Y_l)\mright|_p
\eas
for all $p \in M$ and $Y_1, \dots, Y_l \in \mathfrak{X}(M)$.
\end{remark}

In case of antisymmetric tensors we of course preserve that.

\begin{propositions}{Graded extensions of antisymmetric tensors}{GradedExtensionPlusAntiSymm}
Let $E_1, E_2 \to N$ be real vector bundles of finite rank over a smooth manifold $N$, $F \in \Omega^2(E_1; E_2)$. Then
\ba
F \mleft( A \stackrel{\wedge}{,} B \mright)
&=
-\mleft( -1 \mright)^{km}
F \mleft( B \stackrel{\wedge}{,} A \mright)
\ea 
for all $A \in \Omega^k(N; E_1)$ and $B \in \Omega^m(N; E_2)$ ($k,m \in \mathbb{N}_0$). Similarly extended to all $F \in \Omega^l(E_1; E_2)$.
\end{propositions}

\begin{remark}
\leavevmode\newline
This is a generalization of similar relations just using the Lie algebra bracket $\mleft[ \cdot, \cdot\mright]_{\mathfrak{g}}$ of a Lie algebra $\mathfrak{g}$, see \cite[\S 5, first statement of Exercise 5.15.14; page 316]{hamilton}.
\end{remark}

\begin{proof}
\leavevmode\newline
Trivial by using Eq.~\eqref{CoordExprOfGradedExtension}.
\end{proof}

We also need to know what a Lie algebroid is, a generalization of both, tangent bundles and Lie algebras; this concept will just be defined, refer to the references for thorough discussions of these definitions, especially \cite{mackenzieGeneralTheory} and \cite[\S VII; page 113ff.]{DaSilva}.

\begin{definitions}{Lie algebroid, \newline \cite[\S 3.3, first part of Definition 3.3.1; page 100]{mackenzieGeneralTheory}}{test}
Let $E \to N$ be a real vector bundle of finite rank. Then $E$ is a smooth Lie algebroid if there is a bundle map $\rho: E \to \mathrm{T}N$, called the \textbf{anchor}, and a Lie algebra structure on $\Gamma(E)$ with Lie bracket $\mleft[ \cdot, \cdot \mright]_E$ satisfying
\ba
  \mleft[\mu, f \nu\mright]_E = f \mleft[\mu, \nu\mright]_E + \mathcal{L}_{\rho(\mu)}(f) ~ \nu
\label{eq:E-Leibniz}
\ea
for all $f \in C^\infty(N)$ and $\mu, \nu \in \Gamma(E)$, where $\mathcal{L}_{\rho(\mu)}(f)$ is the action of the vector field $\rho(\mu)$ on the function $f$ by derivation. We will sometimes denote a Lie algebroid by $\mleft( E, \rho, \mleft[ \cdot, \cdot \mright]_E \mright)$.
\end{definitions}

Tangent bundles and bundles of Lie algebras are canonical examples of Lie algebroids, their anchor is the identity and zero, respectively. The important example for us is a mixture of those examples:

\begin{propositions}{Action Lie algebroids, \cite[\S 16.2, Example 5; page 114]{DaSilva}}{ActionLieoidsAreOids}
Let $\mleft(\mathfrak{g}, \mleft[\cdot, \cdot \mright]_{\mathfrak{g}}\mright)$ be some Lie algebra equipped with a Lie algebra action $\gamma: \mathfrak{g} \to \mathfrak{X}(N)$ on a smooth manifold $N$. Then there is a unique Lie algebroid structure on $E = N \times \mathfrak{g}$ such that we have
\ba
\rho(\nu)
&=
\gamma(\nu),
\\
\mleft[\mu, \nu\mright]_E
&=
\mleft[\mu, \nu\mright]_{\mathfrak{g}}
\ea
for all constant sections $\mu, \nu \in \Gamma(E)$. We call this structure \textbf{action Lie algebroid}.
\end{propositions}


\section{Pullbacks of Lie algebroid connections}\label{PullbacksOfEConnections}

We want to formulate infinitesimal gauge transformations by using Lie algebroid connections:

\begin{definitions}{$E$-connection, \cite[variation of Definition 5.2.5; page 186]{mackenzieGeneralTheory}}{Econnection}
Let $E \to N$ be a Lie algebroid over a smooth manifold $N$ and $V \to N$ be a vector bundle over $N$. An $E$-connection on the vector bundle $V$ is an $\mathbb{R}$-bilinear map ${}^E\nabla: \Gamma(E) \times \Gamma(V) \to \Gamma(V)$, $(\nu, v) \mapsto {}^E\nabla_\nu v$, such that 
\ba
{}^E\nabla_{f\nu} v
&=
f ~ {}^E\nabla_\nu v,
\\
{}^E\nabla_{\nu} (fv)
&=
f ~ {}^E\nabla_\nu v
	+ \mathcal{L}_{\rho(\nu)}(f) ~ v
\ea
for all $\nu \in \Gamma(E),$ $v\in \Gamma(V)$ and $f \in C^\infty(N)$.
\end{definitions}

\begin{examples}{Canonically induced $E$-connection, \newline \cite[first example in Example 2.8]{ELeviCivita}}{NablaRhoConnection}
Let $E \to N$ be a Lie algebroid over a smooth manifold $N$ and $V \to N$ be a vector bundle over $N$, equipped with a vector bundle connection $\nabla$. Then define ${}^E\nabla$ on $V$ by 
\ba
{}^E\nabla_\mu
&\coloneqq 
\nabla_{\rho(\mu)}
\ea
for all $\mu \in \Gamma(E)$. This is a canonical example of an $E$-connection which we will denote as $\nabla_\rho$.
\end{examples}

\begin{remark}
\leavevmode\newline
Many notions of vector bundle connections carry over to Lie algebroid connections, like the definition of curvatures $R_{{}^E\nabla}$ and torsions $t_{{}^E\nabla}$ (for $E$-connections on $E$ for the latter); see for example \cite{mackenzieGeneralTheory} for a general introduction, or \cite{ELeviCivita} for notions of $E$-Levi-Civita connections and other related terms of Riemannian geometry. 
\end{remark}

We want to show that a pullback of such connections is in general possible if using the following type of morphism:

\begin{definitions}{Anchor-preserving vector bundle morphism, \newline \cite[\S 4.3, Equation (22); page 157]{mackenzieGeneralTheory}}{DefOfAnchorPreservingStuff}
Let $E_i\stackrel{\pi_i}{\to} N_i$ ($i \in \{1,2\}$) be two Lie algebroids over smooth manifolds $N_i$. Then we say that a vector bundle morphism $\xi: E_1 \to E_2$ over a smooth map $f: N_1 \to N_2$\footnote{That means $\pi_2 \circ \xi = f \circ \pi_1$.} is \textbf{anchor-preserving} if it satisfies
\ba\label{EqFuerAnchorBundleMorphisms}
\mathrm{D}f \circ \rho_{E_1}
&=
\rho_{E_2} \circ \xi.
\ea
\end{definitions}

\begin{remarks}{Notations and base-preserving morphisms}{SomeExtraNotationForAnchorBundleMorphs}
As it is well-known, $\xi$ does not necessarily induce a map $\Gamma(E_1) \to \Gamma(E_2)$ on sections, that depends on how $f$ is structured. However, we have 
\bas
\pi_2 \bigl( \xi(\nu) \bigr)
&=
f\bigl( \underbrace{\pi_1 (\nu)}_{= \mathds{1}_{N_1}} \bigr)
=
f
\eas
for all $\nu \in \Gamma(E_1)$, such that $\xi$ induces a tensor on $\Gamma(E_1) \to \Gamma(f^*E_2)$ (the $C^\infty(N_1)$-linearity follows trivially); see \textit{e.g.}~\cite[paragraph after Propositon 7.10]{meinrenkenlie}, too, or our arguments before Def.~\ref{def:GradingOfProducts}. Thence, as we already argued, $\mathrm{D}f \in \Omega^1(N_1; f^*\mathrm{T}N_2)$, which is also trivially an anchor-preserving vector bundle morphism over $f$. This is why we write equations like Eq.~\eqref{EqFuerAnchorBundleMorphisms} often as
\ba
\mathrm{D}f \circ \rho_{E_1}
&=
(f^*\rho_{E_2}) \circ \xi
\ea
when we view that condition as an equation for sections, in order to emphasize the relationship with the pullback;
recall that $f^*\rho_{E_2}: \Gamma(f^*E_2) \to \Gamma(f^*\mathrm{T}N_2)$. However, sometimes we may also omit the notation of that pullback in this case such that this is an optional notation.
\end{remarks}

Finally let us discuss pullbacks of Lie algebroid connections.

\begin{propositions}{Pullbacks of Lie algebroid connections by anchor-preserving morphisms}{GeneralPullbackAnchorPreserving}
Let $E_i \to N_i$ ($i \in\{1,2\}$) be two Lie algebroids over smooth manifolds $N_i$ with anchors $\rho_{E_i}$, $V \to N_2$ a vector bundle, and ${}^{E_2}\nabla$ an $E_2$-connection on $V$. Also fix an anchor-preserving vector bundle morphism $\xi: E_1 \to E_2$ over a smooth map $f: N_1 \to N_2$. Then there is a unique $E_1$-connection $f^*\mleft( {}^{E_2}\nabla \mright)$ on $f^*V$ with
\ba\label{GeneralPullbackDef}
\mleft(f^*\mleft( {}^{E_2}\nabla \mright)\mright)_\nu (f^*v)
&=
f^*\mleft(
	{}^{E_2}\nabla_{\xi(\nu)} v
\mright)
\ea
for all $v \in \Gamma(V)$ and $\nu \in \Gamma(E_1)$.
\end{propositions}

\begin{remark}
\leavevmode\newline
This result is motivated by \cite[Example 7.7]{meinrenkenlie} where it is shown that there is a 1:1 correspondence of Lie algebroid paths and anchor-preserving morphisms; in literature one usually uses Lie algebroid paths for pullbacks along curves, see also \cite[\S 2, discussion around Definition 2.4]{ELeviCivita}.

We may sometimes rewrite the right-hand side of Eq.~\eqref{GeneralPullbackDef} to
\bas
\mleft(f^*\mleft(
	{}^{E_2}\nabla v
\mright)\mright)\bigl(\xi(\nu)\bigr)
\eas
to emphasise that $\xi(\nu) \in \Gamma(f^*E_2)$, where we view ${}^{E_2}\nabla v$ as a vector bundle morphism $E_2 \to V$, hence, $f^*\mleft( 	{}^{E_2}\nabla v \mright)$ gives a map $\Gamma(f^*E_2) \to \Gamma(f^*V)$.
\end{remark}

\begin{proof}
\leavevmode\newline
The proof is basically the same as for pullbacks of vector bundle connections. The idea is the following: As usual, the idea is that the pullbacks of sections, $f^*v$ ($v \in \Gamma(V)$), generate $\Gamma(f^*V)$. Thus, Eq~\eqref{GeneralPullbackDef} defines the $E_1$-connection uniquely, that is, sections $w$ of $f^*V$ are determined by sums of elements of the form $h \cdot f^*v$, $h \in C^\infty(N_1)$, and by the Leibniz rule any $E_1$-connection $f^*\mleft( {}^{E_2}\nabla \mright)$ satisfying Eq.~\eqref{GeneralPullbackDef} also satisfies
\bas
\mleft(f^*\mleft( {}^{E_2}\nabla \mright)\mright)_{\nu} (h ~ f^*v)
&=
\mathcal{L}_{\rho_{E_1}(\nu)}(h) ~ f^* v
	+ h ~ f^*\mleft( {}^{E_2}\nabla_{\xi(\nu)} v \mright)
\eas
for all $\nu \in \Gamma(E_1)$, such that uniqueness follows by linearity, assuming existence is given; for the existence one can simply take this equation as a possible definition for $f^*\mleft( {}^{E_2}\nabla \mright)$. That is, let $f^*\mleft( {}^{E_2}\nabla \mright)$ locally be defined by
\ba\label{FullPulbackGConnection}
\mleft(f^*\mleft( {}^{E_2}\nabla \mright)\mright)_{\nu} w
&\coloneqq
\mathcal{L}_{\rho_{E_1}(\nu)}\mleft(w^a\mright) ~ f^*e_a
	+ w^a ~ f^*\mleft( {}^{E_2}\nabla_{\xi(\nu)} e_a \mright)
\ea
for all $w = w^a ~ f^*e_a$,
where $\mleft( e_a \mright)_a$ is a local frame of $V$. Linearity in all arguments and the Leibniz rule follow by construction. Also observe that for a function $h \in C^\infty(N_1)$ and $v \in \Gamma(V)$ we can calculate
\ba
f^*\mleft({}^{E_2}\nabla_{\xi(\nu)} (h v)\mright)
&=
(h\circ f) ~ f^*\mleft({}^{E_2}\nabla_{\xi(\nu)}  v\mright)
	+ f^*\Bigl( \underbrace{\mathcal{L}_{(\rho_{E_2}\circ\xi)(\nu)}}
	_{\mathclap{ \stackrel{\text{Def.~\ref{def:DefOfAnchorPreservingStuff}}}{=}~ \mathcal{L}_{ \mleft(\mathrm{D}f \circ \rho_{E_1}\mright)(\nu)}  }}
	(h)\Bigr) ~ f^*v \nonumber
\\\label{ImportantEquationToCheckForPullbacks}
&=
(h\circ f) ~ f^*\mleft({}^{E_2}\nabla_{\xi(\nu)}  v\mright)
	+ \mathcal{L}_{\rho_{E_1}(\nu)}(h \circ f) ~ f^*v
\ea
for all $\nu \in \Gamma(E_1)$, using
\bas
f^*\Bigl(
	\mathcal{L}_{ \mleft(\mathrm{D}f \circ \rho_{E_1}\mright)(\nu)} (h)
\Bigr)
&=
\mleft(f^*\mathrm{d}h\mright)\bigl(\mleft(\mathrm{D}f \circ \rho_{E_1}\mright)(\nu)\bigr)
=
\underbrace{\mleft(f^!\mathrm{d}h\mright)}_{\mathclap{ = \mathrm{d}f^! h }}\bigl( \rho_{E_1}(\nu) \bigr)
=
\mathcal{L}_{\rho_{E_1}(\nu)}(h \circ f).
\eas
Thus, 
\bas
\mleft(f^*\mleft( {}^{E_2}\nabla \mright)\mright)_{\nu} \mleft( f^*v \mright)
&\stackrel{\eqref{FullPulbackGConnection}}{=}
\mathcal{L}_{\rho_{E_1}(\nu)}\mleft(v^a \circ f\mright) ~ f^*e_a
	+ \mleft(v^a \circ f\mright) ~ f^*\mleft( {}^{E_2}\nabla_{\xi(\nu)} e_a \mright)
\stackrel{\eqref{ImportantEquationToCheckForPullbacks}}{=}
f^*\mleft({}^{E_2}\nabla_{\xi(\nu)} v\mright),
\eas
so, Eq.~\eqref{GeneralPullbackDef} is satisfied.
Finally, by Eq.~\eqref{ImportantEquationToCheckForPullbacks} it also follows that \eqref{FullPulbackGConnection} is independent of the chosen frame and, thus, globally defined. To see this, observe that any other frame $\mleft( g_b \mright)_b$ of $E$, intersecting the neighbourhood of $\mleft( e_a \mright)_a$, is given by $e_a = M_a^b g_b$, where $M_a^b$ is a local invertible matrix function on $N$. Then
\bas
w
&=
w^a ~ f^*e_a
=
\mleft(M_a^b \circ f \mright) w^a ~ f^*g_b
\eqqcolon
\tilde{w}^b ~ f^*g_b,
\eas
such that $w^a = \mleft(\mleft( M^{-1} \mright)^a_b \circ f \mright) \tilde{w}^b$, and, thus, as a direct consequence of Eq.~\eqref{ImportantEquationToCheckForPullbacks},
\bas
\mleft(f^*\mleft( {}^{E_2}\nabla \mright)\mright)_{\nu} w
&~~\stackrel{\mathclap{\eqref{FullPulbackGConnection}}}{=}~~
\mathcal{L}_{\rho_{E_1}(\nu)}\mleft(w^a\mright) ~ f^*e_a
	+ w^a ~ f^*\mleft( {}^{E_2}\nabla_{\xi(\nu)} e_a \mright)
\\
&=
\mathcal{L}_{\rho_{E_1}(\nu)}\mleft(\mleft(\mleft( M^{-1} \mright)^a_d \circ f \mright) \tilde{w}^d\mright) ~ f^*\mleft( M_a^b g_b \mright)
\\
&\hspace{1cm}
	+ \mleft(\mleft( M^{-1} \mright)^a_d \circ f \mright) \tilde{w}^d ~ f^*\mleft( {}^{E_2}\nabla_{\xi(\nu)} \mleft( M_a^b g_b \mright) \mright)
\\
&\stackrel{\mathclap{ \text{Eq.~\eqref{ImportantEquationToCheckForPullbacks}} }}{=}\quad~
\mathcal{L}_{\rho_{E_1}(\nu)}\mleft(\tilde{w}^b\mright) ~ f^*g_b
	+ \tilde{w}^b ~ f^*\mleft( {}^{E_2}\nabla_{\xi(\nu)} g_b \mright)
\\
&\hspace{1cm}
	+ \tilde{w}^d ~ \biggl(
		- \mathcal{L}_{\rho_{E_1}(\nu)}\mleft(M^b_a \circ f \mright) ~ \mleft( \mleft( M^{-1} \mright)^a_d \circ f \mright)
\\
& \hphantom{+c\tilde{\mu}^d ~ \biggl(} \hspace{2cm}
		+ \mleft(\mleft( M^{-1} \mright)^a_d \circ f \mright) ~ \mathcal{L}_{\rho_{E_1}(\nu)}\mleft( M^b_a \circ f \mright)
	\biggr) ~ f^*g_b
\\
&=
\mathcal{L}_{\rho_{E_1}(\nu)}\mleft(\tilde{w}^b\mright) ~ f^*g_b
	+ \tilde{w}^b ~ f^*\mleft( {}^{E_2}\nabla_{\xi(\nu)} g_b \mright),
\eas
using formulas of the differential of the inverse like $M ~ \mathrm{d}M^{-1} = - \mathrm{d}M ~ M^{-1}$ (similar for $f^*M = M \circ f$). Hence, Def.~\eqref{FullPulbackGConnection} is frame-independent, and this finishes the proof.
\end{proof}

\begin{remarks}{Essential condition for pullbacks of connections}{ImportantRemarkAboutPullbacks}
Observe that the essential part of the proof is Eq.~\eqref{ImportantEquationToCheckForPullbacks}, everything follows either by this equation or by the standard construction in \eqref{FullPulbackGConnection}. This will be important now because we are going to generalise this statement. To avoid doing the same all over again, we will just refer to this proof and remark, essentially one only needs to check something like Eq.~\eqref{ImportantEquationToCheckForPullbacks}. Eq.~\eqref{ImportantEquationToCheckForPullbacks} essentially proves that the Leibniz rule inherited by ${}^{E_2}\nabla$ is in alignment with the Leibniz rule of $E_1$-connections on $f^*V \to N_1$.

Eq.~\eqref{ImportantEquationToCheckForPullbacks} also motivates why anchor-preserving vector bundle morphisms are precisely the objects one needs in order to provide pullbacks of $E_2$-connections.
\end{remarks}

Using this, we now generalise this notion at a "fixed direction of differentiation".

\begin{corollaries}{Pullbacks of connections just differentiating along one vector field}{VeryGeneralPullbackConnection}
Let $E_i \to N_i$ ($i \in\{1,2\}$) be two Lie algebroids over smooth manifolds $N_i$ and with anchors $\rho_{E_i}$, $V \to N_2$ a vector bundle, and ${}^{E_2}\nabla$ an $E_2$-connection on $V$. Moreover, let $f \in C^\infty(N_1;N_2)$, $\nu_1 \in \Gamma(E_1)$ and $\nu_2 \in \Gamma(f^*E_2)$ such that
\ba\label{WeakAnchorPreserv}
\mathrm{D}f\bigl(\rho_{E_1}(\nu_1)\bigr)
&=
\mleft(f^*\rho_{E_2}\mright)(\nu_2).
\ea

Then there is a unique $\mathbb{R}$-linear operator $\delta_{\nu_1}: \Gamma(f^*V) \to \Gamma(f^*V)$ with
\ba
\delta_{\nu_1}(h s)
&=
\mathcal{L}_{\rho_{E_1}(\nu_1)}(h) ~ s
	+ h ~ \delta_{\nu_1} s,
\\
\delta_{\nu_1} (f^*v)
&=
f^*\mleft(
	{}^{E_2}\nabla_{\nu_2} v
\mright)
\ea
for all $s \in \Gamma(f^*V)$, $v \in \Gamma(V)$ and $h \in C^\infty(N_1)$.
\end{corollaries}

\begin{remarks}{Commutating diagram behind pullbacks}{CommutingDiagramOfPullbacks}
Recall Remark \ref{rem:SomeExtraNotationForAnchorBundleMorphs}, the pullback in $\mleft(f^*\rho_{E_2}\mright)(\nu_2)$ in Eq.~\eqref{WeakAnchorPreserv} is just for emphasizing that $\nu_2$ is a section along $f$; one can omit this in the notation, especially if one views sections like $\nu_2$ as a map $N_1 \to E_2$. Then we can equivalently write
\ba
\mathrm{D}f \circ \rho_{E_1}(\nu_1)
&=
\rho_{E_2} \circ \nu_2,
\ea
that is equivalent to that the following diagram commutes
\begin{center}
	\begin{tikzcd}
		N_1 \arrow{r}{\nu_2} \arrow[d, "\rho_{E_1}(\nu_1)", swap]	& E_2 \arrow{d}{\rho_{E_2}} 
		\\
		\mathrm{T}N_1 \arrow{r}{\mathrm{D}f} & \mathrm{T}N_2
	\end{tikzcd}
\end{center}
\end{remarks}

\begin{remark}\label{JustLieDerivativeForGeneralPullbackAndlineBundle}
\leavevmode\newline
\indent $\bullet$ In general one may want to write $\delta_{\nu_1} = \mleft(f^*\mleft( {}^{E_2}\nabla \mright)\mright)_{\nu_1}$, because it is precisely this by uniqueness if a general pullback is possible. But to avoid confusion about the existence of a general pullback we will stick with $\delta_{\nu_1}$, and it will be clear by context which connection and $\nu_2$ is used for the definition of $\delta_{\nu_1}$.

$\bullet$ In the case of $V = N_2 \times \mathbb{R}$, the trivial line bundle over $N_2$, we canonically use ${}^{E_2}\nabla \coloneqq \nabla^0_{\rho_{E_2}}$, where $\nabla^0 \coloneqq \mathrm{d}$, so, ${}^{E_2}\nabla_{\rho_{E_2}(\nu_2)} = \mathcal{L}_{\rho_{E_2}(\nu_2)}$. Then one can trivially show that, using uniqueness,
\bas
\delta_{\nu_1}
&=
\mathcal{L}_{\rho_{E_1}(\nu_1)}.
\eas
\end{remark}

\begin{proof}[Proof of Cor.~\ref{cor:VeryGeneralPullbackConnection}]
\leavevmode\newline
That is precisely the same proof as in Prop.~\ref{prop:GeneralPullbackAnchorPreserving}; the only difference is just the meaning, $\nu_i$ are fixed sections, but that does not matter in the calculations. Eq.~\eqref{WeakAnchorPreserv} is just the condition about anchor-preservation in the case of a fixed pair of sections, and one uses this equation in the same fashion to how we used an anchor-preserving morphism previously; essentially replace $\nu$ with $\nu_1$ and $\xi(\nu)$ with $\nu_2$ in the proof of Prop.~\ref{prop:GeneralPullbackAnchorPreserving}, and Def.~\eqref{FullPulbackGConnection} is then the definition of $\delta_{\nu_1}$, so, $\delta_{\nu_1}$ plays the role of $\mleft(f^*\mleft( {}^{E_2}\nabla \mright)\mright)_{\nu}$ of the previous proof.
\end{proof}

\section{Infinitesimal gauge transformations}\label{ClassicalGaugeTrafoDiscussion}

We want to express the definition of infinitesimal gauge transformations, as it usually arises in gauge theory, by using Cor.~\ref{cor:VeryGeneralPullbackConnection}. Let us first state how we usually understand infinitesimal gauge transformations in the context of infinitesimal\footnote{That is we assume a gauge or equivalently a trivial principal bundle, too.} Yang-Mills-Higgs gauge theory.

\subsection{Basic setup}\label{ClassicalBasicSetup}

\begin{definitions}{The space of fields}{ClassicSpaceofFieldsAgain}
Let $M$ be a smooth manifold, $W$ a vector space, and $\mathfrak{g}$ a Lie algebra. Then we define the \textbf{space of fields} by
\ba
\mathfrak{M}_{\mathfrak{g}}
&\coloneqq
\mathfrak{M}_{\mathfrak{g}}(M; W)
\coloneqq
\left\{ (\Phi, A)
~\middle|~
\Phi \in C^\infty(M;W) \text{ and } A \in \Omega^1(M; \mathfrak{g})
\right\}.
\ea
\end{definitions}

\begin{definitions}{Infinitesimal gauge transformation of the Higgs field and the field of gauge bosons, \newline \cite[infinitesimal version of Theorem 5.3.9, see also comment afterwards; page 269f.]{hamilton} and \cite[infinitesimal version of Theorem 5.4.4; page 273]{hamilton}}{ClassicTrafos}
Let $M$ be a smooth manifold, $W$ a vector space, and $\mathfrak{g}$ a Lie algebra, equipped with a Lie algebra representation $\psi: \mathfrak{g} \to \mathrm{End}(W)$. Moreover, let $\epsilon \in C^\infty(M; \mathfrak{g})$.

Then we define the \textbf{infinitesimal gauge transformation $\delta_\epsilon \Phi$ of the Higgs field $\Phi \in C^\infty(M;W)$} also as an element of $C^\infty(M; W)$ by
\ba
\delta_\epsilon \Phi
&\coloneqq
\psi(\epsilon)(\Phi).
\ea
The \textbf{infinitesimal gauge transformation $\delta_\epsilon A$ of the field of gauge bosons $A \in \Omega^1(M; \mathfrak{g})$} is defined as an element of $\Omega^1(M; \mathfrak{g})$ by
\ba
\delta_\epsilon A
&\coloneqq
\mleft[ \epsilon, A \mright]_{\mathfrak{g}}
	- \mathrm{d}\epsilon.
\ea
\end{definitions}

With that one can define the infinitesimal gauge transformation of functionals depending on $\mathfrak{M}_{\mathfrak{g}}$.

\begin{definitions}{Infinitesimal gauge transformation of functionals, \newline \cite[motivated by statements like Theorem 7.3.2; page 414ff.]{hamilton}}{ClassFunctionalGaugeTrafoBlag}
Let $M$ be a smooth manifold, $W, K$ vector spaces, and $\mathfrak{g}$ a Lie algebra, equipped with a Lie algebra representation $\psi: \mathfrak{g} \to \mathrm{End}(W)$. Moreover, let $\epsilon \in C^\infty(M; \mathfrak{g})$.

Then we define the \textbf{infinitesimal gauge transformation $\delta_\epsilon L$ of $L: \mathfrak{M}_{\mathfrak{g}}(M; W) \to \Omega^k(M; K)$ ($k \in \mathbb{N}_0$)} as a map $\mathfrak{M}_{\mathfrak{g}}(M; W) \to \Omega^k(M;K)$ by
\ba
\mleft(\delta_\epsilon L\mright)(\Phi, A)
&\coloneqq
\mleft.\frac{\mathrm{d}}{\mathrm{d}t}\mright|_{t=0}
\mleft[ t \mapsto
	L\mleft(
		\Phi + t \delta_\epsilon \Phi,
		A + t \delta_\epsilon A
	\mright)
\mright]
\ea
for $t \in \mathbb{R}$, where $\mathrm{d}/\mathrm{d}t$ is defined using the canonical flat connection on $M \times K \to M$.
\end{definitions}

\subsection{Lie algebroid connections and infinitesimal gauge transformations}\label{ClassicalWithLieAlgebroidConnections}

Given a Lie algebra $\mathfrak{g}$ and a Lie algebra representation $\psi: \mathfrak{g} \to \mathrm{End}(W)$ on a vector space $W$ one has a natural sense of Lie algebra action $\gamma: \mathfrak{g} \to \mathfrak{X}(W)$ given by 
\ba\label{ActionAndRep}
\gamma(X)_v &\coloneqq - \psi(X)(v)
\ea
for all $X \in \mathfrak{g}$ and $v \in W$; see \textit{e.g.}~\cite[generalisation of parts of Example 3.4.2; page 143f.]{hamilton}. By Prop.~\ref{prop:ActionLieoidsAreOids} there is a natural Lie algebroid structure on the trivial vector bundle $E \coloneqq W \times \mathfrak{g}$ over $W$, an action Lie algebroid encoding $\mathfrak{g}$ and $\psi$. Hence, we want to express Def.~\ref{def:ClassFunctionalGaugeTrafoBlag} now using this action Lie algebroid. For this we will use an $E$-connection on $K$, where $K$ was the vector space a functional has values in; given an $E$-connection we will apply Cor.~\ref{cor:VeryGeneralPullbackConnection} to define $\delta_\epsilon$.

Since we want to generalize the following notions to vector bundles later, we already want to be a bit careful in our notation. Using an $E$-connection on $K$ requires to view $K$ as a trivial vector bundle $W \times K$ over $W$. However, the space of functionals $L: \mathfrak{M}_{\mathfrak{g}}(M; W) \to \Omega^k(M; K)$ ($k \in \mathbb{N}_0$) is actually $\Omega^{k,0}\bigl(M \times \mathfrak{M}_{\mathfrak{g}}(M; W); K \bigr)$, where we mean with $(k,0)$ a $k$ degree just with respect to the factor $M$: 
\bas
\Omega^{p,q} \mleft( M \times \mathfrak{M}_{\mathfrak{g}}; K \mright)
\coloneqq
\Gamma\mleft(
	\pi_1^*\mleft(\bigwedge^p \mathrm{T}^*M\mright) \otimes \pi_2^*\mleft(\bigwedge^q \mathrm{T}^*\mathfrak{M}_{\mathfrak{g}}\mright) \otimes K
\mright),
\eas
where $p, q \in \mathbb{N}_0$, and $\pi_1$ and $\pi_2$ are the projections onto the first and second factor of $M \times \mathfrak{M}_{\mathfrak{g}}$, respectively.
In $\Omega^{k,0}\bigl(M \times \mathfrak{M}_{\mathfrak{g}}(M; W); K \bigr)$ the vector space $K$ is viewed as a trivial vector bundle $M \times \mathfrak{M}_{\mathfrak{g}}(M; W) \times K$ over $M \times \mathfrak{M}_{\mathfrak{g}}(M; W)$. A pullback of $W \times K$ using any map from $M \times \mathfrak{M}_{\mathfrak{g}}(M; W)$ to $W$ will be trivially isomorphic to $M \times \mathfrak{M}_{\mathfrak{g}}(M; W) \times K$, and we need to make such a pullback of a given $E$-connection ${}^E\nabla$ on $W \times K$ because we want that a derivation induced by ${}^E\nabla$ acts on functionals $L$. For this we want to use Cor.~\ref{cor:VeryGeneralPullbackConnection} using the following base map:

\begin{definitions}{The evaluation map}{FirstAttemptOfEvaluationMap}
Let $M$ be a smooth manifold, $W$ a vector space, and $\mathfrak{g}$ a Lie algebra. Then we define the \textbf{evaluation map} $\mathrm{ev}: M \times \mathfrak{M}_{\mathfrak{g}}(M; W) \to W$ by
\ba
\mathrm{ev}(p, \Phi, A)
&\coloneqq
\Phi(p)
\ea
for all $(p, \Phi, A) \in M \times \mathfrak{M}_{\mathfrak{g}}(M; W)$.
\end{definitions}

As argued before, we can view the vector space $K$ as a trivial vector bundle over $M \times \mathfrak{M}_{\mathfrak{g}}(M; W)$, but we can do the same for $W$ as base, so, $K$ can also be viewed as trivial vector bundle over $W$, and elements of $K$ are just constant sections of such a bundle. For bookkeeping, let us denote with $\iota_{M \times \mathfrak{M}_{\mathfrak{g}}(M; W)}$ and $\iota_W$ maps $K \hookrightarrow \Gamma(M \times \mathfrak{M}_{\mathfrak{g}}(M; W) \times K)$ and $K \hookrightarrow \Gamma(W \times K)$, respectively, which embed elements of $K$ canonically into the space of constant sections of the trivial bundles $M \times \mathfrak{M}_{\mathfrak{g}}(M; W) \times K$ and $W \times K$, respectively. Then take a functional $L \in \Omega^{k,0}(M \times \mathfrak{M}_{\mathfrak{g}}(M; W); K)$ ($k \in \mathbb{N}_0$) and a basis $\mleft( e_a \mright)_a$ of $K$. Then we can express $L$ as, making use of $\iota_{M \times \mathfrak{M}_{\mathfrak{g}}(M; W)}$,
\bas
L
&=
L^a \otimes \iota_{M \times \mathfrak{M}_{\mathfrak{g}}(M; W)}(e_a),
\eas
where $L^a \in \Omega^{k,0}(M \times \mathfrak{M}_{\mathfrak{g}}(M; W))$. We can trivially identify
\bas
\iota_{M \times \mathfrak{M}_{\mathfrak{g}}(M; W)}(e_a)
&=
\mathrm{ev}^*\bigl( \iota_W (e_a) \bigr)
\eas
because $e_a$ is viewed as a constant section in both trivial vector bundles. Thus, we can also write
\bas
L
&=
L^a \otimes \mathrm{ev}^*\bigl( \iota_W(e_a) \bigr)
\eqqcolon
\iota(L),
\eas
and that interpretation of $L$ we denote as $\iota(L)$ for bookkeeping reasons, called the \textbf{bookkeeping trick}. Observe
\bas
\mathrm{ev}^*\bigl( \iota_W(e_a) \bigr)
&\in
\Gamma(\mathrm{ev}^*(W\times K)),
\eas
therefore we write $\iota(L)$ to give an accentuation on when we view $L$ as an element of $\Omega^{k,0}\bigl(M \times \mathfrak{M}_{\mathfrak{g}}(M; W); \mathrm{ev}^*(W \times K)\bigr)$. We may also write
\bas
\bigl( \iota(L) \bigr)(\Phi, A)
&=
\iota\bigl( L(\Phi, A) \bigr)
=
L^a \otimes \Phi^*\bigl( \iota_W(e_a) \bigr)
\eas
for all $(\Phi, A) \in \mathfrak{M}_{\mathfrak{g}}$.

Keeping that in mind, it is now clearer what we talked about before Def.~\ref{def:FirstAttemptOfEvaluationMap}. We start with a Lie algebroid connection on $W \times K$, and make a pullback with the evaluation map; as we have seen in Section \ref{PullbacksOfEConnections}, this is in general possible if the evaluation map can be lifted to an anchor-preserving vector bundle morphism. However, this is in general not possible, but infinitesimal gauge transformations are not derivations along all possible vector fields in $M \times \mathfrak{M}_{\mathfrak{g}}(M; W)$, just along certain ones; given a parameter $\epsilon$ as in Def.~\ref{def:ClassicSpaceofFieldsAgain}, the vector field is actually given by $\Psi_\epsilon|_{(p, \Phi, A)} \coloneqq (0, \delta_\epsilon \Phi, \delta_\epsilon A)$ for all $(p, \Phi, A) \in M \times \mathfrak{M}_{\mathfrak{g}}(M; W)$. Thence, fixing an $\epsilon$ we just need one vector field, $\Psi_\epsilon$, and for this we can use Cor.~\ref{cor:VeryGeneralPullbackConnection}, avoiding the question about a lift of the evaluation map.

We also introduce the shorter notation $\Psi_\epsilon \coloneqq (\delta_\epsilon \Phi, \delta_\epsilon A)$, especially in order to emphasize that it is a vector field along $\mathfrak{M}_{\mathfrak{g}}(M; W)$ (canonically embedded into $\mathfrak{X}\bigl(M \times \mathfrak{M}_{\mathfrak{g}}(M; W) \bigr)$.

In the context of Cor.~\ref{cor:VeryGeneralPullbackConnection} we have $E_1 \coloneqq \mathrm{T}\bigl( M \times \mathfrak{M}_{\mathfrak{g}}(M;W) \bigr)$, $E_2 \coloneqq E = W \times \mathfrak{g}$ (action Lie algebroid over $W$), and $f \coloneqq \mathrm{ev}$. Then we need Eq.~\eqref{WeakAnchorPreserv}, that is,
\bas
\mathrm{D}\mathrm{ev}( \Psi_\epsilon)
&=
\mleft(\mathrm{ev}^*\rho_{E}\mright)(\nu)
\eas
for some $\nu \in \Gamma(\mathrm{ev}^*E)$.

\begin{corollaries}{Gauge transformation of the Higgs field as condition for pullbacks}{GaugeTrafoAsPullbackCond}
Let $M$ be a smooth manifold, $W$ a vector space, and $\mathfrak{g}$ a Lie algebra, equipped with a Lie algebra representation $\psi: \mathfrak{g} \to \mathrm{End}(W)$. We also denote with $E \coloneqq W \times \mathfrak{g}$ the associated action Lie algebroid over $W$. Then a vector field $\Psi \coloneqq \mleft(0, \Psi^{(\Phi)}, \Psi^{(A)} \mright) \in \mathfrak{X}\bigl(M \times \mathfrak{M}_{\mathfrak{g}}(M;W)\bigr)$ satisfies
\bas
\exists \varepsilon \in \Gamma ( \mathrm{ev}^*E ): ~
\mathrm{D}\mathrm{ev}( \Psi)
&=
- \mleft(\mathrm{ev}^*\rho_{E}\mright)(\varepsilon)
\eas
if and only if
\ba\label{ClassicalGaugeTrafoButABitMoreGeneral}
\exists \varepsilon \in \Gamma ( \mathrm{ev}^*E ): ~
\Psi^{(\Phi)}
&=
- \mleft(\mathrm{ev}^*\rho_{E}\mright)(\varepsilon).
\ea

In this case, for a given pair $(\Phi, A) \in \mathfrak{M}_{\mathfrak{g}}(M;W)$, we call $\delta_\epsilon \Phi \coloneqq \mleft.\Psi^{(\Phi)}\mright|_{(\Phi,A)} = - \mleft(\Phi^*\rho_{E}\mright)(\epsilon)$ the \textbf{infinitesimal gauge transformation of the Higgs field $\Phi$}, where $\epsilon \in \Gamma(\Phi^*E)$ is 	given by $\epsilon_p \coloneqq \varepsilon(p, \Phi, A)$ for all $p \in M$.
\end{corollaries}

\begin{remark}\label{RemarkAboutThatTheCLassicalFormulaCanBeWrittenDifferently}
\leavevmode\newline
Recall that $E$ is an action Lie algebroid, and thus, using Eq.~\eqref{ActionAndRep},
\bas
- \mleft(\Phi^*\rho_{E}\mright)(\epsilon)
&=
\psi(\epsilon)(\Phi),
\eas
which is precisely how we defined the infinitesimal gauge transformation of the Higgs field\footnote{Therefore we introduced the minus sign in front of $\varepsilon$.} in Def.~\ref{def:ClassicTrafos}; also $\Phi^*E \cong M \times \mathfrak{g}$ as trivial vector bundle over $M$ such that $\epsilon \in C^\infty(M;\mathfrak{g})$. Eq.~\eqref{ClassicalGaugeTrafoButABitMoreGeneral} is also a generalization of a similar equation for a gauge transformation given in \cite[paragraph before Equation (10); we have a different sign in $\varepsilon$]{CurvedYMH}. However, $\epsilon$ is here induced by $\varepsilon$ which in general depends on $\mathfrak{M}_{\mathfrak{g}}(M;W)$; this implies that the parameter for the infinitesimal gauge transformation may have a gauge transformation on its own; it is straightforward to extend Def.~\ref{def:ClassicTrafos} to $\varepsilon \in \Gamma(\mathrm{ev}^*E)$ as parameter which we are going to assume in the following without further separate definition, similar for the definition of $\Psi_\epsilon$. This technical nuance and generalisation will be important when we later calculate the commutator of infinitesimal gauge transformations.
\end{remark}

\begin{proof}[Proof of Cor.~\ref{cor:GaugeTrafoAsPullbackCond}]
\leavevmode\newline
Let $\gamma = (\Phi, A): I \to \mathfrak{M}_{\mathfrak{g}}(M;W), t \mapsto \gamma(t)= (\Phi_t, A_t)$ ($I \subset \mathbb{R}$ an open interval containing 0) be the flow of $\mleft(\Psi^{(\Phi)}, \Psi^{(A)} \mright) \in \mathfrak{X}\bigl(\mathfrak{M}_{\mathfrak{g}}(M;W)\bigr)$ through $(\Phi_0, A_0) \in \mathfrak{M}_{\mathfrak{g}}(M;W)$ at $t=0$. Then the local flow of $\Psi$ through $(p, \Phi_0, A_0) \in M \times \mathfrak{M}_E(M;N)$ is given by $(p, \Phi, A)$. Thus,
\bas
\mathrm{D}_{(p, \Phi_0, A_0)}\mathrm{ev}(\Psi)
&=
\mleft. \frac{\mathrm{d}}{\mathrm{d}t} \mright|_{t=0}\mleft(
	\mathrm{ev}(p, \Phi, A)
\mright)
=
\mleft. \frac{\mathrm{d}}{\mathrm{d}t} \mright|_{t=0}\mleft[
	t \mapsto \Phi_t(p)
\mright]
=
\mleft.\mleft(\mleft.\Psi^{(\Phi)}\mright|_{(\Phi_0, A_0)}\mright)\mright|_p
\in \mathrm{T}_{\Phi_0(p)}N.
\eas
Thence, $\mathrm{Dev}(\Psi) = \Psi^{(\Phi)}$, which proves the statement.
\end{proof}

Hence, we see that Eq.~\eqref{WeakAnchorPreserv} is satisfied solely by the definition of the infinitesimal gauge transformation of the Higgs field. Therefore there is a unique operator $\delta_{\Psi_\varepsilon}$ in the sense of Cor.~\ref{cor:VeryGeneralPullbackConnection} for all $\varepsilon \in \Gamma(\mathrm{ev}^*E)$.

\begin{definitions}{Infinitesimal gauge transformation}{InfinitesimalGaugeTrafoClassicAsConnection}
Let $M$ be a smooth manifold, $W, K$ vector spaces, and $\mathfrak{g}$ a Lie algebra, equipped with a Lie algebra representation $\psi: \mathfrak{g} \to \mathrm{End}(W)$. We also denote with $E \coloneqq W \times \mathfrak{g}$ the associated action Lie algebroid over $W$, and let $^{E}\nabla$ be an $E$-connection on the trivial vector bundle $W \times K$ over $W$, and $\Psi_\varepsilon = (\delta_\varepsilon \Phi, \delta_\varepsilon A)$ for an $\varepsilon \in \Gamma(\mathrm{ev}^*E)$.

Then we define the \textbf{infinitesimal gauge transformation $\delta_\varepsilon L$ for $L \in \Omega^{k,0}\bigl(M \times \mathfrak{M}_{\mathfrak{g}}(M; W); K\bigr)$} ($k \in \mathbb{N}_0$) as an element of $\Omega^{k,0}\bigl(M \times \mathfrak{M}_{\mathfrak{g}}(M; W); K\bigr)$ by
\ba\label{DefOfGaugeTrafoWithBookkeep}
\mleft(\delta_\varepsilon L\mright)(Y_1, \dotsc, Y_k)
&\coloneqq
\delta_{\Psi_\varepsilon}\bigl(
	\iota(L)(Y_1, \dotsc, Y_k)
\bigr)
\ea
for all $Y_1, \dotsc, Y_k \in \mathfrak{X}(M)$, where $\delta_{\Psi_\varepsilon}$ is the unique operator given in the context of Cor.~\ref{cor:VeryGeneralPullbackConnection} with respect to ${}^{E}\nabla$ on $V \coloneqq W \times K$, $\nu_1 \coloneqq \Psi_\varepsilon$, $\nu_2 = - \varepsilon$, $E_1 \coloneqq \mathrm{T}\bigl( M \times \mathfrak{M}_{\mathfrak{g}}(M;W) \bigr)$, $E_2 \coloneqq E$, and $f \coloneqq \mathrm{ev}$.
\end{definitions}

\begin{remark}\label{WeshalbKlapptMeineKonstruktion}
\leavevmode\newline
Recall Cor.~\ref{cor:GaugeTrafoAsPullbackCond}, which is needed for Cor.~\ref{cor:VeryGeneralPullbackConnection}, and that $\iota(L)$ was the bookkeeping trick, and, thus,
\bas
\iota(L)(Y_1, \dotsc, Y_k)
&\in
\Gamma(\mathrm{ev}^*(W\times K))
\eas
for all $Y_1, \dotsc, Y_k \in \mathfrak{X}(M)$. Hence, the right hand side is well-defined. 

That $\delta_\varepsilon L$ is an element of $\Omega^{k,0}\bigl(M \times \mathfrak{M}_{\mathfrak{g}}(M; W); K\bigr)$ also follows by construction: Usually one would define
\ba\label{OriginalFormula}
\mleft(\delta_\varepsilon L \mright)\mleft( X_1, \dotsc, X_k \mright)
&\coloneqq
\delta_{\Psi_\varepsilon}\bigl(
	\iota(L)\mleft(X_1, \dotsc, X_k\mright)
\bigr)
	- \sum_{i=1}^k L\mleft( X_1, \dotsc, \mathcal{L}_{\Psi_\varepsilon} X_i, \dotsc, X_k \mright)
\ea
for all $X_1, \dotsc, X_k \in \mathfrak{X}\bigl( M \times \mathfrak{M}_{\mathfrak{g}}(M; W) \bigr)$. As usual, such a definition leads to $C^\infty\bigl( M \times \mathfrak{M}_{\mathfrak{g}}(M; W) \bigr)$-multilinearity, and vector fields of $M \times \mathfrak{M}_{\mathfrak{g}}(M; W)$ are generated by vector fields of $M$ and $\mathfrak{M}_{\mathfrak{g}}(M; W)$, such that we can restrict ourselves to vector fields of $M$ and $\mathfrak{M}_{\mathfrak{g}}(M; W)$. If for example $X_1 = \Psi \in \mathfrak{X}\bigl(\mathfrak{M}_{\mathfrak{g}}(M; W) \bigr) \subset \mathfrak{X}\bigl( M \times \mathfrak{M}_{\mathfrak{g}}(M; W) \bigr)$, then
\bas
\iota(L)\mleft( \Psi, X_2, \dotsc, X_k \mright)
&=
0
\eas
and
\bas
\sum_{i=1}^k L\mleft( X_1, \dotsc, \mathcal{L}_{\Psi_\varepsilon} X_i, \dotsc, X_k \mright)
&=
L( \mathcal{L}_{\Psi_\varepsilon} \Psi, X_2 \dotsc, X_k )
	+ \sum_{i=2}^k \underbrace{L\mleft( \Psi, X_2, \dotsc, \mathcal{L}_{\Psi_\varepsilon} X_i, \dotsc, X_k\mright)}_{=0}
\\
&=
L\bigl( \underbrace{[\Psi_\varepsilon, \Psi]}_{ \mathclap{\in ~ \mathfrak{X}\mleft( \mathfrak{M}_{\mathfrak{g}}(M; W) \mright)} }, X_2 \dotsc, X_k \bigr)
\\
&=
0,
\eas
using $L \in \Omega^{k,0}\bigl(M \times \mathfrak{M}_{\mathfrak{g}}(M; W); K\bigr)$ and $\Psi_\varepsilon \in \mathfrak{X}\bigl( \mathfrak{M}_{\mathfrak{g}}(M; W) \bigr)$. On the other hand if $X_1 = Y \in \mathfrak{X}(M)$, then 
\bas
L( \mathcal{L}_{\Psi_\varepsilon} Y, X_2 \dotsc, X_k )
&=
L\bigl( \underbrace{[\Psi_\varepsilon, Y]}_{ \mathclap{= 0} }, X_2 \dotsc, X_k \bigr)
=
0.
\eas
Using these relations, we can conclude that the non-trivial information of Def.~\eqref{OriginalFormula} is encoded completely on $\mathfrak{X}(M)$ as a $C^\infty(M)$-module,\footnote{Observe that $\mathcal{L}_{\Psi_\varepsilon}(f) = 0$ for all $f \in C^\infty(M)$.} as given in Def.~\eqref{DefOfGaugeTrafoWithBookkeep}. Therefore one can use Def.~\eqref{DefOfGaugeTrafoWithBookkeep} instead and canonically/trivially extend this definition to the "full" form.
%
\end{remark}

In fact, we recover the infinitesimal gauge transformation $\delta_\epsilon$ by using a canonical flat connection:

\begin{theorems}{Recover of classical definition of infinitesimal gauge transformation}{RecoverOfClassicInfgGaugeTrafo}
Let $M$ be a smooth manifold, $W, K$ vector spaces, and $\mathfrak{g}$ a Lie algebra, equipped with a Lie algebra representation $\psi: \mathfrak{g} \to \mathrm{End}(W)$. We also denote with $E \coloneqq W \times \mathfrak{g}$ the associated action Lie algebroid over $W$, and let $^{E}\nabla = \nabla_\rho$ be the $E$-connection\footnote{Recall Ex.~\ref{ex:NablaRhoConnection}.} induced by the canonical flat connection $\nabla$ of the trivial vector bundle $W \times K$ over $W$, and $\Psi_\varepsilon = (\delta_\varepsilon \Phi, \delta_\varepsilon A)$ for an $\varepsilon \in \Gamma(\mathrm{ev}^*E)$.

Then we have
\ba
\mleft(\delta_\varepsilon L\mright)(\Phi, A)
&=
\mleft.\frac{\mathrm{d}}{\mathrm{d}t}\mright|_{t=0}
\mleft[ t \mapsto
	L\mleft(
		\Phi + t \delta_\epsilon \Phi,
		A + t \delta_\epsilon A
	\mright)
\mright]
\ea
for all $L \in \Omega^{k,0}\bigl(M \times \mathfrak{M}_{\mathfrak{g}}(M; W); K\bigr)$ ($k \in \mathbb{N}_0$) and $(\Phi, A) \in \mathfrak{M}_{\mathfrak{g}}(M;W)$, where $\epsilon \coloneqq \varepsilon(\Phi,A)$, $t \in \mathbb{R}$, and $\delta_\varepsilon$ is as defined in Def.~\ref{def:InfinitesimalGaugeTrafoClassicAsConnection} with respect to $\nabla_\rho$.

In other words, we recover Def.~\ref{def:ClassFunctionalGaugeTrafoBlag}, especially if taking an $\varepsilon \in C^\infty(M;\mathfrak{g})$, \textit{i.e.}~a constant $\varepsilon$, "constant" in sense of
\bas
\varepsilon(\Phi,A)
&=
\varepsilon\mleft(\Phi^\prime,A^\prime\mright)
\eas
for all $(\Phi, A), \mleft(\Phi^\prime,A^\prime\mright) \in \mathfrak{M}_{\mathfrak{g}}(M;W)$.
\end{theorems}

\begin{remarks}{$\delta_\varepsilon A$ as transformation of a functional}{BosonsAsFunctionalies}
Observe that $\delta_\varepsilon A$ is here trivially also given by $\delta_\varepsilon \varpi_2$, where $\varpi_2(\Phi, A) \coloneqq A$, the projection onto the second factor in $\mathfrak{M}_{\mathfrak{g}}$. Viewing the field of gauge bosons as the functional $\varpi_2$, one may want to define the infinitesimal gauge transformation of $A$ as the infinitesimal gauge transformation of $\varpi_2$; since $\varpi_2$ is $\mathfrak{g}$-valued, we would have
\bas
\iota(\varpi_2)(Y)
&\in
\Gamma(\mathrm{ev}^*(W \times \mathfrak{g}))
\eas
for all $Y \in \mathfrak{X}(M)$, and, thus, $\iota(A) \coloneqq \iota(\varpi_2)(\Phi, A) \in \Omega^1(M; \Phi^*(W \times \mathfrak{g}))$ for any fixed $\Phi$. For the infinitesimal gauge transformation of the field strength one also applies the bookkeeping trick such that it would have values in $\mathrm{ev}^*(W \times \mathfrak{g})$.
\end{remarks}

\begin{proof}[Proof of Thm.~\ref{thm:RecoverOfClassicInfgGaugeTrafo}]
\leavevmode\newline
Let $\mleft( e_a \mright)_a$ be a basis of $K$, that especially implies
\bas
\nabla \bigl( \iota_W(e_a) \bigr)
&=
0.
\eas
For $L \in \Omega^{k,0}\bigl(M \times \mathfrak{M}_{\mathfrak{g}}(M; W); K\bigr)$ we then write
\bas
\iota(L)
&=
L^a \otimes \mathrm{ev}^*\bigl(\iota_W(e_a)\bigr)
\eas
for $L^a \in \Omega^k(M\times \mathfrak{M}_{\mathfrak{g}}(M;W))$, and,
thus, by using Cor.~\ref{cor:VeryGeneralPullbackConnection},
\bas
\mleft(\delta_\varepsilon L\mright)(Y_1, \dotsc, Y_k)
&=
\delta_{\Psi_\varepsilon}\bigl(
	\iota(L)(Y_1, \dotsc, Y_k)
\bigr)
\\
&=
\mathcal{L}_{\Psi_\varepsilon}\mleft(
	L^a(Y_1, \dotsc, Y_k)
\mright) 
~ \underbrace{\mathrm{ev}^*\bigl(\iota_W(e_a)\bigr)}_
{ \mathclap{= \iota_{M \times \mathfrak{M}_{\mathfrak{g}}(M; W)}(e_a)} }
\\
&\hspace{1cm}
	- \underbrace{\mleft(L^a(Y_1, \dotsc, Y_k) ~ \mathrm{ev}^*\mleft(\nabla_{\rho(\varepsilon)}\bigl( \iota_W(e_a) \bigr) \mright)\mright)}_{=0}
\\
&=
\mleft( 
	\mathcal{L}_{\Psi_{\varepsilon}}\mleft(L^a\mright)
	\otimes \iota_{M \times \mathfrak{M}_{\mathfrak{g}}(M; W)}(e_a)
\mright)(Y_1, \dotsc, Y_k)
\eas
for all $Y_1, \dotsc, Y_k \in \mathfrak{X}(M)$, hence,
\bas
\mleft(\delta_\varepsilon L\mright)(\Phi, A)
&=
	\mleft.\frac{\mathrm{d}}{\mathrm{d}t}\mright|_{t=0}
\mleft[ t \mapsto
	L\mleft(
		\Phi + t \delta_\epsilon \Phi,
		A + t \delta_\epsilon A
	\mright)
\mright]
\eas
for all $(\Phi, A)\in \mathfrak{M}_{\mathfrak{g}}(M;W)$, using that $\mleft.\Psi_{\varepsilon}\mright|_{(\Phi,A)} = (\delta_\epsilon \Phi, \delta_\epsilon A)$, where $\epsilon \coloneqq \varepsilon(\Phi, A)$.
\end{proof}

This concludes this section, we have shown how to write the infinitesimal gauge transformation using Lie algebroid connections. One can even show that the gauge invariance of the Yang-Mills-Higgs Lagrangian can be shown with the same calculation of the previous section if $\varepsilon$ is allowed to depend on $\mathfrak{M}_{\mathfrak{g}}(M;W)$. Such a dependency starts to matter when applying the infinitesimal gauge transformation twice, which we will discuss later in full generality. Let us now shortly discuss what we have learned so far.

First of all, we needed to do the bookkeeping trick. That was due to the Lie algebra action $\gamma$, which acts on $N =W$ and not on $M$, and if one generalizes $\gamma$ to a base-preserving morphism like $\rho$ the more natural construction uses bundles defined over $N$. This was why we needed to make a pullback and to think of functionals as having values in a pullback of a trivial bundle over $N$, especially using $\mathrm{ev}$, or $\Phi \in C^\infty(M;N)$ if fixing a point of $\mathfrak{M}_{\mathfrak{g}}$. For example, we thought of the Lie algebra $\mathfrak{g}$ as a trivial bundle over $M \times \mathfrak{M}_{\mathfrak{g}}$ and $N$, $M \times \mathfrak{M}_{\mathfrak{g}} \times \mathfrak{g}$ and $N \times \mathfrak{g}$, respectively, and it is more suitable to think of $M \times \mathfrak{M}_{\mathfrak{g}} \times \mathfrak{g}$ as $\mathrm{ev}^*(N \times \mathfrak{g})$. The aim of the presented generalised gauge theory in \cite{CurvedYMH} is also to generalise the trivial Lie algebra bundle, especially getting rid of a global trivialisation by replacing it with a general Lie algebroid $E$. Hence, motivated by this section and as an ansatz, we are going to define $E$ in place of $N \times \mathfrak{g}$ later and $\mathrm{ev}^*E$ will replace $M \times \mathfrak{M}_{\mathfrak{g}} \times \mathfrak{g}$. In the same manner other vector spaces may be replaced.

Second, assume we have that general Lie algebroid $E$ now. Then we cannot impose the existence of a canonical flat connection anymore as we did in definitions like Def.~\ref{def:ClassFunctionalGaugeTrafoBlag}; defining $\mathrm{d}/\mathrm{d}t$ using the tangent map would lead to arising horizontal components in the corresponding tangent bundle which may make further calculations more complicated when a functional is used in other functionals, like in contractions using scalar products and metrics, such that one may need to fix a horizontal distribution. Therefore the definition of infinitesimal gauge transformation as provided here is a first step towards a formulation using Lie algebroid connections, \textit{e.g.}~taking a connection $\nabla$ and then defining ${}^{E}\nabla = \nabla_\rho$.

Third, one could argue that one could just look at vector bundle connections $\nabla$ for which there is always a pullback, avoiding the problems discussed in the previous section. However, Lie algebroid connections are clearly more general. Especially when thinking about that the infinitesimal gauge transformations are just certain, not all, vector fields on $\mathfrak{M}_{\mathfrak{g}}$, one might argue why not using a different connection like a Lie algebroid connection which does not lift all possible vector fields on the base manifold. In fact, we are going to take the \textbf{basic connection} later. The basic connection does not necessarily have any notion of a parallel frame, even when it is assumed to be flat, such that it is in general different to a typical flat connection. The advantage of the basic connection will be that it supports the symmetries of gauge theories, leading to more convenient formulas of infinitesimal gauge transformations.

Fourth, the Lie algebra $\mathfrak{g}$ is not only important from an algebraic point of view, but also in sense of a connection besides the field of gauge bosons $A$, playing the role of a "direction of derivative" similar to the tangent bundle when defining typical vector bundle connections. Thus, it makes even more sense to use Lie algebroids in the context of gauge theory. In fact, there is already a formulation of infinitesimal gauge theory using Lie algebroids, called \textbf{Curved Yang-Mills-Higgs Gauge Theory}, see \textit{e.g.} \cite{CurvedYMH} for a short summary, and see the references therein for more elaborated details. My whole Ph.D.~thesis is also devoted to this gauge theory and gives both, a summary and a reformulation; see \cite{MyThesis}.

Last, even though we will not introduce curved Yang-Mills-Higgs gauge theory, we got motivated by Thm.~\ref{thm:RecoverOfClassicInfgGaugeTrafo} to think of the typical formulation for gauge theory as a theory corresponding to an action Lie algebroid over $N = W$ equipped with its canonical flat connection $\nabla$. When we speak of the standard/classical setting or formulas, then we therefore usually mean precisely such Lie algebroids and their connections.

However, although we will not introduce this gauge theory in full details here, we will provide how to formulate infinitesimal gauge transformations as a straightforward generalisation of what we just have seen, useful for the context of theories like curved Yang-Mills-Higgs gauge theories.

\section{Generalisation}\label{GeneralisedGaugeTrafoStuffYeah}

\begin{definitions}{Space of fields}{SpaceOfFields}
Let $M, N$ be two smooth manifolds and $E\to N$ a Lie algebroid. Then we denote the \textbf{space of fields} by
\ba
\mathfrak{M}_E
&\coloneqq
\mathfrak{M}_E(M; N)
\coloneqq
\left\{ (\Phi, A)
~\middle|~
\Phi \in C^\infty(M;N) \text{ and } A \in \Omega^1(M; \Phi^*E)
\right\}
\ea
which we sometimes view as a fibration over $C^\infty(M;N)$
\begin{center}
	\begin{tikzcd}
		\mathfrak{M}_E(M; N) \arrow{d} \\
		C^\infty(M;N)
	\end{tikzcd}
\end{center}
where the projection is given by $\mathfrak{M}_E(M; N) \ni (\Phi, A) \mapsto \Phi$.

We will refer to $A \in \Omega^1(M; \Phi^*E)$ as the \textbf{field of gauge bosons} and $\Phi$ just as a \textbf{physical} or \textbf{Higgs field} of this theory.
\end{definitions}

We want to study the tangent space of $\mathfrak{M}_E$, and for this recall that for each vector bundle $V \stackrel{\pi}{\to} N$ there is also a vector bundle structure for $\mathrm{T}V \stackrel{\mathrm{D}\pi}{\to} \mathrm{T}N$, and the following diagram describes a double vector bundle
\begin{center}
	\begin{tikzcd}
		 \mathrm{T}V \arrow{r}{\mathrm{D}\pi} \arrow{d}{\pi_{\mathrm{T}V}} & \mathrm{T}N \arrow{d}{\pi_{\mathrm{T}N}} \\
		V \arrow[r, "\pi"]& N
	\end{tikzcd}
\end{center}
that is, each horizontal and vertical line is a vector bundle, and the horizontal and vertical scalar multiplications on $\mathrm{T}V$ commute, see \textit{e.g.}~\cite[\S 3ff.]{Highervectorbundles}. Let us shortly recap the vector bundle structure of $\mathrm{T}V \stackrel{\mathrm{D}\pi}{\to} \mathrm{T}N$, following \cite[discussion at the beginning of \S 3.4; page 110ff.]{mackenzieGeneralTheory}: The linear structure at $v \in \mathrm{T}_p N$ ($p \in N$) is basically given by the vertical structure of $V$ prolonged along the fibre $V_p$, but as an affine space whose offset is given by $v$. That is, let $\xi, \eta \in \mathrm{T}V$ with 
\bas
\mathrm{D}_{\pi_{\mathrm{T}V}(\xi)}\pi(\xi)
&=
\mathrm{D}_{\pi_{\mathrm{T}V}(\eta)}\pi(\eta)
\eqqcolon
v,
\eas
and, hence, due to $\pi_{\mathrm{T}N}(v) = p$,
\bas
p
&= 
(\pi \circ \pi_{\mathrm{T}V})(\xi)
=
(\pi \circ \pi_{\mathrm{T}V})(\eta).
\eas
Thus, one can take curves $f,h: I \to V$ ($I \in \mathbb{R}$ an open interval around 0) with
\bas
f(0)
&=
\pi_{\mathrm{T}V}(\xi),
&
\mleft.\frac{\mathrm{d}}{\mathrm{d}t}\mright|_{t=0} f
&=
\xi,
\\
h(0)
&=
\pi_{\mathrm{T}V}(\eta),
&
\mleft.\frac{\mathrm{d}}{\mathrm{d}t}\mright|_{t=0} h
&=
\eta,
\eas
such that
\bas
\pi \circ f = \pi \circ h,
\eas
because the condition on $\xi$ and $\eta$ imply on the base paths $\pi \circ f, \pi \circ h: I \to N$ that
\bas
(\pi\circ f)(0)
&=
p
=
(\pi \circ h)(0),
\\
\mleft.\frac{\mathrm{d}}{\mathrm{d}t}\mright|_{t=0} \bigl( \pi \circ f \bigr)
&=
\mathrm{D}_{\pi_{\mathrm{T}V}(\xi)}(\xi)
=
\mathrm{D}_{\pi_{\mathrm{T}V}(\eta)}(\eta)
=
\mleft.\frac{\mathrm{d}}{\mathrm{d}t}\mright|_{t=0} \bigl( \pi \circ h \bigr).
\eas
Then the addition and scalar multiplication with $\lambda \in \mathbb{R}$ for $\mathrm{T}V \stackrel{\mathrm{D}\pi}{\to} \mathrm{T}N$ is defined by
\bas
\xi \RPlus \eta
&\coloneqq
\mleft.\frac{\mathrm{d}}{\mathrm{d}t}\mright|_{t=0} (f + h),
\\
\lambda \boldsymbol{\cdot} \xi
&\coloneqq
\mleft.\frac{\mathrm{d}}{\mathrm{d}t}\mright|_{t=0} (\lambda h),
\eas
where the addition of curves is well-defined because of $\pi \circ f = \pi \circ h$ which implies $\pi(f+h)= \pi(f) = \pi(h)$; so, one can take the sum of the curves and
\bas
\mathrm{D}\pi(\xi \RPlus \eta)
&=
\mleft.\frac{\mathrm{d}}{\mathrm{d}t}\mright|_{t=0} \bigl( \underbrace{\pi (f+h)}_{= \pi(f)} \bigr)
=
\mathrm{D}\pi(\xi)
=
v.
\eas
In other words, those operations come from interpreting tangent vectors as equivalence classes of curves, assuming there are representatives of the classes sharing the same base path ($\pi \circ f = \pi \circ h$) with which one can do those operations.
It is trivial to show that we have a double vector bundle. The operations of the linear structure in $\mathrm{T}V \stackrel{\pi_{\mathrm{T}V}}{\to} V$ is still denoted in the same manner as usual, and by definition one also gets
\bas
\pi_{\mathrm{T}V}(\xi \RPlus \eta)
&=
\pi_{\mathrm{T}V}(\xi)
	+ \pi_{\mathrm{T}V}(\eta),
\\
\pi_{\mathrm{T}V}(\lambda \boldsymbol{\cdot} \xi)
&=
\lambda ~ \pi_{\mathrm{T}V}(\xi).
\eas
Let us now turn to the tangent space of $\mathfrak{M}_E$.

\begin{propositions}{Tangent space of $\mathfrak{M}_E(M; N)$}{TangentSpaceOfSpaceOfFields}
Let $M, N$ be two smooth manifolds and $E \stackrel{\pi}{\to} N$ a Lie algebroid. Then the tangent space $\mathrm{T}_{(\Phi_0,A_0)} \bigl(\mathfrak{M}_E(M; N)\bigr)$ of $\mathfrak{M}_E(M; N)$ at $(\Phi_0, A_0)$ consists of pairs $(\mathcal{v}, \mathcal{a})$ with $\mathcal{v} \in \Gamma\mleft(\Phi_0^*\mathrm{T}N\mright)$ and $\mathcal{a} \in \Omega^1\mleft(M; \mathcal{v}^*\mathrm{T}E\mright)$, where $\mathcal{v}^*\mathrm{T}E$ is the pullback of $\mathrm{T}E \stackrel{\mathrm{D}\pi}{\to}\mathrm{T}N$ as a vector bundle, viewing $\mathcal{v}$ as a map $M \to \mathrm{T}N$. This pair also satisfies
\ba
\pi_{\mathrm{T}E}(\mathcal{a})
&=
A_0,
\ea
where $\pi_{\mathrm{T}E}$ denotes the projection of the vector bundle $\mathrm{T}E \to E$.
\end{propositions}

\begin{remarks}{Total situation as commuting diagram}{TangentCommutingDiagram}
This implies that we have in total
\begin{center}
	\begin{tikzcd}
		 \mathrm{T}E \arrow{rr}{\mathrm{D}\pi} \arrow[dd, "\pi_{\mathrm{T}E}", swap] && \mathrm{T}N \arrow{dd}{\pi_{\mathrm{T}N}} \\
		&M \arrow[ld, "A_0(Y)", pos=0.3] \arrow{rd}{\Phi_0} \arrow{ru}{\mathcal{v}} \arrow[lu, "\mathcal{a}(Y)", pos=0.2]  \\
		E \arrow[rr, "\pi", swap]&& N
	\end{tikzcd}
\end{center}
for all $(\Phi_0,A_0) \in \mathfrak{M}_E(M;N)$, $(\mathcal{v}, \mathcal{a}) \in \mathrm{T}_{(\Phi_0, A_0)}\bigl( \mathfrak{M}_E(M;N) \bigr)$ and $Y \in \mathfrak{X}(M)$, that is,
\ba
\pi\bigl(A_0(Y)\bigr)
&=
\Phi_0,
\\
\pi_{\mathrm{T}N} (\mathcal{v})
&=
\Phi_0,
\\
\pi_{\mathrm{T}E}(\mathcal{a})
&=
A_0,\label{AGaugeTrafoIsOverA}
\\
\mathrm{D}\pi \bigl( \mathcal{a}(Y) \bigr)
&=
\mathcal{v}\label{HorizontalCompOfDeltaA}
\ea
for all $Y \in \mathfrak{X}(M)$, where the projections of the vector bundles $\mathrm{T}E \to E$ and $\mathrm{T}N \to N$ are denoted by $\pi_{\mathrm{T}E}$ and $\pi_{\mathrm{T}N}$, respectively.
\end{remarks}

\begin{remark}\label{RemarkAboutThatWeStillHaveLinearStructureinDeltaA}
\leavevmode\newline
Especially for Eq.~\eqref{HorizontalCompOfDeltaA} recall the discussion about the double vector bundle structure, also useful for the note about that $\mathcal{v}^*\mathrm{T}E$ is the pullback of $\mathrm{T}E \stackrel{\mathrm{D}\pi}{\to}\mathrm{T}N$. That is, we have
\bas
\mathcal{a}(f Y + h Z)
&=
f \boldsymbol{\cdot} \mathcal{a}(Y)
	\RPlus h \boldsymbol{\cdot} \mathcal{a}(Z)
\eas
for all $Y, Z \in \mathfrak{X}(M)$ and $f, h \in C^\infty(M)$. Therefore also
\bas
\mathrm{D}\pi\bigl( \mathcal{a}(f Y + h Z) \bigr)
&=
\mathrm{D}\pi\bigl( \mathcal{a}(Y) \bigr).
\eas
This is also in alignment with Eq.~\eqref{AGaugeTrafoIsOverA} although this equation is about the vector bundle $\mathrm{T}E \to E$, so,
\bas
\pi_{\mathrm{T}E}\bigl( \mathcal{a}(f Y + h Z) \bigr)
&=
\pi_{\mathrm{T}E}\bigl( f \boldsymbol{\cdot} \mathcal{a}(Y)
	\RPlus h \boldsymbol{\cdot} \mathcal{a}(Z) \bigr)
\\
&=
f ~ \pi_{\mathrm{T}E}\bigl( \mathcal{a}(Y) \bigr)
	+ h ~ \pi_{\mathrm{T}E}\bigl( \mathcal{a}(Z) \bigr)
\\
&=
A_0\bigl( f Y + h Z \bigr).
\eas
\end{remark}

\begin{proof}[Proof of Prop.~\ref{prop:TangentSpaceOfSpaceOfFields}]
\leavevmode\newline
Follows immediately by definition of $\mathfrak{M}_E$; alternatively follow the straightforward calculation in \cite[Proposition 4.1.2]{MyThesis}, that is, just represent tangent vectors by velocities of curves and then everything follows in a straightforward manner.
\end{proof}

Think of $(\mathcal{v}, \mathcal{a})$ as the candidates for the infinitesimal gauge transformations, for which we wrote $(\delta_\varepsilon \Phi, \delta_\varepsilon A)$ previously. But now we cannot assume canonical flat connections which is why the last result shows that we cannot view $(\mathcal{v}, \mathcal{a})$ as an element of $\mathfrak{M}_E(M;N)$ in general, thus, we changed the notation to $(\mathcal{v}, \mathcal{a})$ for now. So, we do not have any canonical horizontal distribution as in the previous classical situation, and therefore let us study the vertical structure first.

Recall that a \textbf{vertical bundle} for fibre bundles $F \stackrel{\pi}{\to} N$ (as \textit{e.g.}~introduced in \cite[\S 5.1.1, for principal bundles, but it is straightforward to extend the definitions]{hamilton}), which is defined as a subbundle $\mathrm{V}F$ of the tangent bundle $\mathrm{T}F \to F$ given as the kernel of $\mathrm{D}\pi : \mathrm{T}F \to \mathrm{T}N$. The fibres $\mathrm{V}_eF$ of $F$ at $e \in F$ are then given by 
\bas
\mathrm{V}_e F
&=
\mathrm{T}_e F_p,
\eas
where $p \coloneqq \pi(e) \in N$ and $F_p$ is the fibre of $F$ at $p$.
Now consider a vector bundle $E \stackrel{\pi}{\to} N$, then $\mathrm{V}_e E = \mathrm{T}_e E_p \cong E_p$ because the fibres are vector spaces.

\begin{propositions}{Vertical bundle of $\mathfrak{M}_E(M; N)$}{VerticalBundleOfFracM}
Let $M, N$ be two smooth manifolds and $E \stackrel{\pi}{\to} N$ a Lie algebroid. Then the vertical bundle of $\mathfrak{M}_E(M; N)$, viewed as a fibration over $C^\infty(M;N)$, is given by
\ba
\mathrm{V}_{(\Phi,A)}\bigl(\mathfrak{M}_E(M; N)\bigr)
&\cong
\left\{
	(\mathcal{v}, \mathcal{a})
~\middle|~
	\mathcal{v}= 0 \in \Gamma(\Phi^*\mathrm{T}N), ~
	\mathcal{a} \in \Omega^1(M; \Phi^*E)
\right\}
\cong
\Omega^1(M; \Phi^*E).
\ea
\end{propositions}

\begin{proof}[Proof of Prop.~\ref{prop:VerticalBundleOfFracM}]
\leavevmode\newline
We have the fibration $\mathfrak{M}_E(M; N) \stackrel{\varpi}{\to} C^\infty(M;N)$, where $\varpi(\Phi, A) \coloneqq \Phi$ for all $(\Phi, A) \in \mathfrak{M}_E(M; N)$. Hence,
\bas
\mathrm{D}_{(\Phi, A)}\varpi(\mathcal{v}, \mathcal{a})
=
\mathcal{v}
\eas
for all $(\mathcal{v}, \mathcal{a}) \in \mathrm{T}_{(\Phi,A)}\mathfrak{M}_E(M; N)$. The kernel of $\mathrm{D}\varpi$ at $(\Phi,A) \in \mathfrak{M}_E(M; N)$ is then given by 
\bas
\mathrm{Ker}\mleft( \mathrm{D}_{(\Phi,A)} \varpi \mright)
&=
\left\{
(\mathcal{v}, \mathcal{a}) \in \mathrm{T}_{(\Phi,A)} \mathfrak{M}_E(M; N)
~\middle|~
\mathcal{v}=0
\right\}.
\eas
By Prop.~\ref{prop:TangentSpaceOfSpaceOfFields}, we then know that $\mathcal{a}$ has values in the vertical bundle $\mathrm{V}E$, that is, for $\mathcal{a}_p(Y_p) \in \mathrm{T}_{A_p(Y_p)} E$ ($p \in M$, $Y \in \mathfrak{X}(M)$) we have
\bas
&&
\mathrm{D}_{A_p(Y_p)}\pi\mleft(\mathcal{a}_p(Y_p)\mright)
&=
0
\\
&\Leftrightarrow&
\mathcal{a}_p(Y_p)
&\in
\mathrm{V}_{A_p(Y_p)} E
\cong
E_{\Phi(p)}.
\eas
Thus, we can view $\mathcal{a}$ equivalently as an element of $\Omega^1(M; \Phi^*E)$, so,
\bas
\mathrm{V}_{(\Phi,A)}\mathfrak{M}_E(M; N)
&\cong
\left\{
	(\mathcal{v}, \mathcal{a})
~\middle|~
	\mathcal{v}= 0 \in \Gamma(\Phi^*\mathrm{T}N), ~
	\mathcal{a} \in \Omega^1(M; \Phi^*E)
\right\}
\cong
\Omega^1(M; \Phi^*E).
\eas
\end{proof}

Thinking again of $(\mathcal{v}, \mathcal{a})$ as $\mleft( \delta_\varepsilon \Phi, \delta_\varepsilon A \mright)$, we see that we can in general not expect that $\delta_\varepsilon A$ is in the same vector space as $A$ since $\delta_\varepsilon \Phi$ will be in general nonzero in physical settings; $\delta_\varepsilon A$ is still vertical but with offset given by $\delta_\varepsilon \Phi$ by Prop.~\ref{prop:TangentSpaceOfSpaceOfFields} and Remark \ref{RemarkAboutThatWeStillHaveLinearStructureinDeltaA}.

Now we need to define at what type of functionals we are going to look at. One key step is to look at $M \times \mathfrak{M}_E(M;N)$ as we did in Def.~\ref{def:FirstAttemptOfEvaluationMap} and afterwards.

\begin{definitions}{Evaluation map of $M \times \mathfrak{M}_E$}{EvaluationMap}
Let $M, N$ be manifolds, and $E \to N$ a Lie algebroid over $N$.
Then we define the \textbf{evaluation map} $\mathrm{ev}$ by
\ba
M \times \mathfrak{M}_E(M;N) &\to N
\nonumber\\
(\Phi,A)
&\mapsto
\mathrm{ev}(p, \Phi, A)
\coloneqq
\Phi(p)
\ea
for all $p\in M$ and $(\Phi, A) \in \mathfrak{M}_E$.
\end{definitions}

\begin{definitions}{Space of functionals in gauge theory}{FunctionalsAsForms}
Let $M, N$ be two smooth manifolds, $E\to N$ a Lie algebroid, and $V \to N$ a vector bundle. Then the \textbf{space of functionals $\mathcal{F}^k_E(M; {}^*V)$} ($k \in \mathbb{N}_0$) is defined as
\ba
\mathcal{F}^k_E(M; {}^*V)
&\coloneqq
\Omega^{k,0}\bigl(M \times \mathfrak{M}_E(M;N); \mathrm{ev}^*V\bigr).
\ea

If $V = N \times \mathbb{R}$ is the trivial line bundle over $N$, then we just write $\mathcal{F}_E^k(M)$ instead of $\mathcal{F}^k_E(M;{}^*V)$.
\end{definitions}

\begin{remark}\label{FunctionalsAreAsUsual}
\leavevmode\newline
As previously, we often write for $L \in \mathcal{F}^k_E(M; {}^*V)$
\bas
\mathfrak{M}_E \ni (\Phi, A)
&\mapsto
L(\Phi, A) 
\coloneqq 
\mleft.L\mright|_{M \times \{\Phi, A\}}
\in \Omega^k(M; \Phi^*V)
\eas
especially when we do not evaluate at $p \in M$. Observe that $L$ acts non-trivially only on $\mathrm{T}M$.
\end{remark}

\begin{examples}{Projection onto the field of gauge bosons}{ProjectionOntoGaugeBosonies}
We have an important and trivial functional $\varpi_2 \in \mathcal{F}^1_E(M; {}^*E)$ given as the projection onto the field of gauge bosons, that is
\ba
\varpi_2(\Phi,A)
&\coloneqq
A
\ea
for all $(\Phi, A) \in \mathfrak{M}_E$.
\end{examples}

\begin{examples}{Tangent map, total differential as functional}{DAsFunctional}
Also the total differential $\mathrm{D}$ can be viewed as a functional. That is $\mathrm{D} \in \mathcal{F}^1_E(M; {}^*\mathrm{T}N)$ by
\ba
\mathrm{D}(\Phi, A)
&\coloneqq
\mathrm{D}\Phi
\in
\Omega^1(M; \Phi^*\mathrm{T}N).
\ea
Hence, when we just write $\mathrm{D}$, then we mean precisely this functional. 
\end{examples}

We have $\mathrm{T}(M \times \mathfrak{M}_E) \cong \pi_1^*\mathrm{T}M \oplus \pi_2^*\mathrm{T}\mathfrak{M}_E$, where $\pi_1$ and $\pi_2$ are the projections onto the first and second factor in $M \times \mathfrak{M}_E$, respectively. If we speak for example about $\mathrm{T}M$, especially sections thereof, $\mathfrak{X}(M)$, then we mean their canonical embedding as a subalgebra of $\mathfrak{X}(M \times \mathfrak{M}_E)$; so, $X \in \mathfrak{X}(M)$ is also viewed as an element of $\mathfrak{X}(M\times \mathfrak{M}_E)$ but constant along $\mathfrak{M}_E$. For vector bundle morphisms defined on $\mathrm{T}(M \times \mathfrak{M}_E)$ we then also mean that forms restricted onto $\mathrm{T}M$ extend to maps acting on $\mathfrak{X}(M)$.

\begin{remarks}{Notions on $\mathcal{F}^k_E$ and further pullbacks with $\mathrm{ev}$}{NotionsOnFunctionals}
By Def.~\ref{def:FunctionalsAsForms}, we recover typical notions on the space of functionals, notions like wedge products, Def.~\ref{def:GradingOfProducts} and contractions \textit{etc.}~by restricting notions on $\Omega^\bullet(M \times \mathfrak{M}_E)$ and $\Omega^\bullet(M \times \mathfrak{M}_E; \mathrm{ev}^*V)$ to $\Omega^{\bullet,0}(M \times \mathfrak{M}_E)$ and $\Omega^{\bullet,0}(M \times \mathfrak{M}_E; \mathrm{ev}^*V)$, respectively, where $V$ is a vector bundle over $N$. Hence, we will not need to define all those notions in that setting, and, especially, $\Gamma(\mathrm{ev}^*V)$ is therefore generated by elements of the form $\mathrm{ev}^*v$, where $v \in \Gamma(V)$. 

\hspace{0.3cm} Now assume we have a vector bundle connection $\nabla$ on $V$, then $\mathrm{ev}^*\nabla$ is a connection on $\mathrm{ev}^*V$. We want to restrict the exterior covariant derivative related to that connection just to vector fields on $M$. Observe for all $X\in \mathfrak{X}(M) \subset \mathfrak{X}(M \times \mathfrak{M}_E)$, with flow $\gamma$ in $M$ through a $p \in M$, $(t, p) \mapsto \gamma_t(p)$ ($t \in I$ for some open interval $I$ in $\mathbb{R}$ containing 0),
\ba\label{DGleichDev}
\mathrm{D}_{(p, \Phi, A)} \mathrm{ev} (X)
&=
\mleft.\frac{\mathrm{d}}{\mathrm{d}t}\mright|_{t=0}
\bigl(
	\mathrm{ev} \circ (\gamma(p), \Phi, A)
\bigr)
=
\mleft.\frac{\mathrm{d}}{\mathrm{d}t}\mright|_{t=0}
\bigl(
	(\Phi \circ \gamma )(p) 
\bigr)
=
\mathrm{D}_p \Phi (X)
\ea
for all $(p, \Phi, A) \in M \times \mathfrak{M}_E$, where $(\gamma(p), \Phi, A)$ is the flow of $X \in \mathfrak{X}(M)$ at $(p, \Phi, A)$, viewed as an element of $\mathfrak{X}(M \times \mathfrak{M}_E)$. So, the pushforward of $X$ with $\mathrm{ev}$ at $(\Phi,A)$ is the same as the pushforward of $X$ with $\Phi$, thus
\bas
\mleft(\mathrm{ev}^* \nabla\mright)_{X_{(p, \Phi, A)}}
&=
\mleft( \Phi^*\nabla \mright)_{X_p}
\eas
for all $(p, \Phi, A)$,
viewing $X$ as an element of $\mathfrak{X}(M \times \mathfrak{M}_E)$ on the left hand side and as an element of $\mathfrak{X}(M)$ on the right hand side. Hence, we then also have
\bas
\mleft.\bigl(\mleft(\mathrm{ev}^* \nabla\mright)_{X} v\bigr)\mright|_{(p, \Phi, A)}
=
\mleft.\mleft(\mleft( \Phi^*\nabla \mright)_{X_p} v|_{(\Phi, A)} \mright) \mright|_p
\eas
for all $v \in \Gamma(\mathrm{ev}^*V)$,
since $X$ does not differentiate along $\mathfrak{M}_E$, and viewing $v|_{(\Phi, A)} \coloneqq [p \mapsto v|_{(p, \Phi, A)}]$ as an element of $\Gamma(\Phi^*V)$ on the right hand side.
Therefore this naturally leads on one hand to an exterior covariant derivative on the space of functionals by restricting $\mathrm{ev}^*\nabla$ to $\mathrm{T}M$ because then the exterior covariant derivative of $\mleft.\mleft(\mathrm{ev}^*\nabla\mright)\mright|_{\mathrm{T}M}$ clearly restricts to $\mathcal{F}^\bullet_E(M; {}^*V)$, and on the other hand
\bas
\mleft.\mleft(\mathrm{d}^{\mleft.\mleft(\mathrm{ev}^*\nabla\mright)\mright|_{\mathrm{T}M}} L\mright)\mright|_{(\Phi,A)} 
&=
\mathrm{d}^{\Phi^*\nabla}\bigl( L(\Phi, A) \bigr),
\eas
also recall Remark \ref{FunctionalsAreAsUsual}.

\hspace{0.3cm} Similarly, one shows for the pullback $\mathrm{ev}^!\omega$ of forms $\omega \in \Omega^k(N; V)$ that
\bas
\mleft. \mleft(\mathrm{ev}^! \omega\mright)\mright|_{(p,\Phi,A)}
\mleft( X_1, \dotsc, X_k \mright)
&=
\mleft. \mleft(\Phi^! \omega\mright)\mright|_{p}
\mleft( X_1, \dotsc, X_k \mright)
\eas
for all $X_1, \dotsc, X_k \in \mathfrak{X}(M)$. Hence, also the $\mathrm{ev}$-pullback of forms restricts to a $\Phi$-pullback of forms when fixing $(\Phi,A)$ and just evaluating at vector fields along $M$.
\end{remarks}

Therefore we define pullback functionals as in the following definition.

\begin{definitions}{Pullbacks as functionals}{PullbacksAsFunctionals}
Let $M, N$ be smooth manifolds, $E \to N$ a Lie algebroid, and $V \to N$ a vector bundle. For all $\omega \in \Gamma\mleft( V \mright)$ we define its \textbf{pullback functional ${}^*v$} as an element of $\mathcal{F}^0_E(M; {}^*V)$ by
\ba
{}^*v
&\coloneqq
\mathrm{ev}^*v.
\ea

For a vector bundle connection $\nabla$ on $V$ we define the \textbf{pullback connection ${}^*\nabla$ (to functionals)} by
\ba
{}^*\nabla
&\coloneqq
\mleft.\mleft(\mathrm{ev}^*\nabla\mright)\mright|_{\mathrm{T}M}.
\ea
Its induced exterior covariant derivative $\mathrm{d}^{{}^*\nabla}$ we view as an exterior covariant derivative on the space of functionals, especially
\ba
\mathrm{d}^{{}^*\nabla}:
\mathcal{F}^k_E(M; {}^*V)
&\to
\mathcal{F}^{k+1}_E(M; {}^*V)
\ea
for all $k \in \mathbb{N}_0$.

For all $\omega \in \Omega^k(N;V)$  ($k \in \mathbb{N}_0$) we define similarly its \textbf{form-pullback functional ${}^!\omega$} as an element of $\mathcal{F}_E^k(M; {}^*V)$ by
\ba
{}^!\omega
&\coloneqq
\mleft.\mleft(\mathrm{ev}^!\omega\mright)\mright|_{\bigwedge^k \mathrm{T}M}.
\ea
\end{definitions}

\begin{remarks}{}{}
Observe that
\ba
\mleft.({}^*v)(\Phi,A)\mright|_p
&\coloneqq
(\mathrm{ev}^*v)|_{(p, \Phi, A)}
=
\Phi^*v|_p
\ea
for all $(p, \Phi, A) \in M \times \mathfrak{M}_E$. Especially, $({}^*v)(\Phi, A) = \Phi^*v$, similarly to what we already pointed out for ${}^!w$ and ${}^*\nabla$ in Remark \ref{rem:NotionsOnFunctionals}. By construction, and as argued in Rem.~\ref{rem:NotionsOnFunctionals}, we also get
\ba
\mleft( \mathrm{d}^{{}^*\nabla} L \mright)(\Phi,A)
&=
\mathrm{d}^{\Phi^*\nabla}\bigl( L(\Phi,A) \bigr)
\ea
for all $L \in \mathcal{F}^k_E(M; {}^*V)$ ($k \in \mathbb{N}_0$) and $(\Phi, A) \in \mathfrak{M}_E(M;N)$. 

We can also locally write, using a frame $\mleft( e_a \mright)_a$ of $V$,
\ba
L
&=
L^a \otimes {}^*e_a,
\ea
using that $\mathrm{ev}$-pullbacks generate $\Gamma(\mathrm{ev}^*V)$,
where $L^a \in \mathcal{F}^k_E(M) = \Omega^{k,0}(M \times \mathfrak{M}_E)$ (restriction on open neighbourhood omitted).
\newline\newline
The first calculation of Remark \ref{rem:NotionsOnFunctionals} also shows that we have
\bas
\mathrm{D}
&=
\mathrm{Dev}|_{\mathrm{T}M}
\eas
as functionals, where we view $\mathrm{Dev}|_{\mathrm{T}M}$ as an element of $\mathcal{F}^1_E(M; {}^*\mathrm{T}N)$ given by Eq.~\eqref{DGleichDev}. This implies that we can apply Eq.~\eqref{EqPullBackFormelFuerVerschiedeneDefinitionen}, that is,
\bas
{}^!\omega
&=
\mleft.\mleft(\mathrm{ev}^!\omega\mright)\mright|_{\bigwedge^k \mathrm{T}M}
\stackrel{\eqref{EqPullBackFormelFuerVerschiedeneDefinitionen}}{=}
\frac{1}{k!}~
\mleft(\mathrm{ev}^*\omega\mright)\mleft(\mathrm{Dev}|_{\mathrm{T}M} \stackrel{\wedge}{,} \dotsc \stackrel{\wedge}{,} \mathrm{Dev}|_{\mathrm{T}M} \mright)
=
\frac{1}{k!}~
\mleft({}^*\omega\mright)\mleft(\mathrm{D} \stackrel{\wedge}{,} \dotsc \stackrel{\wedge}{,} \mathrm{D} \mright)
\eas
for all $\omega \in \Omega^k(N;V)$ ($k \in \mathbb{N}_0$). We are going to use this very often by just giving reference to Eq.~\eqref{EqPullBackFormelFuerVerschiedeneDefinitionen}.
\end{remarks}

\begin{examples}{Anchor as functional}{AnchorAsFunctional}
Recall Ex.~\ref{ex:ProjectionOntoGaugeBosonies}; the anchor gives also rise to a functional, especially needed for the minimal coupling. $({}^*\rho)(\varpi_2)$ is a functional in $\mathcal{F}^1_E(M; {}^*\mathrm{T}N)$, that is
\bas
\bigl(({}^*\rho)(\varpi_2)\bigr)(\Phi, A)
&=
(\Phi^*\rho)(A)
\eas
for all $(\Phi, A) \in \mathfrak{M}_E(M;N)$.
\end{examples}

\subsection{Infinitesimal gauge transformations}\label{InfinitesimalGaugeTransformation}

Let us now turn to the definition of the infinitesimal gauge transformation in this general setting, and it will be mostly a straightforward generalisation of Section \ref{ClassicalGaugeTrafoDiscussion}.

\subsubsection{Infinitesimal gauge transformation of the Higgs field}

Let us now make the first step towards the set of vector fields inducing the derivation of infinitesimal gauge transformations. In order to allow any type of connection we assume for now another Lie algebroid $B$ where the connection will later be inherited from, $B$ may or may not be the same as $E$.

\begin{definitions}{Vector fields along Lie algebroid paths}{VectorFieldAlongEPaths}
Let $M, N$ be two smooth manifolds and $\mleft(E, \rho_E, \mleft[ \cdot,\cdot \mright]_E \mright)$, $\mleft(B, \rho_B, \mleft[ \cdot,\cdot \mright]_B \mright)$ two Lie algebroids over $N$. For $(\Phi, A) \in \mathfrak{M}_E(M; N)$ we define $\mathrm{T}^B_{(\Phi,A)}\mathfrak{M}_E(M; N)$ as a subspace of $\mathrm{T}_{(\Phi,A)}\mathfrak{M}_E(M; N)$ by
\ba
\mathrm{T}^B_{(\Phi,A)}\mathfrak{M}_E(M; N)
&\coloneqq
\left\{ (\mathcal{v}, \mathcal{a}) \in \mathrm{T}_{(\Phi,A)}\bigl(\mathfrak{M}_E(M; N)\bigr)
~\middle|~
\exists \epsilon \in \Gamma(\Phi^*B):~
\mathcal{v} = - (\Phi^*\rho_B)(\epsilon)
\right\}.
\ea
The set of sections with values in these subspaces, called the set of \textbf{vector fields along $B$-paths}, is denoted by $\mathfrak{X}^B\bigl(\mathfrak{M}_E(M; N)\bigr)$.
\end{definitions}

\begin{remark}\label{NotASubalgebraXB}
\leavevmode\newline
As images of the pullback of the anchor, it is clear that $\mathrm{T}^B_{(\Phi,A)}\bigl(\mathfrak{M}_E(M; N)\bigr)$ and $\mathfrak{X}^B\bigl(\mathfrak{M}_E(M; N)\bigr)$ are subspaces of $\mathrm{T}_{(\Phi,A)}\bigl(\mathfrak{M}_E(M; N)\bigr)$ and $\mathfrak{X}\bigl(\mathfrak{M}_E(M; N)\bigr)$, respectively.

For all $\Psi \in \mathfrak{X}^B(\mathfrak{M})$ there is by definition then an $\varepsilon \in \mathcal{F}^0_E(M; {}^*B) = \Gamma(\mathrm{ev}^*B)$ such that 
\ba\label{GaugeTrafoVektor}
\Psi
&=
\mleft( -({}^*\rho_B )(\varepsilon), \mathfrak{a} \mright)
\ea
where $({}^*\rho_B)(\varepsilon)$ is an element of $\mathcal{F}^0_E(M; {}^*\mathrm{T}N)$ given by $\mathfrak{M}_E(M; N) \ni (\Phi, A) \mapsto (\Phi^*\rho_B)(\varepsilon(\Phi, A))$, and $\mathfrak{a}$ is a map defined on $\mathfrak{M}_E(M; N)$ such that $\Psi|_{(\Phi,A)}$ is a tangent vector for all $(\Phi, A) \in \mathfrak{M}_E(M; N)$ as characterized in Prop.~\ref{prop:TangentSpaceOfSpaceOfFields}. We will study $\mathfrak{a}$ in more detail later, but now it will not be important. We will write $\Psi \eqqcolon \Psi_\varepsilon$ to emphasize the relationship with an $\varepsilon \in \mathcal{F}^0_E(M; {}^*B)$. For a given $\varepsilon$ there can be several $\Psi_\varepsilon$ as long as we do not fix $\mathfrak{a}$. Moreover, since $\varepsilon \in \mathcal{F}^0_E(M; {}^*B)$ we cannot expect in general that $\mathfrak{X}^B\bigl(\mathfrak{M}_E(M; N)\bigr)$ is a subalgebra of $\mathfrak{X}\bigl(\mathfrak{M}_E(M; N)\bigr)$. One may be able to show that if just allowing $\varepsilon = {}^*b$ ($b \in \Gamma(B)$), but since those more general $\varepsilon$ can have  very general dependencies on $(\Phi,A) \in \mathfrak{M}_E(M;N)$ one cannot expect a sub-algebraic behaviour at this point. We will come back to this after we will have defined the infinitesimal gauge transformation for the field of gauge bosons.

By construction, the flows of those vector fields carry the structure of what is called a Lie algebroid path; we will not need this notion in this paper, and hence we are not going to show this property of the flows. If you are interested into that, then see \cite[Corollary 4.3.3]{MyThesis}. Hence, the name of these vector fields.
\end{remark}

As before there is a relationship similar to Cor.~\ref{cor:VeryGeneralPullbackConnection}, which summarizes the whole motivation of our construction; also recall Remark \ref{rem:CommutingDiagramOfPullbacks}.

\begin{corollaries}{Infinitesimal gauge transformation as condition for allowing pullbacks}{CoolesCommutingDiagramForHiggsTrafosStuff}
Let $M, N$ be two smooth manifolds and $\mleft(E, \rho_E, \mleft[ \cdot,\cdot \mright]_E \mright)$, $\mleft(B, \rho_B, \mleft[ \cdot,\cdot \mright]_B \mright)$ two Lie algebroids over $N$, and $\varepsilon \in \mathcal{F}^0_E(M; {}^*B)$. Then $\Psi \in \mathfrak{X}\bigl(\mathfrak{M}_E(M;N)\bigr)$ is an element of $\mathfrak{X}^B\bigl(\mathfrak{M}_E(M; N)\bigr)$ if and only if there is an $\varepsilon \in \mathcal{F}^0_E(M; {}^*B)$ such that the following diagram commutes
\begin{center}
	\begin{tikzcd}
		M \times \mathfrak{M}_E(M;N) \arrow{r}{-\varepsilon} \arrow{d}{(0, \Psi)}	& B \arrow{d}{\rho_B} 
		\\
		\mathrm{T}\bigl( M \times \mathfrak{M}_E(M;N) \bigr) \arrow{r}{\mathrm{Dev}} & \mathrm{T}N
	\end{tikzcd}
\end{center}
that is
\ba
\mathrm{Dev} \circ (0, \Psi)
&=
-\rho_B \circ \varepsilon,
\ea
where $(0, \Psi) \in \mathfrak{X}(M) \times \mathfrak{X}\bigl(\mathfrak{M}_E(M; N)\bigr)$ is the canonical embedding of $\Psi$ as a vector field on $M \times \mathfrak{M}_E(M; N)$.
\end{corollaries}

\begin{proof}
\leavevmode\newline
The same fashion as for Cor.~\ref{cor:GaugeTrafoAsPullbackCond}, \textit{i.e.}~we can derive $\mathrm{Dev} \circ (0, \Psi) = \Psi^{(\Phi)}$, if writing $\Psi \eqqcolon \mleft( \Psi^{(\Phi)}, \Psi^{(A)} \mright)$. Then $\Psi \in \mathfrak{X}\bigl(\mathfrak{M}_E(M;N)\bigr)$ if and only if $\mathrm{Dev} \circ (0, \Psi) = \Psi^{(\Phi)} = - \rho_B \circ \varepsilon$ for an $\varepsilon \in \mathcal{F}^0_E(M; {}^*B)$.
\end{proof}

Similar to before, we define the first component of vector fields along $B$-paths as the infinitesimal gauge transformation of the Higgs field.

\begin{definitions}{Infinitesimal gauge transformation of $\Phi$}{VariationenOfAundPhi}
Let $M, N$ be two smooth manifolds, $\mleft(E, \rho_E, \mleft[ \cdot,\cdot \mright]_E \mright)$, $\mleft(B, \rho_B, \mleft[ \cdot,\cdot \mright]_B \mright)$ two Lie algebroids over $N$, and $\varepsilon \in \mathcal{F}^0_E(M; {}^*B)$. For a $(\Phi, A) \in \mathfrak{M}_E(M; N)$ we define the \textbf{infinitesimal gauge transformation $\delta^B_{\varepsilon(\Phi, A)} \Phi$ of $\Phi$ along $\varepsilon(\Phi, A)$} as an element of $\Gamma (\Phi^*\mathrm{T}N)$ by
\ba\label{EqVariationOfHiggsField}
\delta^B_{\varepsilon(\Phi, A)} \Phi
&\coloneqq
\bigl( -\mleft({}^*\rho_B\mright)(\varepsilon) \bigr) (\Phi, A)
=
- \mleft( \Phi^* \rho_B \mright)\bigl(\varepsilon(\Phi, A)\bigr),
\ea
shortly denoted as $\delta^B_\varepsilon \Phi \coloneqq - \mleft({}^*\rho_B\mright)(\varepsilon) \in \mathcal{F}^0_E(M; {}^*\mathrm{T}N)$.

In the case of $E=B$ we just write $\delta_\varepsilon \Phi \coloneqq - ({}^*\rho)(\varepsilon)$.
\end{definitions}

\begin{remark}\label{RemUeberVariationVonHiggs}
\leavevmode\newline
As already mentioned in Remark \ref{RemarkAboutThatTheCLassicalFormulaCanBeWrittenDifferently}, Eq.~\eqref{EqVariationOfHiggsField} is also a generalization of a similar equation for a gauge transformation given in \cite[paragraph before Equation (10); we have a different sign in $\varepsilon$]{CurvedYMH}, and it generalizes Def.~\ref{def:ClassicTrafos}.
\end{remark}

That immediately leads to:

\begin{propositions}{Parametrised variations of functionals}{VariationVonSkalarZeugsEasyPeasy}
Let $M, N$ be two smooth manifolds, $\mleft(E, \rho_E, \mleft[ \cdot,\cdot \mright]_E \mright)$, $\mleft(B, \rho_B, \mleft[ \cdot,\cdot \mright]_B \mright)$ two Lie algebroids over $N$, $V \to N$ a vector bundle, ${}^B\nabla$ a $B$-connection on $V$, and $\Psi_\varepsilon \in\mathfrak{X}^B(\mathfrak{M}_E(M; N))$ for $\varepsilon \in \mathcal{F}^0_E(M; {}^*B)$. Then there is a unique $\mathbb{R}$-linear map $\delta_{\Psi_\varepsilon}: \mathcal{F}_E^\bullet(M;{}^*V) \to \mathcal{F}_E^\bullet(M;{}^*V)$ with
\ba\label{PullBackVariation}
\delta_{\Psi_\varepsilon} \mleft( {}^* v \mright)
&=
- {}^*\mleft({}^B\nabla_{\varepsilon} v \mright),
\\
\iota_Y \delta_{\Psi_\varepsilon}
&=
\delta_{\Psi_\varepsilon} \iota_Y
\label{VertauschungMitVerjuengungVonEichtrafo}\\
\delta_{\Psi_\varepsilon}(f \wedge L)
&=
\mathcal{L}_{\Psi_\varepsilon}(f) \wedge L
	+ f \wedge \delta_{\Psi_\varepsilon} (L), \label{LeibnizForGauging}
\ea
for all $Y \in \mathfrak{X}(M)$, $v \in \Gamma(V)$, $L \in \mathcal{F}_E^k(M; {}^*V)$, and $f \in \mathcal{F}^m_E(M)$ ($k, m \in \mathbb{N}_0$), where $\mathcal{F}_E^\bullet(M;{}^*V) \coloneqq \bigoplus_{l\in \mathbb{N}_0} \mathcal{F}^l_E(M; {}^*V)$ while $\delta_{\Psi_\varepsilon}$ keeps a given degree invariant.
\end{propositions}

\begin{remark}
\leavevmode\newline
Since the notation of $\delta_{\Psi_\varepsilon}$ does not emphasize the used connection, we will often roughly write: \textbf{For the functional space $\mathcal{F}^\bullet_E(M;{}^*V)$ let $\delta_{\Psi_\varepsilon}$ be the unique operator of Prop.~\ref{prop:VariationVonSkalarZeugsEasyPeasy}, using ${}^B\nabla$ as a $B$-connection on $V$}, where $\bullet$ denotes an arbitrary degree.
\end{remark}

\begin{proof}[Proof of Prop.~\ref{prop:VariationVonSkalarZeugsEasyPeasy}]
\leavevmode\newline
That is a trivial consequence of Cor.~\ref{cor:CoolesCommutingDiagramForHiggsTrafosStuff} and Cor.~\ref{cor:VeryGeneralPullbackConnection}, that is, we have a unique $\mathbb{R}$-linear operator $\delta_{\Psi_\varepsilon}: \mathcal{F}^0_E(M; {}^*V) \to \mathcal{F}^0_E(M; {}^*V)$ such that
\bas
\delta_{\Psi_\varepsilon}(h s)
&=
\mathcal{L}_{\Psi_\varepsilon}(h) ~ s
	+ h ~ \delta_{\Psi_\varepsilon} s,
\\
\delta_{\Psi_\varepsilon}\underbrace{({}^*v)}_{\mathclap{ = \mathrm{ev}^*v }}
&=
-{}^*\mleft( {}^B\nabla_\varepsilon v \mright)
\eas
for all $s \in \Gamma(\mathrm{ev}^*V) = \mathcal{F}^0_E(M; {}^*V), h \in C^\infty(M \times \mathfrak{M}_E)$, and $v \in \Gamma(V)$. Eq.~\eqref{VertauschungMitVerjuengungVonEichtrafo} and linearity uniquely extends this operator to $\mathcal{F}^\bullet_E(M; {}^*V)$, that is,
\bas
\mleft(\delta_{\Psi_\varepsilon} L\mright)(Y_1, \dotsc, Y_k)
&\coloneqq
\delta_{\Psi_\varepsilon}\bigl( L(Y_1, \dotsc, Y_k)\bigr)
\eas
for all $L \in \mathcal{F}^k_E(M; {}^*V)$ and $Y_1, \dotsc, Y_k \in \mathfrak{X}(M)$; similar to Def.~\ref{def:InfinitesimalGaugeTrafoClassicAsConnection} this is well-defined (recall also the remark after Def.~\ref{def:InfinitesimalGaugeTrafoClassicAsConnection}). Hence, this definition is not in violation with the desired Leibniz rule. The Leibniz rule in Eq.~\eqref{LeibnizForGauging} then just follows by this and the Leibniz rule inherited by Cor.~\ref{cor:VeryGeneralPullbackConnection}.
\end{proof}

\begin{remark}\label{RemLeibnizeRegelaufProdukteWeshalbEConnectionNichtWichtigIst}
\leavevmode\newline
\indent $\bullet$ Given by Remark \ref{JustLieDerivativeForGeneralPullbackAndlineBundle}, for $V = N \times \mathbb{R}$ we always take the canonical flat $B$-connection, \textit{i.e.}~the canonical flat vector bundle connection $\nabla^0 = \mathrm{d}$ and then ${}^B\nabla \coloneqq \nabla^0_{\rho_B}$ such that
\bas
\delta_{\Psi_\varepsilon}
&=
\mathcal{L}_{\Psi_\varepsilon}.
\eas
Thus, 
\ba
\delta_{\Psi_\varepsilon} \mathrm{d}
&=
\mathcal{L}_{\Psi_\varepsilon} \mathrm{d}
=
\mathrm{d} \mathcal{L}_{\Psi_\varepsilon}
=
\mathrm{d} \delta_{\Psi_\varepsilon}, \label{eqVariationVertauschtMitDifferential}
\ea
since coordinates on $\mathfrak{M}_E(M; N)$ and $M$ are independent, where $\mathrm{d}$ is the de-Rham differential on the factor $M$. The Leibniz rule for $\delta_{\Psi_\varepsilon}$ can be then rewritten to
\ba
\delta_{\Psi_\varepsilon}(f \wedge L)
&=
\delta_{\Psi_\varepsilon}(f) \wedge L
	+ f \wedge \delta_{\Psi_\varepsilon} (L).
\ea

$\bullet$ For dual bundles $V^*$ we canonically take the dual connection to ${}^B\nabla$ in order to have Leibniz rules as usual.\footnote{The definition of a dual Lie algebroid connection is as usual.} That also means the following (still keeping the same notation): Let $L \in \mathcal{F}^k_E(M; {}^*V)$ and $T \in \mathcal{F}^0_E(M; {}^*(V^*))$, then in a frame $\mleft( e_a \mright)_a$ of $V$ and $\mleft( f^a \mright)_a$ of $V^*$, $f^b(e_a) = \delta^b_a$, we locally write $L = L^a \otimes {}^*e_a$ and $T = T_b \cdot {}^*f^b$, where $L^a \in \mathcal{F}^k_E(M)$ and $T_b \in \mathcal{F}^0_E(M)$. Then with these conventions, including the previous bullet point,
\ba
\delta_{\Psi_\varepsilon} (T(L))
&=
\delta_{\Psi_\varepsilon} \underbrace{\mleft(
	T_a L^a
\mright)}_{\in \mathcal{F}^k_E(M)}
=
\mathcal{L}_{\Psi_\varepsilon} \mleft(
	T_a L^a
\mright)
=
\mathcal{L}_{\Psi_\varepsilon} (T_a) ~ L^a
	+ T_a ~ \mathcal{L}_{\Psi_\varepsilon}(L^a),
\ea
hence, one achieves an independence of the chosen ${}^B\nabla$ as expected and similar to vector bundle connections. The connections only get important in explicit calculations when applying the Leibniz rule as in
\bas
\delta_{\Psi_\varepsilon} (T(L))
&=
\mleft(\delta_{\Psi_\varepsilon} T\mright)(L)
	+ T\mleft(\delta_{\Psi_\varepsilon} L\mright),
\eas
which can be trivially proven since we took a dual connection. The result would not change of course.
\end{remark}

This recovers the classical idea of infinitesimal gauge transformation, \textit{i.e.}~it is a Lie derivative of components with respect to flat connections; also recall Thm.~\ref{thm:RecoverOfClassicInfgGaugeTrafo}.

\begin{theorems}{Parametrised variations in the flat case}{NewFormulaRecoversOldGaugeTrafoYay}
Let $M, N$ be two smooth manifolds, $\mleft(E, \rho_E, \mleft[ \cdot,\cdot \mright]_E \mright)$, $\mleft(B, \rho_B, \mleft[ \cdot,\cdot \mright]_B \mright)$ two Lie algebroids over $N$, and $V \to N$ a trivial vector bundle. Also let $\nabla$ be the canonical flat connection of $V$, $\Psi_\varepsilon \in \mathfrak{X}^B\bigl( \mathfrak{M}_E(M;N) \bigr)$ for an $\varepsilon \in \mathcal{F}^0_E(M; {}^*B)$ and for $\mathcal{F}^\bullet_E(M; {}^*V)$ let $\delta_{\Psi_\varepsilon}$ be the unique operator of Prop.~\ref{prop:VariationVonSkalarZeugsEasyPeasy}, using ${}^B\nabla \coloneqq \nabla_{\rho_B}$ as a $B$-connection on $V$.

Then we have
\ba
\delta_{\Psi_\varepsilon} L
&=
\mleft(\mathcal{L}_{\Psi_\varepsilon}L^a\mright) \otimes {}^*e_a
\ea
for all $L \in \mathcal{F}^\bullet_E(M; {}^*V)$, where $\mleft( e_a \mright)_a$ is a global constant frame of $V$.
\end{theorems}

\begin{proof}
\leavevmode\newline
That is basically the same proof as in Thm.~\ref{thm:RecoverOfClassicInfgGaugeTrafo}. Take a global constant frame $\mleft( e_a \mright)_a$ of $V$, then
\bas
\nabla e_a &= 0,
\eas
and therefore
\bas
(\Phi^*\nabla)(\Phi^*e_a)
&=
\Phi^!(\nabla e_a)
=
0
\eas
for all $\Phi \in C^\infty(M;N)$. Hence, $({}^*\nabla)({}^*e_a) = {}^!(\nabla e_a) = 0$, such that, using the Leibniz rule,
\bas
\delta_{\Psi_\varepsilon} L
&=
\mleft(\mathcal{L}_{\Psi_\varepsilon}L^a\mright) \otimes {}^*e_a.
\eas
\end{proof}

\subsubsection{Infinitesimal gauge transformation of the field of gauge bosons}

As argued before, we can write $\Psi_\varepsilon = \mleft( -({}^*\rho_B )(\varepsilon), \mathfrak{a} \mright)$ (Eq.~\eqref{GaugeTrafoVektor}), and we want to identify its first and second component as the gauge transformation of $\Phi$ and $A$, respectively. Right now $\mathfrak{a}$ is just fixed by Prop.~\ref{prop:TangentSpaceOfSpaceOfFields} such that it is very arbitrary; as in the standard setting of gauge theory, we want to fix it now.

One of the arguments in the standard formulation is given by looking at the transformation of the minimal coupling; we will do the same. Let us recall what that argument was: Again, let $N =W$ be a vector space, and $E = N \times \mathfrak{g}$ an action Lie algebroid associated to a Lie algebra $\mathfrak{g}$ whose Lie algebra action is induced by a Lie algebra representation $\psi: \mathfrak{g} \to \mathrm{End}(W)$. Then, for an $\epsilon \in C^\infty(M; \mathfrak{g})$, we have the infinitesimal gauge transformation $\delta_\epsilon \Phi = \psi(\epsilon)(\Phi)$ for all $\Phi \in C^\infty(M;W)$; recall Def.~\ref{def:ClassicTrafos}. The minimal coupling $\mathfrak{D}$ is then defined by $\mathfrak{D}(\Phi, A) \coloneqq \mathfrak{D}^A \Phi \coloneqq \mathrm{d}\Phi + \psi(A)(\Phi)$, where $A \in \Omega^1(M; \mathfrak{g})$; as reference see \textit{e.g.}~\cite[Definition 5.9.3; page 292; Definition 7.5.5 \textit{et seq.}; page 426]{hamilton}. The (infinitesimal) gauge transformation of $A$ is then chosen in such a way that it is an element of $\Omega^1(M; \mathfrak{g})$, and such that one gets for the infinitesimal gauge transformation of the minimal coupling
\ba\label{StandardArgumenFuerDieMinimaleKopplungImBabyFall}
\mleft(\delta_\epsilon \mathfrak{D} \mright)(\Phi, A)
= 
\psi(\epsilon) \mleft( \mathfrak{D}^A \Phi \mright)
\ea
among the category of gauge theories. In order to do something similar, we need two ingredients: The definition of the minimal coupling in our general setting, and the basic connection. Let us start with the former.

\begin{definitions}{Minimal coupling, \cite[Eq.~(3), $\Phi$ is denoted as $X$ there]{CurvedYMH}}{MinimalCoupling}
Let $M, N$ be smooth manifolds and $E \to N$ a Lie algebroid. Then we define the \textbf{minimal coupling $\mathfrak{D}$} as an element of $\mathcal{F}_E^1(M; {}^*\mathrm{T}N)$ by
\ba\label{MinimalCouplingInKurz}
\mathfrak{D}
&\coloneqq
\mathrm{D}
	- ({}^*\rho)(\varpi_2).
\ea

We also write
\ba
\mathfrak{D}^A \Phi
&\coloneqq
\mathfrak{D}(\Phi, A)
=
\mathrm{D}\Phi
	- \mleft( \Phi^*\rho\mright)(A)
\ea
for all $\Phi \in C^\infty(M;N)$ and $A \in \Omega^1(M; \Phi^*E)$, and we say that \textbf{$\Phi$ is minimally coupled to $A$}.
\end{definitions}

\begin{remark}
\leavevmode\newline
Restricting this to the standard situation gives back the standard definition: Assume $N = W$ where $W$ is a vector space, $E = W \times \mathfrak{g}$ an action Lie algebroid over $W$,
whose action is induced by a Lie algebra representation $\psi: \mathfrak{g} \to \mathrm{End}(W)$. Then the minimal coupling is
\bas
\mleft.\mathfrak{D}^A \Phi\mright|_p
&=
\mleft.\mathrm{d}_p\Phi^\alpha \otimes \Phi^*\partial_\alpha\mright|_p
	+ \psi\bigl(A_p(Y)\bigr)\bigl(\Phi(p)\bigr)
\eas
for all $(p, \Phi, A) \in M \times \mathfrak{M}_E(M;W)$ and $Y \in \mathrm{T}_pM$,
where we use some global coordinates $\mleft(\partial_\alpha\mright)_\alpha$ of $W$ and Eq.~\eqref{ActionAndRep}. Now we make use of the canonical identification of $W$'s tangent spaces with $W$ itself, especially, $v_\alpha = \partial_\alpha$ for some basis $\mleft( v_\alpha \mright)_\alpha$ on $W$. Then the first summand is clearly $\mathrm{d}\Phi^\alpha \otimes \Phi^*\partial_\alpha = \iota(\mathrm{d}\Phi)$ (the bookkeeping trick with respect to a fixed point $(\Phi, A)$). Hence, also here we arrive at the classical definition.
\end{remark}

In order to study the infinitesimal gauge transformation of $\mathfrak{D}$ we need to fix a Lie algebroid connection on $\mathrm{T}N$ because we want to use Prop.~\ref{prop:VariationVonSkalarZeugsEasyPeasy}; we will use the basic connection.

\begin{definitions}{Basic connection, \cite[Definition 2.9]{basicconn}}{CanonicalBasicConnection}
Let $E \to N$ be a Lie algebroid over a smooth manifold $N$, and let $\nabla$ be a vector bundle connection on $E$. We then define the \textbf{basic connection (induced by $\nabla$)} as a pair of $E$-connections, one on $E$ itself and the other one on $\mathrm{T}N$, both denoted by $\nabla^{\mathrm{bas}}$.
\begin{enumerate}
\item \textbf{(Basic $E$-connection on $E$)}
\newline The basic connection on $E$ is defined as the conjugate of $\nabla_\rho$, that is,
\ba
\nabla^{\mathrm{bas}}_\mu \nu \coloneqq [\mu, \nu]_E + \nabla_{\rho(\nu)} \mu
\ea
for all $\mu, \nu \in \Gamma(E)$
\item \textbf{(Basic $E$-connection on $\mathrm{T}N$)}
\newline The basic connection on $\mathrm{T}N$ is defined by
\ba
\nabla^{\mathrm{bas}}_\mu X \coloneqq [\rho(\mu), X] + \rho\left( \nabla_X \mu \right)
\ea
for all $\mu \in \Gamma(E)$ and $X \in \mathfrak{X}(N)$
\end{enumerate}
\end{definitions}

\begin{remark}
\leavevmode\newline
In the following we often just write of the "basic connection" or $\nabla^{\mathrm{bas}}$, while we then always mean both connections. It should be clear by context which of both connections we mean then. Similar for its curvature $R_{\nabla^\mathrm{bas}}$; but the torsion $t_{\nabla^{\mathrm{bas}}}$ will only denote the torsion for the basic connection on $E$ since only on $E$ the torsion is formulated.

It is also trivial to see that we have
\ba
\rho\circ\nabla^{\mathrm{bas}}
&=
\nabla^{\mathrm{bas}} \circ \rho.
\ea
\end{remark}

It would exceed to discuss the basic connection here in full details; see for example \cite{basicconn}, or \cite{CurvedYMH} and \cite{MyThesis} if you are interested into its relations with gauge theory, although you will see some relations in the following.

We especially also need its related notion of the \textbf{basic curvature}; not to be confused with the curvature of the basic connection.

\begin{definitions}{Basic curvature, \cite[Definition 2.10]{basicconn}}{basiccurvature}
Let $E \to N$ be a Lie algebroid over a smooth manifold $N$, and let $\nabla$ be a connection on $E$. The \textbf{basic curvature $R^{\mathrm{bas}}_\nabla$} is then defined as an element of $\Gamma\left(\bigwedge^2E^* \otimes \mathrm{T}^*N \otimes E \right)$ by
\ba
R^{\mathrm{bas}}_\nabla(\mu, \nu) X
&\coloneqq
\nabla_X\mleft(\mleft[\mu, \nu\mright]_E\mright) 
	- \mleft[ \nabla_X \mu, \nu \mright]_E 
	- \mleft[ \mu, \nabla_X \nu \mright]_E 
	- \nabla_{\nabla^{\mathrm{bas}}_\nu X} \mu 
	+ \nabla_{\nabla^{\mathrm{bas}}_\mu X} \nu,
\ea
where $\mu, \nu \in \Gamma(E)$ and $X \in \mathfrak{X}(N)$.
\end{definitions}

\begin{remark}\label{PropsOfBasicConnection}
\leavevmode\newline
\indent $\bullet$ It is a straight-forward task to check that the basic curvature is a tensor; see also the mentioned references.

$\bullet$ Although this paper will not explain certain aspects of the basic curvature, we want to cite several important relations. As stated in \cite{basicconn} one may think of this as $\nabla_X([\mu, \nu]_E) - [ \nabla_X \mu, \nu ]_E - [ \mu, \nabla_X \nu ]_E$ which is a measure of the derivation property of $\nabla$ w.r.t. $[\cdot, \cdot]_E$, but corrected in such a way that it is tensoriel in all arguments. For a zero anchor the basic curvature would be equivalent to $\nabla_X([\mu, \nu]_E) - [ \nabla_X \mu, \nu ]_E - [ \mu, \nabla_X \nu ]_E$ since then the basic connection on $\mathrm{T}N$ is identically zero.

$\bullet$ As one can also check by straight-forward calculation, one can show
\begin{enumerate}
\item The curvature of $\nabla^{\mathrm{bas}}$ on $E$ is equal to $- R_\nabla^{\mathrm{bas}}(\cdot, \cdot) \circ \rho$,
\item The curvature of $\nabla^{\mathrm{bas}}$ on $\mathrm{T}N$ is equal to $- \rho \circ R^{\mathrm{bas}}_\nabla$,
\end{enumerate}
see \textit{e.g.}~\cite[Proposition 2.11]{basicconn}. Also,
\bas
R_\nabla^{\mathrm{bas}}(\mu, \nu)X 
&= \left( \nabla_X t_{\nabla^{\mathrm{bas}}} \right)(\mu, \nu) 
- R_\nabla(\rho(\mu), X) \nu + R_\nabla(\rho(\nu), X) \mu,
\eas
see for example \cite[Equation (9)]{CurvedYMH} and \cite[generalization of second statement of the first proposition in \S 4.6]{blaomTangentBundleAsLieGroup}.
\end{remark}

Let us go back to the discussion around Eq.~\eqref{StandardArgumenFuerDieMinimaleKopplungImBabyFall}. We want to use the basic connection as the Lie algebroid connection behind $\delta_\varepsilon$, and we now do not assume a second Lie algebroid anymore as we did in the discussion around Prop.~\ref{prop:VariationVonSkalarZeugsEasyPeasy}, hence, $E=B$. We start with a reinterpretation of Eq.~\eqref{StandardArgumenFuerDieMinimaleKopplungImBabyFall} in the context of Section \ref{ClassicalGaugeTrafoDiscussion}.

\begin{corollaries}{Gauge transformation of the minimal coupling in the standard framework}{EichtrafovonDAPHIinClassicIstBabyEinfach}
Let $N=W$ be a vector space, $E = N \times \mathfrak{g}$ be an action Lie algebroid of a Lie algebra $\mathfrak{g}$ whose action is induced by a Lie algebra representation $\psi: \mathfrak{g} \to \mathrm{End}(W)$, $E$ is also equipped with its canonical flat connection $\nabla$. Also let $\Psi_\varepsilon \in \mathfrak{X}^E(\mathfrak{M}_E(M; N))$ for an $\varepsilon \in \mathcal{F}^0_E(M; {}^*E)$ and for the functional space $\mathcal{F}^\bullet_E(M; {}^*\mathrm{T}N)$ let $\delta_{\Psi_\varepsilon}$ be the unique operator of Prop.~\ref{prop:VariationVonSkalarZeugsEasyPeasy}, using $\nabla^{\mathrm{bas}}$ as $E$-connection on $\mathrm{T}N$. Then we have
\ba\label{EichtrafovonDAPHIinClassicIstBabyEinfachDieAequivalenz}
\bigl(\delta_{\Psi_\varepsilon} \mathfrak{D}\bigr)(\Phi, A)
&=
0
&\Leftrightarrow&&
\bigl(\delta_{\Psi_\varepsilon} \mathfrak{D}^\alpha \bigr)(\Phi, A)
&=
\mleft( \psi\bigl(\varepsilon(\Phi, A)\bigr) \mleft( \mathfrak{D}^A \Phi \mright) \mright)^\alpha
\ea
for all $(\Phi, A) \in \mathfrak{M}_E(M; N)$ and $\alpha \in \{1, \dotsc, \mathrm{dim}(W)\}$,
where the components are with respect to global coordinate vector fields $\mleft( \partial_\alpha \mright)_\alpha$ on $W$, and where we used the canonical trivializations $\mathrm{T}W \cong W\times W$ and $\Phi^*\mathrm{T}W \cong M \times W$ such that $\mathfrak{D}^A \Phi$ can be viewed as an element of $\Omega^1(M; W)$.
\end{corollaries}

\begin{proof}
\leavevmode\newline
Let $\mleft( e_a \mright)_a$ be a global and constant frame of $E$ and $\partial_\alpha$ coordinate vector fields on $N$, then we can write $\mathfrak{D} = \mathfrak{D}^\alpha \otimes {}^*\partial_\alpha$, and, thus, by the Leibniz rule and with $\epsilon \coloneqq \varepsilon(\Phi, A)$
\ba
\bigl(\delta_{\Psi_\varepsilon} \mathfrak{D}^\alpha\bigr)(\Phi, A)
	- \bigl( \underbrace{\mleft( \delta_{\Psi_\varepsilon} \mathfrak{D}\mright)}
	_{ \mathclap{ = \delta_{\Psi_\varepsilon} \mleft(\mathfrak{D}^\alpha\mright) \otimes {}^*\partial_\alpha
		+ \mathfrak{D}^\alpha \otimes \delta_{\Psi_\varepsilon} \mleft({}^*\partial_\alpha\mright) } }
	(\Phi, A) \bigr)^\alpha
&=
	- \biggl( \mleft( \mathfrak{D}^A \Phi \mright)^\beta \otimes \underbrace{\mleft(\delta_{\Psi_\varepsilon} \mleft( {}^* \partial_\beta \mright)\mright) (\Phi, A)}_{\mathclap{\stackrel{\text{Prop.~\ref{prop:VariationVonSkalarZeugsEasyPeasy}}}{=} - \Phi^* \mleft( \nabla^{\mathrm{bas}}_\epsilon \partial_\beta\mright)}} \biggr)^\alpha
\nonumber \\
&=
\epsilon^a ~ \Phi^*\mleft( 
	- \partial_\beta\rho_a^\alpha
	+ \rho^\alpha\mleft( \nabla_{\partial_\beta} e_a \mright) 
\mright) ~ \mleft( \mathfrak{D}^A \Phi \mright)^\beta \label{CompsVonDMinimalAlsErstes}
\ea
for all $\alpha$.
Let us write $\partial_\alpha = \partial/\partial w^\alpha$ for some coordinates $\mleft( w^\alpha \mright)_\alpha$ on $W$. Then by Eq.~\eqref{ActionAndRep},
\ba\label{eqAbleitungVomAnkerGibtRepraesentierung}
- \partial_\beta\bigl[ w \mapsto \rho_a^\alpha(w) \bigr]
&=
- \partial_\beta\bigl[ w \mapsto \gamma_a^\alpha(w) \bigr]
=
\partial_\beta\mleft[ w \mapsto \bigl(\psi(e_a)(w)\bigr)^\alpha \mright]
=
\bigl( \psi(e_a) \bigr)^\alpha_\beta
\ea
for $w \in W$, because the differential is then just the differential of a matrix vector-product $W \ni w \mapsto \psi(e_a)(w)$. Since $\nabla$ is the canonical flat connection, constant sections are parallel, thus, we get in total
\bas
\mleft(\delta_{\Psi_\varepsilon} \mathfrak{D}^\alpha\mright)(\Phi, A)
	- \bigl( \mleft( \delta_{\Psi_\varepsilon} \mathfrak{D}\mright) (\Phi, A) \bigr)^\alpha
&=
\epsilon^a ~ \Phi^*\underbrace{\bigl( \psi(e_a) \bigr)^\alpha_\beta}_{\mathclap{\text{const.}}} ~ \mleft( \mathfrak{D}^A \Phi \mright)^\beta
=
\mleft( \psi(\epsilon) \mleft( \mathfrak{D}^A \Phi \mright) \mright)^\alpha
\eas
for all $\alpha$, having $\epsilon \in C^\infty(M; \mathfrak{g})$ and $\mathfrak{D}^A \Phi \in \Omega^1(M;W)$. That shows that we have 
\bas
\mleft(\delta_{\Psi_\varepsilon} \mathfrak{D}^\alpha\mright)(\Phi, A)
&=
\mleft( \psi(\epsilon) \mleft( \mathfrak{D}^A \Phi \mright) \mright)^\alpha
\eas
if and only if 
\bas
\delta_{\Psi_\varepsilon} \mathfrak{D}
&=
0.
\eas
\end{proof}

The right equation in the Equivalence \eqref{EichtrafovonDAPHIinClassicIstBabyEinfachDieAequivalenz} describes precisely the components of the expected infinitesimal gauge transformation of the minimal coupling in the standard formulation of gauge theory, and it is no coincidence that this is equivalent to $\delta_{\Psi_\varepsilon} \mathfrak{D} = 0$ if using the basic connection: The basic connection on $E$ and $\mathrm{T}N$ can be seen as a generalization of the adjoint and Lie algebra representation $\psi$, respectively. See for example \cite{CurvedYMH} or \cite[Lemma 4.3.12]{MyThesis}.

Hence, when using the basic connection, we want that $\delta_{\Psi_\varepsilon} \mathfrak{D} = 0$ such that we can recover the classical formula in sense of Cor.~\ref{cor:EichtrafovonDAPHIinClassicIstBabyEinfach}. To study this and later results we need several auxiliary results, recall also Ex.~\ref{ex:ProjectionOntoGaugeBosonies}, \ref{ex:DAsFunctional} and \ref{ex:AnchorAsFunctional}.

\begin{lemmata}{Several identities related to variations with the basic connection}{VariationsIdentities}
Let $M, N$ be two smooth manifolds, $E \to N$ a Lie algebroid over $N$, $\nabla$ a connection on $E$, and $\Psi_\varepsilon \in \mathfrak{X}^E(\mathfrak{M}_E(M; N))$ for an $\varepsilon \in \mathcal{F}^0_E(M; {}^*E)$. For both functional spaces, $\mathcal{F}^\bullet_E(M; {}^*E)$ and $\mathcal{F}^\bullet_E(M; {}^*\mathrm{T}N)$, let $\delta_{\Psi_\varepsilon}$ be the unique operator of Prop.~\ref{prop:VariationVonSkalarZeugsEasyPeasy}, using $\nabla^{\mathrm{bas}}$ as $E$-connection on $E$ and $\mathrm{T}N$, respectively. Then
\ba
\delta_{\Psi_\varepsilon} \mathrm{D}
&=
- \mleft( {}^*\rho \mright) \bigl( {}^*\nabla \varepsilon \bigr), \label{DPhiVariation}
\\
\delta_{\Psi_\varepsilon} \mleft({}^*\rho\mright)
&=
0, \label{eqPhiRhoDieGeileSauIstnichtVariiert}
\\
\delta_{\Psi_\varepsilon} \bigl( ({}^*\rho)(\varpi_2) \bigr)
&=
\mleft( {}^* \rho \mright) \bigl( \delta_{\Psi_\varepsilon} \varpi_2 \bigr),
\label{eqRhoAVariation}
\\
\delta_{\Psi_\varepsilon} \mleft( {}^!\mleft(\nabla \mu \mright) \mright)
&=
- \biggl(
	{}^!\mleft(\nabla^{\mathrm{bas}}_\varepsilon \nabla \mu \mright)
	+ {}^*\mleft( \nabla_{({}^*\rho)\mleft( ({}^*\nabla) \varepsilon \mright)} \mu \mright)
\biggr) \label{EqVariationVonFormenBrrrr}
\ea
for all $\mu \in \Gamma(E)$, where we view $\nabla \mu$ as an element of $\Omega^1(N; E)$.
\end{lemmata}

\begin{remark}
\leavevmode\newline
Regarding the notation for Eq.~\eqref{EqVariationVonFormenBrrrr}, let us shortly write down what it is for each $(\Phi, A) \in \mathfrak{M}_E(M; N)$,
\bas
\mleft(\delta_{\Psi_\varepsilon} \mleft( {}^!\mleft(\nabla \mu \mright) \mright)\mright)(\Phi, A)
&=
- \biggl(
	\Phi^!\mleft(\nabla^{\mathrm{bas}}_{\epsilon} \mleft( \nabla \mu \mright)\mright)
	+ \Phi^*\mleft( \nabla_{(\Phi^*\rho)\mleft( (\Phi^*\nabla) \epsilon \mright)} \mu \mright)
\biggr)
\eas
where $\epsilon \coloneqq \varepsilon(\Phi, A)$ and we view terms like $\nabla \mu$ as elements of $\Omega^1(N; E)$. When $\varepsilon = {}^*\nu$ for a $\nu \in \Gamma(E)$, then $(\Phi^*\nabla ) (\Phi^*\nu) = \Phi^!(\nabla \nu)$ by definition of the pullback connection, so, $({}^*\nabla) ({}^*\nu) = {}^!(\nabla \nu)$. Thus, we can then write
\ba\label{EqVariationVonFormenBrrrrVereinfacht}
\delta_{\Psi_{{}^*\nu}} \mleft( {}^!\mleft(\nabla \mu \mright) \mright)
&=
- {}^!\mleft(
	\nabla^{\mathrm{bas}}_\nu \nabla \mu
	+ \nabla_{\rho(\nabla \nu)} \mu
\mright).
\ea
\end{remark}

\begin{proof}[Proof for Lemma \ref{lem:VariationsIdentities}]
\leavevmode\newline
In the following $\mleft( e_a \mright)_a$ denotes a local frame of $E$, and $\partial_\alpha$ are local coordinate vector fields on $N$, and $(\Phi, A) \in \mathfrak{M}_E(M; N)$. Regarding $\varepsilon \in \mathcal{F}^0_E(M; {}^*E)$ we also write $\epsilon \coloneqq \varepsilon(\Phi, A)$.

$\bullet$ For Eq.~\eqref{DPhiVariation} we write locally
\bas
\mathrm{D} \Phi
&=
\mathrm{d} \Phi^\alpha \otimes \Phi^* \partial_\alpha,
\eas
where we view $(\Phi,A) \mapsto \Phi^\alpha$ as an element of $\mathcal{F}^0_E(M)$ (on an open subset of $M$), such that by $\delta_\varepsilon \Phi = - ({}^*\rho) (\varepsilon)$, and by using $\mathrm{d} \delta_{\Psi_\varepsilon} = \delta_{\Psi_\varepsilon} \mathrm{d}$ and $\delta_{\Psi_\varepsilon} = \mathcal{L}_{\Psi_\varepsilon}$ on $\mathcal{F}^0_E(M)$ (recall the discussion around Eq.~\eqref{eqVariationVertauschtMitDifferential}),
\bas
\mleft(\delta_\varepsilon \mathrm{d} \mleft[ (\Phi, A) \mapsto \Phi^\alpha \mright]\mright)(\Phi, A)
&=
\mleft(\mathrm{d} \mathcal{L}_{\Psi_\varepsilon} \mleft[ (\Phi, A) \mapsto \Phi^\alpha \mright]\mright)(\Phi, A)
=
- \mathrm{d} \mleft( \mleft( \rho^\alpha_a \circ \Phi \mright) ~ \epsilon^a \mright)
\eas
then by Eq.~\eqref{PullBackVariation} and the Leibniz rule of $\delta_{\Psi_\varepsilon}$
\bas
\mleft(\delta_{\Psi_\varepsilon} \mathrm{D} \mright)(\Phi, A)
&=
- \mathrm{d} \mleft( \mleft( \rho^\alpha_a \circ \Phi \mright) ~ \epsilon^a \mright) \otimes \Phi^* \partial_\alpha
	- \mathrm{d} \Phi^\alpha \otimes \epsilon^a~ \Phi^* \mleft( \nabla^{\mathrm{bas}}_{e_a}\partial_\alpha \mright) \\
&=
- \Bigl( \underbrace{\mathrm{d} \mleft( \rho^\alpha_a \circ \Phi \mright)}
	_{\mathclap{= ~ \mleft(\partial_\beta \rho^\alpha_a \circ \Phi \mright) ~ \mathrm{d} \Phi^\beta}}
 ~ \epsilon^a
	+ \mleft( \rho^\alpha_a \circ \Phi \mright) ~ \mathrm{d}\epsilon^a \Bigr) \otimes \Phi^* \partial_\alpha \\
&\hspace{1cm}
	- \mathrm{d} \Phi^\alpha \otimes \epsilon^a~ \Phi^* \mleft( 
	- \partial_\alpha \rho^\beta_a ~ \partial_\beta
	+ \rho \mleft( \nabla_{\partial_\alpha} e_a \mright)
	  \mright) \\
&=
- \mleft( \rho^\alpha_a \circ \Phi \mright) ~ \mathrm{d}\epsilon^a \otimes \Phi^* \partial_\alpha
	- \mathrm{d} \Phi^\beta \otimes \epsilon^b~ \mleft( \rho^\alpha_a \circ \Phi \mright) ~ \mleft(\omega_{b\beta}^a \circ \Phi \mright)
	~\Phi^*\partial_\alpha \\
&=
- \mleft( \rho^\alpha_a \circ \Phi \mright) \mleft(
\mathrm{d}\epsilon^a
	+ \epsilon^b~ \mleft(\omega_{b\beta}^a \circ \Phi \mright) ~\mathrm{d} \Phi^\beta
\mright) \otimes \Phi^*\partial_\alpha \\
&=
- \mleft( \Phi^*\rho \mright)\bigl( \mleft( \Phi^* \nabla\mright) \epsilon \bigr).
\eas

$\bullet$ By Eq.~\ref{PullBackVariation},
\bas
\delta_{\Psi_\varepsilon} \mleft( {}^*\rho \mright)
&=
- {}^*\mleft( \nabla^{\mathrm{bas}}_\varepsilon \rho \mright),
\eas
and by $\rho \circ \nabla^{\mathrm{bas}} = \nabla^{\mathrm{bas}} \circ \rho$ we get
\bas
\mleft( \nabla^{\mathrm{bas}} \rho \mright)(\mu)
&=
\nabla^{\mathrm{bas}}\mleft( \rho(\mu) \mright)
	- \rho\mleft( \nabla^{\mathrm{bas}}\mu \mright)
=
0
\eas
for all $\mu \in \Gamma(E)$. Hence,
\bas
\delta_{\Psi_\varepsilon} \mleft( {}^*\rho \mright)
&=
0.
\eas

$\bullet$ By the Leibniz rule and the previous result we also have
\bas
\delta_{\Psi_\varepsilon} \bigl( ({}^*\rho)(\varpi_2) \bigr)
&=
\mleft( {}^* \rho \mright) \bigl( \delta_{\Psi_\varepsilon} \varpi_2 \bigr).
\eas

$\bullet$ We view terms like $\nabla \mu$ as elements of $\Omega^1(N; E)$ for all $\mu \in \Gamma(E)$, $\mathfrak{X}(N) \ni Y \mapsto (\nabla \mu)(X) = \nabla_X \mu$, and therefore we can use the Leibniz rule on ${}^!(\nabla \mu) = \bigl({}^*(\nabla \mu)\bigr)(\mathrm{D}) = {}^*\mleft(\nabla_{\mathrm{D}} \mu\mright)$, \textit{i.e.}~due to
\bas
\Phi^!(\nabla \mu) &= \bigl(\Phi^*(\nabla \mu)\bigr)(\mathrm{D}\Phi)
\eas
we can view ${}^!(\nabla \mu)$ as a contraction of the functionals ${}^*(\nabla \mu)$ and $\mathrm{D}$. Hence,
\bas
\delta_{\Psi_\varepsilon} \mleft( {}^!\mleft(\nabla \mu \mright) \mright)
&=
\bigl( \delta_{\Psi_\varepsilon} ({}^*(\nabla \mu))\bigr)(\mathrm{D})
	+ {}^*\mleft(\nabla_{\delta_{\Psi_\varepsilon} \mathrm{D}} \mu\mright)
\\
&\stackrel{\mathclap{\text{Eq.~\eqref{PullBackVariation}}}}{=}~~~~
-\mleft({}^*\mleft(\nabla^{\mathrm{bas}}_\varepsilon \nabla \mu \mright)\mright)(\mathrm{D})
	+ {}^*\mleft(\nabla_{\delta_{\Psi_\varepsilon} \mathrm{D}} \mu\mright)
\\
&\stackrel{\mathclap{\text{Eq.~\eqref{DPhiVariation}}}}{=}~~~~
- \biggl(
	{}^!\mleft(\nabla^{\mathrm{bas}}_\varepsilon \nabla \mu \mright)
	+ {}^*\mleft( \nabla_{({}^*\rho)\mleft( ({}^*\nabla) \varepsilon \mright)} \mu \mright)
\biggr).
\eas
\end{proof}

Let us now fix the gauge transformation of $A$ using these results. Recall that we write $\Psi = \Psi_\varepsilon$ for a $\Psi \in \mathfrak{X}^E(\mathfrak{M}_E(M; N))$, where $\varepsilon \in \mathcal{F}^0_E(M;{}^*E)$ such that we can write (recall Eq.~\eqref{GaugeTrafoVektor})
\bas
\Psi_\varepsilon
&=
\mleft( -({}^*\rho_B )(\varepsilon), \mathfrak{a} \mright)
\eas
where $\mathfrak{a}$ is a map on $\mathfrak{M}_E(M; N)$ such that $\Psi|_{(\Phi,A)}$ is a tangent vector for all $(\Phi, A) \in \mathfrak{M}_E(M; N)$, \textit{i.e.}~satisfying the diagram of Prop.~\ref{prop:TangentSpaceOfSpaceOfFields} for all $(\Phi, A)$. For a given $\varepsilon$ such a $\Psi_\varepsilon$ is in general not unique. We denote also the following with respect to a local frame $\mleft( e_a \mright)_a$ of $E$ and local coordinate functions $\mleft(\partial_\alpha\mright)_\alpha$ on $N$ we have
\bas
\mleft[ e_b, e_c \mright]_E
&=
C^a_{bc} e_a, &
\nabla e_b
&=
\omega^a_b \otimes e_a, &
\nabla_{\partial_\alpha} e_b
&=
\omega^a_{b\alpha} ~ e_a,
\eas
\textit{i.e.}~$C^a_{bc}$ denote the structure functions of $\mleft[\cdot, \cdot\mright]_E$ and $\omega^a_b$ the connection 1-forms of $\nabla$.

\begin{propositions}{Gauge transformation of the field of gauge bosons}{VariationOfA}
Let $M, N$ be two smooth manifolds, $E \to N$ a Lie algebroid over $N$, $\nabla$ a connection on $E$, $\varepsilon \in \mathcal{F}^0_E(M; {}^*E)$, and for the functional space $\mathcal{F}^\bullet_E(M; {}^*E)$ let $\delta_{\Psi_\varepsilon}$ be the unique operator of Prop.~\ref{prop:VariationVonSkalarZeugsEasyPeasy}, using $\nabla^{\mathrm{bas}}$ as $E$-connection on $E$ and any $\Psi_\varepsilon \in \mathfrak{X}^E\bigl( \mathfrak{M}_E(M;N) \bigr)$. Then there is a unique $\Psi_\varepsilon \in \mathfrak{X}^E\bigl(\mathfrak{M}_E(M; N)\bigr)$ such that
\ba\label{EichtrafoVonANochmal}
\delta_{\Psi_\varepsilon} \varpi_2
&=
- ({}^*\nabla) \varepsilon.
\ea
Locally with respect to a given frame $\mleft( e_a \mright)_a$
\ba
\mleft(\delta_{\Psi_\varepsilon} \varpi_2^a\mright)(\Phi, A)
&=
\mleft( C^a_{bc} \circ \Phi \mright) ~\epsilon^b A^c
	+ \mleft(\omega^a_{b\alpha} \circ \Phi \mright) ~ \mleft( \rho^\alpha_c \circ \Phi \mright)~\epsilon^b A^c
	- \mathrm{d}\epsilon^a - \epsilon^b ~ \Phi^!\mleft(\omega^a_{b} \mright)
\nonumber \\ \label{eqGaugeTrafoOfAacomps}
&=
\mleft( \epsilon^b A^c \otimes \Phi^*\mleft( \nabla^{\mathrm{bas}}_{e_b} e_c\mright)
	- \mleft(\Phi^*\nabla\mright)\epsilon \mright)^a
\ea
for all $(\Phi, A) \in \mathfrak{M}_E(M; N)$, where $\epsilon \coloneqq \varepsilon(\Phi, A)$.

Moreover, if we also have $\alpha, \beta \in \mathbb{R}$ and $\vartheta \in \mathcal{F}^0_E(M; {}^*E)$, then
\ba\label{LinearityOfPsiEpsilon}
\Psi_{\alpha \varepsilon + \beta \vartheta}
=
\alpha \Psi_\varepsilon + \beta \Psi_\vartheta,
\ea
where the vector fields are the ones uniquely given by Eq.~\eqref{EichtrafoVonANochmal}.
\end{propositions}

\begin{proof}[Proof of Prop.~\ref{prop:VariationOfA}]
\leavevmode\newline
Since it is about a vector field on $\mathfrak{M}_E(M; N)$, we will classify $\Psi_\varepsilon$ by its flow: We denote its flow through a fixed point $(\Phi_0, A_0) \in \mathfrak{M}_E(M; N)$ by $\gamma: I \to \mathfrak{M}_E(M; N)$, $t \mapsto \gamma(t) \eqqcolon (\Phi_t, A_t) \in \mathfrak{M}_E(M; N)$, where $I$ is an open interval of $\mathbb{R}$ containing 0, and we write $\Psi|_{\gamma(t)} = \mleft( - (\Phi_t^*\rho)(\epsilon_t), \mathcal{a}_t \mright) \in \mathrm{T}^E_{(\Phi_t, A_t)}\mathfrak{M}_E(M; N)$, where $\epsilon_t \coloneqq \varepsilon(\Phi_t, A_t)\in \Gamma(\Phi^*_tE)$, and $\mathcal{a}_t$ is a morphism $\mathrm{T}M \to \mathrm{T}E$ satisfying the diagram in Prop.~\ref{prop:TangentSpaceOfSpaceOfFields}. So, we have a curve $\gamma$ with 
\bas
\gamma(0) &= (\Phi_0, A_0), \\
\frac{\mathrm{d}}{\mathrm{d}t} \gamma
&=
\Psi|_{\gamma(t)}
=
\mleft( - (\Phi_t^*\rho)(\epsilon_t), \mathcal{a}_t \mright).
\eas
$(\Phi_0, A_0)$ and $- (\Phi_t^*\rho)(\epsilon_t)$ are fixed, and we show that Eq.~\eqref{EichtrafoVonANochmal} will fix $\mathcal{a}_t$. Without loss of generality let us assume that everything is small and local enough such that we have frames and coordinates, like a frame $\mleft(e_a\mright)_a$ of $E$.\footnote{One could even fix a point $p \in M$ because we just need an interval for $t$ for $\mathrm{d}/\mathrm{d}t$.}
Making use of Prop.~\ref{prop:VariationVonSkalarZeugsEasyPeasy}, we get
\bas
\mleft(\delta_{\Psi_\varepsilon} \varpi_2 \mright) (\Phi_t, A_t)
&=
\mleft.\mathcal{L}_{\Psi_\varepsilon} \mleft(\varpi^a_2\mright)\mright|_{(\Phi_t, A_t)} \otimes \Phi^*_t e_a
	- A_t^a \otimes \Phi^*_t\mleft( \nabla^{\mathrm{bas}}_{\epsilon_t} e_a \mright).
\eas
Eq.~\eqref{EichtrafoVonANochmal} does then hold if and only if
\bas
&\mleft.\mathcal{L}_{\Psi_\varepsilon} \mleft(\varpi^a_2\mright)\mright|_{(\Phi_t, A_t)} \otimes \Phi^*_t e_a
\\
&=
\epsilon_t^b A_t^c \otimes \Phi^*_t\mleft( \nabla^{\mathrm{bas}}_{e_b} e_c \mright)
	- \mleft(\Phi_t^* \nabla\mright) \epsilon_t
\\
&=
\mleft(
\mleft( C^a_{bc} \circ \Phi_t \mright) ~\epsilon_t^b A_t^c
	+ \mleft(\omega^a_{b\alpha} \circ \Phi_t \mright) ~ \mleft( \rho^\alpha_c \circ \Phi_t \mright)~\epsilon_t^b A_t^c
	- \mathrm{d}\epsilon_t^a - \epsilon_t^b ~ \Phi_t^!\mleft(\omega^a_{b} \mright)
\mright) \otimes \Phi^*_t e_a
\eas
which gives Eq.~\eqref{eqGaugeTrafoOfAacomps} at $t=0$. By the definition of $\gamma$ and the Lie derivative we also get
\bas
\mleft.\mathcal{L}_{\Psi_\varepsilon} \mleft(\varpi^a_2\mright)\mright|_{(\Phi_t, A_t)} 
&=
\frac{\mathrm{d}}{\mathrm{d}t} \mleft(\varpi^a_2\circ\gamma\mright)
=
\frac{\mathrm{d}}{\mathrm{d}t} \mleft[ t \mapsto A^a_t \mright],
\eas
and, thus,
\ba\label{DiffEqFuerAComp}
\frac{\mathrm{d}}{\mathrm{d}t} \mleft[ t \mapsto A^a_t \mright]
&=
\mleft( C^a_{bc} \circ \Phi_t \mright) ~\epsilon^b_t A_t^c
	+ \mleft(\omega^a_{b\alpha} \circ \Phi_t \mright) ~ \mleft( \rho^\alpha_c \circ \Phi_t \mright)~\epsilon^b_t A_0^c
	- \mathrm{d}\epsilon^a - \epsilon^b_t ~ \Phi_t^!\mleft(\omega^a_{b} \mright).
\ea
So, Eq.~\eqref{EichtrafoVonANochmal} is equivalent to a set of coupled differential equations: We have a curve $\gamma(t) = (\Phi_t, A_t)$, with $\Phi_{t=0} = \Phi_0$ and 
\bas
\frac{\mathrm{d}}{\mathrm{d}t} [t \mapsto \Phi_t]
&=
- (\Phi_t^*\rho) (\epsilon_t),
\eas
and $A_{t=0} = A_0$, while
\bas
\mathcal{a}_t
&=
\frac{\mathrm{d}}{\mathrm{d}t} \mleft[ t \mapsto A_t \mright]
=
\frac{\mathrm{d}}{\mathrm{d}t} \mleft[ t \mapsto A_t^a \otimes \Phi^*_t e_a \mright].
\eas
$t \mapsto \Phi_t$ and $t \mapsto A^a_t$ are uniquely given by this system and the differential equation \eqref{DiffEqFuerAComp}, and, so, $t \mapsto A_t = A^a_t \otimes \Phi^*_t e_a$ is uniquely given, too. Hence, $\mathcal{a}_t$ is unique, and, thus, $\Psi_\varepsilon$. Alternatively, the differential equations for $\mathrm{d}/\mathrm{d}t ~ \Phi$ and $\mathrm{d}/\mathrm{d}t ~ A^a$ are the action of the vector field $\Psi_\varepsilon$ on the coordinates of $\mathfrak{M}_E$ (along the flow line $\gamma$), and therefore defining $\Psi_\varepsilon$.

The linearity of $\psi_\varepsilon$ in $\varepsilon$ over $\mathbb{R}$ simply follows by the linearity given in the differential equations above: Define $\Theta \coloneqq \alpha \Psi_\varepsilon + \beta \Psi_\vartheta$ for $\alpha, \beta \in \mathbb{R}$ and $\vartheta \in \mathcal{F}^0_E(M; {}^*E)$, where $\Psi_\varepsilon$ and $\Psi_\vartheta$ are the unique vector fields as given above, \textit{i.e.}~$\delta_{\Psi_\varepsilon} \varpi_2 = - ({}^*\nabla) \varepsilon$ and $\delta_{\Psi_\vartheta} \varpi_2 = - ({}^*\nabla) \vartheta$, respectively. Observe that $\Theta \in \mathfrak{X}^E\bigl(\mathfrak{M}_E(M; N)\bigr)$, where the component along the "$\Phi$-direction" is by definition given by
\bas
-\alpha ~ ({}^*\rho)(\varepsilon) - \beta ~ ({}^*\rho)(\vartheta)
&=
- ({}^*\rho)(\alpha \varepsilon + \beta \vartheta),
\eas
then, using the linearity of Eq.~\eqref{DiffEqFuerAComp} in $\varepsilon$,
\bas
\delta_\Theta \varpi_2
&=
\mathcal{L}_\Theta (\varpi_2^a) \otimes {}^*e_a
	- \varpi_2^a \otimes {}^*\mleft(\nabla^{\mathrm{bas}}_{\alpha \varepsilon + \beta \vartheta} e_a\mright)
\\
&=
\mleft( \alpha \mathcal{L}_{\Psi_{\varepsilon}}  + \beta \mathcal{L}_{\Psi_\vartheta} \mright) (\varpi_2^a) \otimes {}^*e_a
	- \varpi_2^a \otimes {}^*\mleft(\nabla^{\mathrm{bas}}_{\alpha \varepsilon + \beta \vartheta} e_a\mright)
	\\
&\stackrel{\mathclap{ \text{ Eq.~\eqref{DiffEqFuerAComp}} }}{=}~~~
\mathcal{L}_{\Psi_{\alpha \varepsilon + \beta \vartheta}} (\varpi_2^a) \otimes {}^*e_a
	- \varpi_2^a \otimes {}^*\mleft(\nabla^{\mathrm{bas}}_{\alpha \varepsilon + \beta \vartheta} e_a\mright)
\\
&=
\delta_{\Psi_{\alpha \varepsilon + \beta \vartheta}} \varpi_2.
\eas
By the shown uniqueness of vector fields like $\Psi_{\alpha \varepsilon + \beta \vartheta}$, we get
\bas
\Theta
&=
\Psi_{\alpha \varepsilon + \beta \vartheta}.
\eas
\end{proof}

\begin{remark}\label{RemDifferentVersionsOfGaugeTrafos}
\leavevmode\newline
Eq.~\eqref{eqGaugeTrafoOfAacomps} is also \textit{e.g.}~defined in \cite[Eq.~(10); opposite sign of $\varepsilon$]{CurvedYMH}, but in this reference it was not known how a coordinate-free version can look like. This equation recovers the standard formula of the infinitesimal gauge transformation of $A$.
In order to see why this restricts to the standard formula, let us look again at the standard setting: When $E= N \times \mathfrak{g}$ is an action Lie algebroid with Lie algebra $\mathfrak{g}$, equipped with its canonical flat connection $\nabla$, then we get the classical formula of gauge transformation by using a constant frame $\mleft( e_a \mright)_a$ for $E$, \textit{i.e.}
\bas
\mleft(\delta_{\Psi_\varepsilon} \varpi_2^a\mright)(\Phi, A)
&=
\Phi^*C^a_{bc} ~ \epsilon^b A^c - \mathrm{d}\epsilon^a 
=
\mleft(
	\mleft[ \epsilon \stackrel{\wedge}{,} A \mright]_{\mathfrak{g}}
	- \mathrm{d}^{\Phi^*\nabla} \epsilon 
\mright)^a
\eas
for all $(\Phi, A) \in \mathfrak{M}_E(M; N)$,
because $\omega^a_b = 0$ and $\Phi^*C^a_{bc} = C^a_{bc} = \text{const.}$, the structure constants of $\mathfrak{g}$. We can understand $\epsilon$ as an element of $C^\infty(M; \mathfrak{g})$ as usual in the standard setting. That is precisely the typical formula of the classical setting as in Def.~\ref{def:ClassicTrafos}, because $\Phi^*\nabla$ is the canonical flat connection of $\Phi^*E \cong M \times \mathfrak{g}$ such that $\mathrm{d}^{\Phi^*\nabla} = \mathrm{d}$.
\end{remark}

Using such a $\Psi_\varepsilon$ results into an infinitesimal gauge transformation of the minimal coupling as in Cor.~\ref{cor:EichtrafovonDAPHIinClassicIstBabyEinfach}.

\begin{propositions}{Infinitesimal gauge transformation of the minimal Coupling}{InfinitesimalGaugeTrafoOfMinimalCoupleSmiley}
Let $M, N$ be two smooth manifolds, $E \to N$ a Lie algebroid over $N$, $\nabla$ a connection on $E$, and $\varepsilon \in \mathcal{F}^0_E(M; {}^*E)$ together with the unique $\Psi_\varepsilon \in \mathfrak{X}^E(\mathfrak{M}_E(M; N))$ as given in Prop.~\ref{prop:VariationOfA}. For both functional spaces, $\mathcal{F}^\bullet_E(M; {}^*E)$ and $\mathcal{F}^\bullet_E(M; {}^*\mathrm{T}N)$, let $\delta_{\Psi_\varepsilon}$ be the unique operator of Prop.~\ref{prop:VariationVonSkalarZeugsEasyPeasy}, using $\nabla^{\mathrm{bas}}$ as $E$-connection on $E$ and $\mathrm{T}N$, respectively.

Then we have
\ba
\delta_{\Psi_\varepsilon} \mathfrak{D}
&=
0.
\ea
\end{propositions}

\begin{remark}
\leavevmode\newline
We already have derived the variation of the components of $\mathfrak{D}$, for this recall the general calculation for Eq.~\eqref{CompsVonDMinimalAlsErstes}:
Let $\mleft( e_a \mright)_a$ be a local frame of $E$ and $\partial_\alpha$ coordinate vector fields on $N$, then we can write $\mathfrak{D} = \mathfrak{D}^\alpha \otimes {}^*\partial^\alpha$, and, thus, with $\epsilon \coloneqq \varepsilon(\Phi, A)$, 
\ba
\bigl(\delta_{\Psi_\varepsilon} \mathfrak{D}^\alpha\bigr)(\Phi, A)
&=
\epsilon^a ~ \Phi^*\mleft( 
	- \partial_\beta\rho_a^\alpha
	+ \rho^\alpha\mleft( \nabla_{\partial_\beta} e_a \mright) 
\mright) ~ \mleft( \mathfrak{D}^A \Phi \mright)^\beta.
\ea
That is precisely the same formula as given in \cite[Eq.~(12), different sign for $\epsilon$ there]{CurvedYMH}, but there only the formula for the components was known.
\end{remark}

\begin{proof}[Proof of Prop.~\ref{prop:InfinitesimalGaugeTrafoOfMinimalCoupleSmiley}]
\leavevmode\newline
This quickly follows by Lemma \ref{lem:VariationsIdentities}, especially Eq.~\eqref{DPhiVariation} and \eqref{eqRhoAVariation},
\bas
\delta_{\Psi_\varepsilon} \mathfrak{D}
&=
\delta_{\Psi_\varepsilon} \bigl( \mathrm{D} - ({}^*\rho)(\varpi_2) \bigr)
=
-({}^*\rho)({}^*\nabla\varepsilon)
	- \mleft( {}^* \rho \mright) \bigl( \delta_{\Psi_\varepsilon} \varpi_2 \bigr)
\stackrel{\text{Prop.~\ref{prop:VariationOfA}}}{=}
0.
\eas
\end{proof}

By this result and Cor.~\ref{cor:EichtrafovonDAPHIinClassicIstBabyEinfach} we define the following.

\begin{definitions}{Infinitesimal gauge transformation of gauge bosons}{GaugeTrafoOfA}
Let $M, N$ be two smooth manifolds, $E \to N$ a Lie algebroid over $N$, $\nabla$ a connection on $E$, and $\varepsilon \in \mathcal{F}^0_E(M; {}^*E)$ together with the unique $\Psi_\varepsilon \in \mathfrak{X}^E\bigl(\mathfrak{M}_E(M; N)\bigr)$ as given in Prop.~\ref{prop:VariationOfA}. For the functional space $\mathcal{F}^\bullet_E(M; {}^*E)$ let $\delta_{\Psi_\varepsilon}$ be the unique operator of Prop.~\ref{prop:VariationVonSkalarZeugsEasyPeasy}, using $\nabla^{\mathrm{bas}}$ as $E$-connection on $E$.

For a $(\Phi,A) \in \mathfrak{M}_E(M; N)$ we define the \textbf{infinitesimal gauge transformation $\delta_{\varepsilon(\Phi, A)} A$ of $A$} as an element of $\Omega^1(M; \Phi^*E)$ by
\ba
\delta_{\varepsilon(\Phi, A)} A
&\coloneqq
\mleft( \delta_{\Psi_\varepsilon} \varpi_2 \mright)(\Phi, A)
=
- (\Phi^*\nabla) \bigl( \varepsilon(\Phi, A) \bigr),
\ea
shortly denoted as $\delta_\varepsilon A \coloneqq \delta_{\Psi_\varepsilon} \varpi_2 = -({}^*\nabla) \varepsilon$. Given a local frame $\mleft( e_a \mright)_a$ of $E$, we also similarly define $\delta_\varepsilon A^a \coloneqq \delta \varpi_2^a$.
\end{definitions}

\begin{remark}\label{WhyNablaBasPartOne}
\leavevmode\newline
As discussed in Remark \ref{RemDifferentVersionsOfGaugeTrafos} we have seen that $\delta_\varepsilon A^a$ (using a frame $\mleft( e_a \mright)_a$ of $E$) recovers the classical formula of the infinitesimal gauge transformation. However, the total formula, $\delta_\varepsilon A$, does not recover it which is no problem due to that the variation of the Lagrangian just depends on the variation of the components and its variation is independent on whether or not one uses the basic connection in the definition of $\delta_\varepsilon$; for a rigorous discussion about the (generalised) Lagrangian and its gauge invariance see for example \cite{CurvedYMH} and \cite[\S 4]{MyThesis}.

Alternatively, one could use $\nabla_\rho$ as $E$-connection on $E$ instead of $\nabla^{\mathrm{bas}}$ for the definition of $\delta_{\Psi_\varepsilon}$; especially because of results like Thm.~\ref{thm:NewFormulaRecoversOldGaugeTrafoYay} and Thm.~\ref{thm:RecoverOfClassicInfgGaugeTrafo}, which imply that one recovers classical formulas when $\nabla$ is additionally flat.\footnote{A flat connection is locally canonically flat with respect to the trivialization given by a parallel frame; later we will also see that then $E$ is locally an action algebroid and $\nabla$ its canonical flat connection, if $\nabla$ is flat and has vanishing basic curvature.} When using $\nabla_\rho$, the same $\Psi_\varepsilon$ leads to
\ba
\delta_{\Psi_\varepsilon} \varpi_2
&=
-({}^*t_{\nabla_\rho})(\varepsilon, \varpi_2) - ({}^*\nabla)\varepsilon,
\ea
where $t_{\nabla_\rho}$ is the torsion of $\nabla_\rho$. As we have seen before, $\nabla$ will be the canonical flat connection in the standard setting such that then $\delta_{\Psi_\varepsilon} A^a = \mleft(\delta_{\Psi_\varepsilon} A \mright)^a$ by flatness and Thm.~\ref{thm:NewFormulaRecoversOldGaugeTrafoYay}. With similar calculations as before one also shows that the variation of the components, $\delta_{\Psi_\varepsilon} \varpi_2^a$, recovers the classical formula of the infinitesimal gauge transformation of the field of gauge bosons, thus, $\delta_{\Psi_\varepsilon} \varpi_2$ would restrict to the classical formula in the standard setting, too. Hence, $\nabla_\rho$ would look like the canonical choice, not $\nabla^{\mathrm{bas}}$. But we will later see that while $\nabla_\rho$ is in general not flat, $\nabla^{\mathrm{bas}}$ will be flat after applying a reasonable condition, such that only for the basic connection the infinitesimal gauge transformations in form of the operator $\delta_{\Psi_\varepsilon}$ will give rise to a Lie algebra in general. Moreover, we are not going to fix any separate connection on $\mathrm{T}N$ which would be identified with a canonical flat connection in the standard situation, such that the only canonical choice for $\mathrm{T}N$ is the basic connection; using the basic connections also for $E$-valued tensors is then in alignment to $\mathrm{T}N$-valued tensors.
\end{remark}

\subsubsection{Infinitesimal gauge transformation of functionals}

Hence, we finally arrived at defining the infinitesimal gauge transformation of functionals.

\begin{definitions}{Infinitesimal gauge transformation of functionals}{TotalInfGaugeTrafoYayy}
Let $M, N$ be two smooth manifolds, $E \to N$ a Lie algebroid over $N$, $V \to N$ a vector bundle, $\nabla$ a connection on $E$, ${}^E\nabla$ an $E$-connection on $V$, and $\varepsilon \in \mathcal{F}^0_E(M; {}^*E)$ together with the unique $\Psi_\varepsilon \in \mathfrak{X}^E(\mathfrak{M}_E(M; N))$ as given uniquely in Prop.~\ref{prop:VariationOfA}. For the functional space $\mathcal{F}^\bullet_E(M; {}^*V)$ let $\delta_{\Psi_\varepsilon}$ be the unique operator as in Prop.~\ref{prop:VariationVonSkalarZeugsEasyPeasy}, using ${}^E\nabla$ as $E$-connection on $V$.

Then we define the \textbf{infinitesimal gauge transformation $\delta_\varepsilon L$ of $L \in \mathcal{F}^\bullet_E(M; {}^*V)$} as an element of $\mathcal{F}^\bullet_E(M; {}^*V)$ by
\ba
\delta_\varepsilon L
&\coloneqq
\delta_{\Psi_\varepsilon} L.
\ea

For $V=E$ or $V= \mathrm{T}N$ we take ${}^E\nabla = \nabla^{\mathrm{bas}}$ on $E$ and $\mathrm{T}N$, respectively; for all further tensor spaces constructed of $E$ and $\mathrm{T}N$, like their duals, we take the canonical extensions of the basic connection.
\end{definitions}

\begin{remark}
\leavevmode\newline
In the following we will have just one connection $\nabla$ on $E$ and ${}^E\nabla$ on $V$ given. Without mentioning it further, we always use these connections for the definition of $\delta_\varepsilon$ because it should be clear by context.
\end{remark}

We can quickly list two properties about $\delta_\varepsilon$.

\begin{corollaries}{Linearity in $\varepsilon$}{DeltaEpsilonIstLinearInEpsilon}
Let us assume the same as for Def.~\ref{def:TotalInfGaugeTrafoYayy}. Then
\ba
\delta_{\alpha \varepsilon + \beta \vartheta}
&=
\alpha \delta_\varepsilon
	+ \beta \delta_\vartheta
\ea
for all $\alpha, \beta \in \mathbb{R}$ and $\varepsilon, \vartheta \in \mathcal{F}^0_E(M; {}^*E)$.
\end{corollaries}

\begin{proof}
\leavevmode\newline
Let $k \in \mathbb{N}_0$, $L \in \mathcal{F}^k_E(M; {}^*V)$ and $\mleft( e_a \mright)_a$ a local frame of $V$. Then, using Eq.~\eqref{LinearityOfPsiEpsilon} and the Leibniz rule,
\bas
\delta_{\alpha \varepsilon + \beta \vartheta} L
&=
\underbrace{\mathcal{L}_{\Psi_{\alpha \varepsilon + \beta \vartheta}} L^a}
_{\mathclap{ \stackrel{Eq.~\eqref{LinearityOfPsiEpsilon}}{=} \mathcal{L}_{\alpha \Psi_\varepsilon + \beta \Psi_\vartheta} } }
 \otimes ~ {}^*e_a
	- L^a \otimes {}^*\mleft( {}^E\nabla_{\alpha \varepsilon + \beta \vartheta} e_a \mright)
\\
&=
\alpha ~ \mleft(
	\mathcal{L}_{\Psi_\varepsilon} L^a \otimes {}^*e_a
	- L^a \otimes {}^*\mleft( {}^E\nabla_\varepsilon e_a \mright)
\mright)
	+ \beta ~ \mleft(
	\mathcal{L}_{\Psi_\vartheta} L^a \otimes {}^*e_a
	- L^a \otimes {}^*\mleft( {}^E\nabla_\vartheta e_a \mright)
\mright)
\\
&=
\mleft(\alpha \delta_\varepsilon
	+ \beta \delta_\vartheta\mright) L,
\eas
where vector fields like $\Psi_\varepsilon$ are given by Def.~\ref{def:TotalInfGaugeTrafoYayy}.
\end{proof}

\begin{corollaries}{Independence of $\nabla$}{WennVonAUnabhaengigDannAuchVonNabla}
Let us assume the same as for Def.~\ref{def:TotalInfGaugeTrafoYayy}, and let $L \in \mathcal{F}^k_E(M; {}^*V)$ ($k \in \mathbb{N}_0$) be independent of $A$, \textit{i.e.}~$L(\Phi,A) = L(\Phi, A^\prime)$ for all $(\Phi,A), (\Phi, A^\prime) \in \mathfrak{M}_E(M; N)$.

Then the definition of $\delta_\varepsilon L$ is independent of $\nabla$.\footnote{But not of ${}^E\nabla$, so, if ${}^E\nabla = \nabla^{\mathrm{bas}}$, then there is still the dependency on $\nabla$ in the role of ${}^E\nabla$.}
\end{corollaries}

\begin{proof}
\leavevmode\newline
Let $\mleft( e_a \mright)_a$ be a local frame of $V$, and write $L = L^a \otimes {}^*e_a$, then, using that $\delta_\varepsilon = \mathcal{L}_{\Psi_\varepsilon}$ on $\mathcal{F}^k_E(M)$ (recall Remark \ref{RemLeibnizeRegelaufProdukteWeshalbEConnectionNichtWichtigIst}, and $\Psi_\varepsilon$ is given by Def.~\ref{def:TotalInfGaugeTrafoYayy}),
\bas
\delta_\varepsilon L
&=
\mathcal{L}_{\Psi_\varepsilon} L^a \otimes {}^*e_a
	- L^a \otimes {}^*\mleft({}^E\nabla_\varepsilon e_a \mright).
\eas
The second summand is already independent of $\nabla$, so, let us look at the first summand. Recall that $\Psi_\varepsilon$ contains two components, the first is the differentiation along the "$\Phi$-direction", given by $-({}^*\rho)(\varepsilon)$, and the second for the "$A$-direction", fixed by Prop.~\ref{prop:VariationOfA} using $\nabla$. Due to the independence of $L$ with respect to $A$ we can conclude that $L^a$ must be independent of $A$ since ${}^* e_a$ is already independent of $A$, thus,
\bas
\mathcal{L}_{\Psi} L^a
&=
\mathcal{L}_{\Psi^\prime} L^a
\eas
for all $\Psi, \Psi^\prime \in \mathfrak{X}(\mathfrak{M}_E(M; N))$ whose first component, the derivative along "$\Phi$"-coordinates, coincide. Hence, regardless which connection $\nabla$ we choose to fix the second component of $\Psi_\varepsilon$ the definition of $\delta_\varepsilon L$ will be unaffected by this choice.
\end{proof}

\subsection{Commutator of gauge transformations}\label{CommutatorOfGaugeTrafos}

We want to calculate
\bas
\delta_\vartheta \delta_\varepsilon - \delta_\varepsilon \delta_\vartheta
\eas
for all $\varepsilon, \vartheta \in \mathcal{F}^0_E(M; {}^*E)$, and we want a behaviour similar to representations; thence, we need a bracket on $\mathcal{F}^0_E(M; {}^*E)$. For $\Phi \in C^\infty(M;N)$, $\Phi^*E$ is in general not a Lie algebroid, see \cite[\S 3.2ff.]{meinrenkensplitting} or \cite[\S 7.4; page 42ff.]{meinrenkenlie} about conditions on $\Phi$ which imply a natural Lie algebroid structure on $\Phi^*E$. Therefore we cannot expect to have a Lie bracket on sections of $\Phi^*E$. The essential problem is that we do not have an anchor on $\Phi^*E\to M$ in general such that one cannot try to construct first a bracket on pullbacks of sections and then to canonically extend such a bracket, and this problem extends to $\mathcal{F}^0_E(M; {}^*E)$. But there is a better object measuring a "bracket-like" behaviour on this functional space; we will see at the end that this will be actually a Lie bracket.

\begin{definitions}{Pre-bracket on $\mathcal{F}^0_E(M; {}^*E)$}{PrebracketonPullbackLiealgebroid}
Let $M, N$ be smooth manifolds, $E \to N$ a Lie algebroid, and $\nabla$ a connection on $E$.

Then we define the \textbf{pre-bracket $\llbracket \cdot, \cdot \rrbracket: \mathcal{F}^0_E(M; {}^*E) \times \mathcal{F}^0_E(M; {}^*E) \to \mathcal{F}^0_E(M; {}^*E)$} by
\ba
\llbracket \vartheta, \varepsilon \rrbracket
&\coloneqq
\delta_\varepsilon \vartheta - \delta_\vartheta \varepsilon - \bigl( {}^*t_{\nabla^{\mathrm{bas}}} \bigr)\mleft( \vartheta, \varepsilon \mright)
\ea
for all $\varepsilon, \vartheta \in \mathcal{F}^0_E(M; {}^*E)$.
\end{definitions}

\begin{remark}\label{IdeaOfPrebracket}
\leavevmode\newline
Given an $E$-connection ${}^E\nabla$ on $E$, Lie brackets can be expressed as
\bas
\mleft[ \mu, \nu \mright]_E
&=
{}^E\nabla_\mu \nu
	- {}^E\nabla_\nu \mu
	- t_{{}^E\nabla}(\mu, \nu)
\eas
for all $\mu, \nu \in \Gamma(E)$. Recall that $\delta$ is strongly related to a certain pullback of $\nabla^{\mathrm{bas}}$; then the idea of the pre-bracket is to use the right-hand side as a definition. Since we know under which conditions and how to make pullbacks of $E$-connections and tensors, we circumvent the problem of defining a Lie bracket and anchor on a pullback bundle.
\end{remark}

Let us study this bracket.

\begin{propositions}{Properties of the pre-bracket}{PropertiesOfThePreBracket}
Let $M, N$ be smooth manifolds, $E \to N$ a Lie algebroid, and $\nabla$ a connection on $E$.

Then we have
\ba\label{DeltaIstZumGlueckANtisymm}
\llbracket \cdot, \cdot \rrbracket &\textit{ is antisymmetric}, \\
\llbracket \cdot, \cdot \rrbracket &\textit{ is $\mathbb{R}$-bilinear}, \\
\mleft \llbracket {}^*\mu, {}^*\nu \mright \rrbracket
&=
{}^*\bigl( \mleft[ \mu, \nu \mright]_E \bigr) \label{EqLieKlammerAufPullBackSections}
\ea
for all $\varepsilon, \vartheta \in \mathcal{F}^0_E(M; {}^*E)$, $f \in \mathcal{F}^0_E(M)$, $\mu, \nu \in \Gamma(E)$, and, when expressing everything with respect to a pull-back of a local frame $\mleft( e_a \mright)_a$ of $E$, we get
\ba\label{EqDeltaInFrameKoord}
\llbracket \vartheta, \varepsilon \rrbracket
&=
\delta_{\varepsilon} \vartheta^a ~ {}^*e_a
	- \delta_{\vartheta} \varepsilon^a ~ {}^*e_a
	+ \vartheta^a \varepsilon^b ~  {}^*\bigl( 
	\mleft[ e_a, e_b \mright]_E
	\bigr)
\ea
for all $\vartheta, \varepsilon \in \mathcal{F}^0_E(M; {}^*E)$.

Moreover, $\llbracket \vartheta, \varepsilon \rrbracket$ is independent of the chosen connection $\nabla$ when both, $\varepsilon$ and $\vartheta$, are independent of $A$, that is, $\varepsilon(\Phi, A) = \varepsilon (\Phi, A^\prime)$ for all $(\Phi, A), (\Phi, A^\prime) \in \mathfrak{M}_E(M;N)$; similar for $\vartheta$.
\end{propositions}

\begin{remark}\label{ClassicalCommutatorRemark}
\leavevmode\newline
Eq.~\eqref{EqLieKlammerAufPullBackSections} and \eqref{EqDeltaInFrameKoord} emphasize that we have a suitable candidate in $\llbracket\cdot, \cdot\rrbracket$ as bracket. The latter actually proves that $\llbracket \cdot, \cdot \rrbracket$ is independent of the choice about whether or not one uses the basic connection to define $\delta_\varepsilon$ because the infinitesimal gauge transformation of scalar-valued functionals is just a Lie derivative, see later also Remark \ref{RemarkBracketIsVeryIndependent}. Similar to how one can express $\mleft[ \cdot, \cdot \mright]_E$ using Lie algebroid connections as in Remark \ref{IdeaOfPrebracket}, but $\mleft[ \cdot, \cdot \mright]_E$ is of course independent of any choice of Lie algebroid connection.

Let $E = N \times \mathfrak{g}$ be an action Lie algebroid, the usual relationship in classical gauge theory is for $\varepsilon, \vartheta \in C^\infty(M;\mathfrak{g})$ that
\bas
\mleft[ \delta^{\mathrm{clas}}_\varepsilon, \delta^{\mathrm{clas}}_\vartheta \mright]A
&=
- \delta^{\mathrm{clas}}_{\mleft[\varepsilon, \vartheta\mright]_{\mathfrak{g}}}A,
\eas
where $\delta^{\mathrm{clas}}_\varepsilon$ is given by Def.~\ref{def:ClassFunctionalGaugeTrafoBlag}, and the negative sign on the right hand side is due to our choice of sign with respect to $\varepsilon$, which we prove later in full generality. This shows that we want that $\llbracket \cdot, \cdot \rrbracket$ generalizes and recovers $\mleft[\cdot, \cdot\mright]_{\mathfrak{g}}$. As we discussed, we apply the "bookkeeping trick" to formulate infinitesimal gauge transformations, also recall Def.~\ref{def:InfinitesimalGaugeTrafoClassicAsConnection} and Thm.~\ref{thm:RecoverOfClassicInfgGaugeTrafo}. That is, for a constant frame $\mleft( e_a \mright)_a$ of $E$, we have the "bookkeeping trick" $\iota(\varepsilon)$ given by
\bas
\iota(\varepsilon)
&=
\varepsilon^a ~ {}^*e_a,
\eas
hence, the bookeeping trick is essentially a frame-dependent embedding of the functionals given in the classical gauge theory into $\mathcal{F}^\bullet_E$. $\varepsilon^a$ are in this case only functions depending on $M$, but not on $\mathfrak{M}_E(M;N)$, especially, $\delta^{\mathrm{clas}}_\vartheta \varepsilon^a = 0$. By Eq.~\eqref{EqDeltaInFrameKoord} we then have
\bas
\bigl\llbracket \iota(\vartheta), \iota(\varepsilon) \bigr\rrbracket
&=
\vartheta^a \varepsilon^b ~ {}^*\bigl( 
	\mleft[ e_a, e_b \mright]_{\mathfrak{g}}
	\bigr)
=
\iota\mleft(\mleft[ \vartheta, \varepsilon \mright]_{\mathfrak{g}}\mright),
\eas
which is precisely what we want and expect of a generalized bracket.
\end{remark}

\begin{proof}[Proof of Prop.~\ref{prop:PropertiesOfThePreBracket}]
\leavevmode\newline
The antisymmetry is clear, and the bilinearity follows by the linearity of $\delta_\varepsilon$ for all $\varepsilon \in \mathcal{F}^0_E(M; {}^*E)$, recall Cor.~\ref{cor:DeltaEpsilonIstLinearInEpsilon}. We have
\bas
\bigl( {}^*t_{\nabla^{\mathrm{bas}}} \bigr)\mleft( {}^*\mu, {}^*\nu \mright)
&=
{}^*\mleft( \bigl( t_{\nabla^{\mathrm{bas}}} \bigr)\mleft( \mu, \nu \mright) \mright)
=
{}^*\mleft( 
	\nabla^{\mathrm{bas}}_\mu \nu 
	- \nabla^{\mathrm{bas}}_\nu \mu 
	- \mleft[ \mu, \nu \mright]_E 
\mright)
\eas
for all $\mu, \nu \in \Gamma(E)$, and
\bas
\delta_{{}^*\nu} \mleft( {}^*\mu \mright)
&=
- {}^*\mleft( \nabla^{\mathrm{bas}}_\nu \mu \mright),
\eas
therefore
\bas
\mleft\llbracket {}^*\mu, {}^*\nu \mright\rrbracket
&=
{}^*\mleft( \nabla^{\mathrm{bas}}_\mu \nu \mright)
	- {}^*\mleft( \nabla^{\mathrm{bas}}_\nu \mu \mright)
	- {}^*\mleft( 
	\nabla^{\mathrm{bas}}_\mu \nu 
	- \nabla^{\mathrm{bas}}_\nu \mu 
	- \mleft[ \mu, \nu \mright]_E 
\mright)
=
{}^*\bigl( \mleft[ \mu, \nu \mright]_E \bigr),
\eas
which proves Eq.~\eqref{EqLieKlammerAufPullBackSections}.
For $\varepsilon, \vartheta \in \mathcal{F}^0_E(M;{}^*E)$ we have, with respect to a frame $\mleft( e_a \mright)_a$ of $E$,
\bas
\delta_\vartheta \varepsilon
&=
\delta_{\vartheta} \varepsilon^a ~ {}^*e_a
	- \varepsilon^a \vartheta^b ~ {}^*\mleft(\nabla^{\mathrm{bas}}_{e_b} e_a\mright),
\eas
and so
\bas
\llbracket\vartheta, \varepsilon\rrbracket
&=
\delta_{\varepsilon} \vartheta^a ~ {}^*e_a
	- \vartheta^a \varepsilon^b ~ {}^*\mleft(\nabla^{\mathrm{bas}}_{e_b} e_a\mright)
	- \delta_{\vartheta} \varepsilon^a ~ {}^*e_a
	+ \varepsilon^a \vartheta^b ~ {}^*\mleft(\nabla^{\mathrm{bas}}_{e_b} e_a\mright)
\\
&\hspace{1cm}
	- \vartheta^a \varepsilon^b ~ {}^*\mleft( 
	\nabla^{\mathrm{bas}}_{e_a} e_b 
	- \nabla^{\mathrm{bas}}_{e_b} e_a 
	- \mleft[ e_a, e_b \mright]_E
	\mright) \\
&=
\delta_{\varepsilon} \vartheta^a ~ {}^*e_a
	- \delta_{\vartheta} \varepsilon^a ~ {}^*e_a
	+ \vartheta^a \varepsilon^b ~ 
	{}^*\bigl( 
		\mleft[ e_a, e_b \mright]_E
	\bigr).
\eas
This expression for $\llbracket\vartheta, \varepsilon\rrbracket$ shows that its value is independent of the chosen $\nabla$, when the functionals $\varepsilon = \varepsilon^a \otimes {}^*e_a$ and $\vartheta= \vartheta^a \otimes {}^*e_a$ are independent of $A$, since then also their components with respect to $\mleft({}^*e_a\mright)_a$ are independent of $A$ because ${}^*e_a$ is already independent of $A$. Then apply Cor.~\ref{cor:WennVonAUnabhaengigDannAuchVonNabla}.
\end{proof}

\begin{corollaries}{$\llbracket\cdot, \cdot\rrbracket$ a Lie bracket on the pull-backs of $\Gamma(E)$}{DeltaIstEineLieklammerAufPullbACkSections}
Let $M, N$ be smooth manifolds, $E \to N$ a Lie algebroid, and $\nabla$ a connection on $E$.

Then the restriction of $\llbracket \cdot, \cdot \rrbracket$ on pullback functionals is a Lie bracket.
\end{corollaries}

\begin{proof}
\leavevmode\newline
The antisymmetry, the bilinearity over $\mathbb{R}$ and the closedness follow by Prop.~\ref{prop:PropertiesOfThePreBracket}, the same also for the Jacobi identity by observing
\bas
\bigl\llbracket {}^*\mu, \mleft\llbracket {}^*\nu, {}^*\eta \mright\rrbracket \bigr\rrbracket
&\stackrel{\text{Eq.~\eqref{EqLieKlammerAufPullBackSections}}}{=}
\mleft\llbracket {}^*\mu, {}^*\bigl(\mleft[ \nu,\eta \mright]_E\bigr) \mright\rrbracket
\stackrel{\text{Eq.~\eqref{EqLieKlammerAufPullBackSections}}}{=}
{}^* \bigl( \mleft[ \mu, \mleft[ \nu, \eta \mright]_E \mright]_E\bigr)
\eas
for all $\mu, \nu, \eta \in \Gamma(E)$. Hence, the Jacobiator of the restriction of $\llbracket \cdot, \cdot \rrbracket$ on pullback functionals is given by the pullback of the Jacobiator of $\mleft[ \cdot, \cdot \mright]_E$, the latter is of course zero.
\end{proof}

We will see that $\llbracket\cdot, \cdot\rrbracket$ is actually always a Lie bracket, but for proving this we do not want to show the Jacobi identity directly, due to how we constructed it we rather are going to use the equivalence of Bianchi identities of curvatures with Jacobi identities of brackets; recall any proof of the first Bianchi identity of a curvature. Hence, let us define the curvature we are interested into.

\begin{definitions}{Curvature of infinitesimal gauge transformations}{ErsteKruemmungsFormelFuerEichtrafos}
Let $M, N$ be smooth manifolds, $E \to N$ a Lie algebroid, $V \to N$ a vector bundle, $\nabla$ a connection on $E$, and ${}^E\nabla$ an $E$-connection on $V$.

Then we define the \textbf{curvature $R_{\delta}$} by
\ba
\mathcal{F}^0_E(M;{}^*E) \times \mathcal{F}^0_E(M;{}^*E) \times \mathcal{F}^k_E(M;{}^*V) &\to \mathcal{F}^k_E(M;{}^*V)
\nonumber \\
(\vartheta, \varepsilon, L) &\mapsto R_{\delta}(\vartheta, \varepsilon)L
\coloneqq
	\delta_\vartheta \delta_\varepsilon L 
	- \delta_\varepsilon \delta_\vartheta L 
	+ \delta_{\llbracket\vartheta, \varepsilon \rrbracket} L
\ea
for all $\vartheta, \varepsilon \in \mathcal{F}^0_E(M; {}^*E)$ and $L \in \mathcal{F}^k_E(M;{}^*V)$ ($k \in \mathbb{N}_0$ arbitrary).

In alignment to Def.~\ref{def:GaugeTrafoOfA} we denote $R_\delta(\cdot, \cdot) A \coloneqq R_\delta (\cdot, \cdot) \varpi_2$, and $R_\delta(\cdot, \cdot)A^a \coloneqq R_\delta(\cdot, \cdot)\varpi_2^a$ with respect to a frame $\mleft(e_a\mright)_a$ of $E$.
\end{definitions}

\begin{remark}
\leavevmode\newline
The sign in front of the third term depends on which sign one takes in the definition of $\delta_\varepsilon$. Changing the sign $\varepsilon$ in the definitions of the gauge tranformations would lead to a minus sign in front of the third summand.
\end{remark}

Using a frame of $E$ we can apply the Leibniz rule; since the curvature related to derivations is as expected again a derivation:

\begin{corollaries}{Relationships between curvatures}{RelationShipsOfCurvatures}
Let $M, N$ be smooth manifolds, $E \to N$ a Lie algebroid, $V \to N$ a vector bundle, $\nabla$ a connection on $E$, and ${}^E\nabla$ an $E$-connection on $V$. Then locally
\ba
R_\delta(\cdot, \cdot)L
&=
R_\delta(\cdot, \cdot)L^a \otimes {}^*e_a
	+ L^a \otimes {}^*\bigl( R_{{}^E\nabla}(\cdot, \cdot)e_a \bigr) 
\ea
for all $L \in \mathcal{F}^k_E(M; {}^*V)$ ($k \in \mathbb{N}_0$), where $\mleft( e_a \mright)_a$ is a local frame of $E$ and viewing $R_{{}^E\nabla}(\cdot, \cdot)e_a$ as an element of $\Omega^2(E;E)$.
\end{corollaries}

\begin{proof}
\leavevmode\newline
Let us first study terms like $R_{\delta}(\vartheta, \varepsilon)\mleft({}^*h\mright)$ for $\varepsilon, \vartheta \in \mathcal{F}^0_E(M;{}^*E)$ and $h \in \Gamma(V)$, using a local frame $\mleft( e_a \mright)_a$ of $E$,
\bas
\delta_\vartheta \delta_\varepsilon ({}^*h)
&=
- \delta_\vartheta \mleft(
	\varepsilon^a ~ {}^*\mleft({}^E\nabla_{e_a} h\mright)
\mright)
=
- \delta_\vartheta \varepsilon^a ~ {}^*\mleft({}^E\nabla_{e_a} h\mright)
	+ \varepsilon^a \vartheta^b ~ {}^*\mleft({}^E\nabla_{e_b} {}^E\nabla_{e_a} h\mright),
\eas
and
\bas
\delta_{\llbracket\vartheta, \varepsilon\rrbracket} ({}^*h)
~~~~&\stackrel{\mathclap{\text{Eq.~\eqref{EqDeltaInFrameKoord}}}}{=}~~~~
- \mleft(
	\delta_{\varepsilon} \vartheta^a
	- \delta_{\vartheta} \varepsilon^a
	+ \vartheta^b ~ \varepsilon^c ~  \mleft({}^*\bigl( 
	\mleft[ e_b, e_c \mright]_E
	\bigr)\mright)^a
\mright) ~ {}^*\mleft({}^E\nabla_{e_a} h\mright)
\\
&=
\delta_\vartheta \varepsilon^a ~ {}^*\mleft({}^E\nabla_{e_a} h\mright)
	- \delta_\varepsilon \vartheta^a ~ {}^*\mleft({}^E\nabla_{e_a} h\mright)
	- \varepsilon^a \vartheta^b ~ {}^*\mleft( {}^E\nabla_{\mleft[ e_b, e_a \mright]_E} h \mright),
\eas
in total
\bas
R_{\delta}(\vartheta, \varepsilon)\mleft({}^*h\mright)
&=
\varepsilon^a \vartheta^b ~ {}^* \underbrace{\mleft(
	{}^E\nabla_{e_b} {}^E\nabla_{e_a} h
	- {}^E\nabla_{e_a} {}^E\nabla_{e_b} h
	- {}^E\nabla_{\mleft[ e_a, e_b \mright]_E} h
\mright)}_{R_{{}^E\nabla}(e_b,e_a)h}
=
\mleft(
	{}^*\bigl( R_{{}^E\nabla}(\cdot,\cdot)h \bigr)
\mright)(\vartheta, \varepsilon).
\eas
Therefore we arrive at
\bas
R_\delta(\vartheta, \varepsilon)\mleft(L^a \otimes {}^*e_a\mright)
&=
\delta_\vartheta \delta_\varepsilon L^a \otimes {}^*e_a
	+ \delta_\varepsilon L^a \otimes \delta_\vartheta\mleft({}^*e_a\mright)
	+ \delta_\vartheta L^a \otimes \delta_\varepsilon\mleft({}^*e_a\mright)
	+ L^a \otimes \delta_\vartheta \delta_\varepsilon \mleft({}^*e_a\mright)
\\
&\hspace{1cm}
	- (\vartheta \leftrightarrow \varepsilon)
\\
&\hspace{1cm}
	+ \delta_{\llbracket\vartheta, \varepsilon\rrbracket} L^a \otimes {}^*e_a
	+ L^a \otimes \delta_{\llbracket\vartheta, \varepsilon\rrbracket} {}^*e_a
\\
&=
R_\delta(\vartheta, \varepsilon)L^a \otimes {}^*e_a
	+ L^a \otimes R_\delta(\vartheta, \varepsilon) ({}^*e_a)
\\
&=
R_\delta(\cdot, \cdot)L^a \otimes {}^*e_a
	+ L^a \otimes \mleft({}^*\bigl( R_{{}^E\nabla}(\cdot, \cdot)e_a \bigr) \mright)(\vartheta, \varepsilon)
\eas
for all $L \in \mathcal{F}^k_E(M; {}^*V)$.
\end{proof}

Keep in mind that $R_\delta$ is not a typical curvature, for example $\delta_\varepsilon$ is not $C^\infty$-linear with respect to $\varepsilon$, such that it is not immediately clear whether this curvature is a tensor in all arguments, so, we need to prove this if we want to simplify calculations. We are first focusing on $R_\delta(\cdot, \cdot) A$.

\begin{propositions}{$R_{\delta}$ is a tensor}{WirHabenEinenTensorBeiderTrafoKruemmung}
Let $M, N$ be smooth manifolds, $E \to N$ a Lie algebroid, and $\nabla$ a connection on $E$.

Then $R_{\delta}(\cdot, \cdot)A$ is an anti-symmetric tensor, \textit{i.e.}~anti-symmetric and $\mathcal{F}^0_E(M)$-bilinear, and we have
\ba\label{SplittingVonDerEichtrafo}
R_{\delta}(\varepsilon, \vartheta)A
&=
R_{\delta}(\varepsilon, \vartheta)A^a
	\otimes {}^*e_a
	+ \mleft({}^* R_{\nabla^{\mathrm{bas}}} \mright)( \varepsilon, \vartheta) A
\ea
for all $\varepsilon, \vartheta \in \mathcal{F}^0_E(M; {}^*E)$.
\end{propositions}

\begin{proof}
\leavevmode\newline
$\bullet$ The antisymmetry is clear by Prop.~\ref{prop:PropertiesOfThePreBracket}. Fix a local frame $\mleft( e_a \mright)_a$ of $E$, then we have
\bas
\delta_\vartheta \delta_{f\varepsilon}A
~~~~&\stackrel{\mathclap{\text{Def.~\ref{def:GaugeTrafoOfA}}}}{=}~~~~
- \delta_\vartheta \bigl( \mleft({}^*\nabla \mright) (f \varepsilon) \bigr) \\
&=
- \delta_\vartheta \bigl( \mathrm{d}f \otimes \varepsilon 
+ f ~ \mleft({}^*\nabla \mright) \varepsilon \bigr) \\
&=
- \delta_\vartheta \mathrm{d} f \otimes \varepsilon
	- \mathrm{d} f \otimes \delta_\vartheta \varepsilon
	- \delta_\vartheta f ~ \mleft({}^*\nabla \mright) \varepsilon 
	- f \delta_\vartheta \bigl(\mleft({}^*\nabla \mright) \varepsilon \bigr)
\\
&=
- \delta_\vartheta \mathrm{d} f \otimes \varepsilon
	- \mathrm{d} f \otimes \delta_\vartheta \varepsilon^a ~ {}^*e_a
	+ \mathrm{d} f \otimes \varepsilon^a \vartheta^b ~ {}^*\mleft(\nabla^{\mathrm{bas}}_{e_b} e_a \mright)
	- \delta_\vartheta f ~ \mleft({}^*\nabla \mright) \varepsilon 
	+ f \delta_\vartheta \delta_\varepsilon A
\eas
for all $\vartheta, \varepsilon \in \mathcal{F}^0_E(M; {}^*E)$ and $f \in \mathcal{F}^0_E(M)$, and
\bas
-\delta_{f\varepsilon} \delta_\vartheta A
&=
\delta_{f\varepsilon} \bigl( \mleft( {}^*\nabla \mright) \vartheta \bigr) \\
&=
\delta_{f\varepsilon} \mleft( \mathrm{d}\vartheta^a \otimes {}^*e_a + \vartheta^b ~ {}^! \mleft( \nabla e_b \mright) \mright) \\
&\stackrel{\mathclap{\text{Eq.~\eqref{EqVariationVonFormenBrrrr}}}}{=}~~~~
\delta_{f\varepsilon} \mathrm{d}\vartheta^a \otimes {}^*e_a 
	- \mathrm{d}\vartheta^a \otimes f \varepsilon^b~ {}^*\mleft( \nabla^{\mathrm{bas}}_{e_b} e_a \mright) \\
&\hspace{1cm}
	+ \delta_{f\varepsilon} \vartheta^b ~ {}^!\mleft( \nabla e_b \mright) 
	- f \vartheta^b ~ {}^!\mleft( \nabla^{\mathrm{bas}}_{\varepsilon} (\nabla e_b) \mright)
	- \vartheta^b ~\underbrace{{}^*\mleft( \nabla_{({}^*\rho)\mleft( ({}^*\nabla) (f \varepsilon) \mright)} e_b \mright)}
	_{\mathclap{= \mathrm{d}f \otimes {}^*\mleft( \nabla_{\mleft({}^*\rho\mright)(\varepsilon)} e_b \mright)
		+ f \cdot \underbrace{(\dotsc)}_{\mathclap{\text{indep. of }f}}
		}}
\\
&=
\delta_{f\varepsilon} \mathrm{d}\vartheta^a \otimes {}^*e_a 
	+ \delta_{f\varepsilon} \vartheta^b ~ {}^! \mleft( \nabla e_b \mright) 
	- \vartheta^b \varepsilon^a\mathrm{d}f \otimes {}^*\mleft( \nabla_{\rho(e_a)} e_b \mright)
	+ f \cdot \underbrace{(\dotsc)}_{\mathclap{\text{independent of }f}}.
\eas
Since we want to check the tensorial property, we can ignore the terms proportional to $f$; 
we also have
\bas
\delta_{\llbracket\vartheta, f \varepsilon\rrbracket} A
&=
\mleft( {}^*\nabla \mright)  \mleft( \llbracket f \varepsilon, \vartheta \rrbracket \mright) \\
&\stackrel{\mathclap{\text{Eq.~\eqref{EqDeltaInFrameKoord}}}}{=}~~~~
\mleft( {}^*\nabla \mright)
\mleft(
		\delta_\vartheta f ~ \varepsilon
		+ f \delta_\vartheta \varepsilon^a ~ {}^*e_a 
		- \delta_{f \varepsilon} \vartheta^b ~ {}^*e_b
		+ f \varepsilon^a \vartheta^b~ {}^*\mleft( \mleft[ e_a, e_b \mright]_E \mright)
\mright)
\\
&\stackrel{\mathclap{\text{Eq.~\eqref{eqVariationVertauschtMitDifferential}}}}{=}~~~~
\delta_\vartheta \mathrm{d} f \otimes \varepsilon
	+ \delta_\vartheta f ~ ({}^*\nabla)\varepsilon
	+ \mathrm{d} f \otimes \delta_\vartheta \varepsilon^a ~ {}^*e_a 
	- \delta_{f \varepsilon}\mathrm{d} \vartheta^b \otimes {}^*e_b
	- \delta_{f \varepsilon} \vartheta^b ~ {}^!\mleft( \nabla e_b \mright)
\\
&\hspace{1cm}~~~~
	+ \varepsilon^a \vartheta^b~\mathrm{d}f \otimes {}^*\mleft( \mleft[ e_a, e_b \mright]_E \mright)
	+ f \cdot \underbrace{(\dotsc)}_{\mathclap{\text{independent of }f}}.
\eas
Hence, we get in total
\bas
R_{\delta}(\vartheta, f \varepsilon)A
&=
\varepsilon^a \vartheta^b \mathrm{d} f \otimes {}^*\underbrace{\mleft( 
	\nabla^{\mathrm{bas}}_{e_b} e_a
	- \nabla_{\rho(e_a)} e_b
	+ \mleft[ e_a, e_b \mright]_E
\mright)}_{= \nabla^{\mathrm{bas}}_{e_b} e_a - \nabla^{\mathrm{bas}}_{e_b} e_a = 0}
	+ f \cdot \underbrace{(\dotsc)}_{\mathclap{\text{independent of }f}}
\\
&=
f \cdot \underbrace{(\dotsc)}_{\mathclap{\text{independent of }f}}
\eas
for all $\vartheta, \varepsilon \in \mathcal{F}^0_E(M; {}^*E)$ and $f \in \mathcal{F}^0_E(M)$. Using the antisymmetry proves that $R_{\delta}(\cdot, \cdot)A$ is a tensor because the shown equation also holds for $f \equiv 1$ such that the remaining terms in the $f$-independent bracket are precisely giving rise to $R_{\delta}(\vartheta, \varepsilon)A$.

$\bullet$ Eq.~\eqref{SplittingVonDerEichtrafo} just follows by Cor.~\ref{cor:RelationShipsOfCurvatures}.
\end{proof}

Due to the tensorial behaviour, we can study $R_{\delta}(\cdot, \cdot)A$ just with respect to pullback functionals, such that the notations and calculations can be simplified; also recall Def.~\ref{def:basiccurvature}.

\begin{theorems}{Curvature of the infinitesimal gauge transformation measured by the basic curvature}{CurvatureOfBasicStuffIsEquivalentForGaugeTrafoCurvature}
Let $M, N$ be smooth manifolds, $E \to N$ a Lie algebroid, and $\nabla$ a connection on $E$. Then
\ba
R_{\delta}({}^*\mu, {}^*\nu)A
&=
- {}^!\mleft( R^{\mathrm{bas}}_\nabla(\mu, \nu) \mright)
\ea
for all $\mu, \nu \in \Gamma(E)$, viewing $R^{\mathrm{bas}}_\nabla(\mu, \nu)$ as an element of $\Omega^1(N;E)$.
\end{theorems}

\begin{remark}
\leavevmode\newline
\indent $\bullet$ One can then derive with Eq.~\eqref{EqPullBackFormelFuerVerschiedeneDefinitionen} that
\bas
{}^!\mleft( R^{\mathrm{bas}}_\nabla(\mu, \nu) \mright)
&=
\mleft({}^*\mleft( R^{\mathrm{bas}}_\nabla(\mu, \nu) \mright)\mright) \mathrm{D}
=
\mleft({}^* R^{\mathrm{bas}}_\nabla\mright) ({}^*\mu, {}^*\nu) \mathrm{D},
\eas
viewing $\mathrm{D}$ as an element of $\mathcal{F}^1_E(M; {}^*\mathrm{T}N)$; recall Ex.~\ref{ex:DAsFunctional}.
Using that $R_\delta(\cdot, \cdot) A$ is tensorial and that pullbacks are generators as usual, we get
\bas
R_{\delta}(\varepsilon, \vartheta)A
&=
- \mleft({}^* R^{\mathrm{bas}}_\nabla\mright) (\varepsilon, \vartheta) \mathrm{D}
\eas
for all $\varepsilon, \vartheta \in \mathcal{F}^0_E(M; {}^*E)$.

$\bullet$ One could also view this theorem as a physical interpretation of the basic curvature.
\end{remark}

\begin{proof}[Proof of Thm.~\ref{thm:CurvatureOfBasicStuffIsEquivalentForGaugeTrafoCurvature}]
\leavevmode\newline
We have
\bas
\delta_{{}^*\mu} \mleft( \delta_{{}^*\nu} A \mright)
&=
- \delta_{{}^*\mu} \mleft( {}^!\mleft( \nabla \nu \mright) \mright)
\\
&\stackrel{\mathclap{\text{Eq.~\eqref{EqVariationVonFormenBrrrrVereinfacht}}}}{=}~~~~
{}^! \mleft(
	\nabla^{\mathrm{bas}}_\mu \mleft( \nabla \nu \mright)
	+ \nabla_{\rho\mleft(\nabla \mu \mright)} \nu
\mright),
\eas
and
\bas
\mleft(\nabla^{\mathrm{bas}}_\mu \mleft( \nabla \nu \mright)
	+ \nabla_{\rho\mleft(\nabla \mu \mright)} \nu\mright)(Y)
&=
\nabla^{\mathrm{bas}}_\mu \nabla_Y \nu
	- \nabla_{\nabla^{\mathrm{bas}}_\mu Y} \nu
	+ \nabla_{\rho\mleft(\nabla_Y \mu \mright)} \nu
\\
&=
\mleft[ \mu, \nabla_Y \nu \mright]_E
	+ \nabla_{\rho\mleft( \nabla_Y \nu \mright)} \mu
	- \nabla_{\mleft[ \rho(\mu), Y \mright]} \nu
\eas
for all $Y \in \mathfrak{X}(M)$. In total we would then look at the pull-back of the following form, also using Eq.~\eqref{EqLieKlammerAufPullBackSections},
\bas
&\mleft(
	\nabla^{\mathrm{bas}}_\mu \mleft( \nabla \nu \mright)
	+ \nabla_{\rho\mleft(\nabla \mu \mright)} \nu
	- \nabla^{\mathrm{bas}}_\nu \mleft( \nabla \mu \mright)
	- \nabla_{\rho\mleft(\nabla \nu \mright)} \mu
	- \nabla \mleft( \mleft[ \mu, \nu \mright]_E \mright)
\mright)(Y)
\\
&=
\mleft[ \mu, \nabla_Y \nu \mright]_E
	+ \nabla_{\rho\mleft( \nabla_Y \nu \mright)} \mu
	- \nabla_{\mleft[ \rho(\mu), Y \mright]} \nu
	-	\mleft[ \nu, \nabla_Y \mu \mright]_E
	- \nabla_{\rho\mleft( \nabla_Y \mu \mright)} \nu
	+ \nabla_{\mleft[ \rho(\nu), Y \mright]} \mu
	- \nabla_Y \mleft( \mleft[ \mu, \nu \mright]_E \mright)
\\
&=
- \mleft(
	\nabla_Y\mleft(\mleft[\mu, \nu\mright]_E\mright) 
	- \mleft[ \nabla_Y \mu, \nu \mright]_E 
	- \mleft[ \mu, \nabla_Y \nu \mright]_E 
	- \nabla_{\nabla^{\mathrm{bas}}_\nu Y} \mu 
	+ \nabla_{\nabla^{\mathrm{bas}}_\mu Y} \nu
\mright)
\\
&\stackrel{\mathclap{\text{Def.~\ref{def:basiccurvature}}}}{=}~~~~
- R^{\mathrm{bas}}_\nabla(\mu, \nu)Y.
\eas
Therefore we arrive at
\bas
R_{\delta}({}^*\mu, {}^*\nu)A
&=
- {}^!\mleft( R^{\mathrm{bas}}_\nabla(\mu, \nu)Y \mright).
\eas
\end{proof}

We get immediately the following statement.

\begin{corollaries}{Flat infinitesimal gauge transformation}{FlatnessVonEichtrafos}
Let $M, N$ be smooth manifolds, $E \to N$ a Lie algebroid, and $\nabla$ a connection on $E$ with $R^{\mathrm{bas}}_\nabla=0$. Then
\ba\label{DieKruemmungIstNullVonDenEichtrafosGeeeeil}
R_{\delta}(\cdot, \cdot)A
&=
0.
\ea
With respect to a frame $\mleft( e_a \mright)_a$ of $E$ we then also have
\ba\label{CoordFuerEichKruemmungsRegel}
R_\delta (\cdot, \cdot) A^a
&=
0
\ea
for all $\mu, \nu \in \Gamma(E)$.
\end{corollaries}

\begin{remark}\label{RemarkUeberNablaRhoCurvatureForGauegTrafo}
\leavevmode\newline
\indent $\bullet$ This discussion, especially Cor.~\ref{cor:FlatnessVonEichtrafos} and Thm.~\ref{thm:CurvatureOfBasicStuffIsEquivalentForGaugeTrafoCurvature}, are generalizations of statements in \cite[especially Prop.~8 and Thm.~1]{EichtrafoKruemmungUrspruenglich} and \cite[especially Eq.~9, 10 and 11; there the $S$ denotes the basic curvature]{mayerlieAuchEichtrafoStuff}.\footnote{The sign of $\varepsilon$ in the gauge transformations there is the opposite of our sign.} In both of these works a coordinate-free formulation of $\delta_\varepsilon A$ was not known, just $\delta_\varepsilon A^a$. It was known that $\delta_\varepsilon A^a$ is dependent on coordinates, but not how it can be written/defined such that it is again an element of $\Omega^1(M; \Phi^*E)$. \cite{EichtrafoKruemmungUrspruenglich} tries to formulate infinitesimal gauge transformations in a covariant way with a completely different approach by assuming a weaker form of equality,\footnote{The closure of the algebra of gauge transformations is formulated as an on-shell condition.} but only for a special situation and only for $\varepsilon$ as an element of $\Phi^*(\Gamma(E))$ (\textit{i.e.}~they only looked at pullback functionals, when we express that in our language). \cite{mayerlieAuchEichtrafoStuff} looks at the set $\Gamma(\Phi^*E)$ for $\varepsilon$ but assumes that $\varepsilon^a$ is independent of $\Phi$ and $A$ which is clearly a coordinate-dependent description, because a change of the pull-back frame would introduce a $\Phi$-dependency of the components $\varepsilon^a$ (in our words, they choose a coordinate-dependent embedding of $\Gamma(\Phi^*E)$ as functionals). In one way or the other, both works arrive at Eq.~\eqref{CoordFuerEichKruemmungsRegel}, but only evaluated at pullback functionals, that is, $R_\delta({}^*\mu, {}^*\nu)A^a=0$ for all $\mu, \nu \in \Gamma(E)$.

What we provide is a coordinate-independent and -free definition of such infinitesimal gauge transformations such that these give a closed algebra; see also the following theorems. Moreover, we have generalized Eq.~\eqref{CoordFuerEichKruemmungsRegel} in form of Eq.~\eqref{DieKruemmungIstNullVonDenEichtrafosGeeeeil}, in sense of not only assuming pullback functionals by defining the pre-bracket $\llbracket \cdot, \cdot \rrbracket$.

$\bullet$ Recall Remark \ref{WhyNablaBasPartOne}: One could also take $\nabla_\rho$ to define $\delta_\varepsilon$. It has the advantage that then $\delta_\varepsilon A$ directly restricts to the standard formula when restricting ourselves to the classical setting. When defining and calculating $R_\delta$ in a similar manner,
we also get Eq.~\eqref{SplittingVonDerEichtrafo} where the curvature-term will be replaced with the curvature of $\nabla_\rho$ due to Cor.~\ref{cor:RelationShipsOfCurvatures}. Therefore one needs to impose at least flatness of $\nabla_\rho$ in order to get a similar result like Eq.~\eqref{DieKruemmungIstNullVonDenEichtrafosGeeeeil}; actually, one can check that one still needs a vanishing basic curvature as an additional condition, too.
This leads to an extra condition, while the flatness of the basic connection is implied by the vanishing of the basic curvature by Remark \ref{PropsOfBasicConnection}. 

However, we will see as expected that one just needs Eq.~\eqref{CoordFuerEichKruemmungsRegel} for the closure of the algebra of the vector fields like $\Psi_\varepsilon$ defined by Prop.~\ref{prop:VariationOfA}, and Eq.~\eqref{CoordFuerEichKruemmungsRegel} is independent on whether one uses the basic connection to define $\delta_\varepsilon$.
\end{remark}

\begin{proof}[Proof of Cor.~\ref{cor:FlatnessVonEichtrafos}]
\leavevmode\newline
That is a trivial consequence of Thm.~\ref{thm:CurvatureOfBasicStuffIsEquivalentForGaugeTrafoCurvature} and Prop.~\ref{prop:WirHabenEinenTensorBeiderTrafoKruemmung}, using $R_\nabla^{\mathrm{bas}}=0$ (and that then the basic connection is flat by the last bullet point of Remark \ref{PropsOfBasicConnection}) and that $R_\delta(\cdot, \cdot)A$ is $\mathcal{F}^0_E(M)$-bilinear such that one just needs to look at pullback functionals.
\end{proof}

These results motivate even further why we use the basic connection to define the infinitesimal gauge transformation. Moreover, $R^{\mathrm{bas}}_\nabla=0$ is actually a condition which one needs for gauge invariance of the Lagrangian; see \cite[Eq.~(9)]{CurvedYMH} and \cite[\S 4.4ff., especially Thm.~4.4.3 and Thm.~4.7.5]{MyThesis}. Hence, we also have this condition in the standard formulation of gauge theory such that it is not a newly imposed and a reasonable condition; there is also the following well-known result:

\begin{theorems}{Relation of the basic curvature and action Lie algebroids, \newline \cite[discussion around Eq.~(9)]{CurvedYMH}, \cite[Prop.~2.12]{basicconn}, and \cite[\S 2.5, Theorem A]{blaomTangentBundleAsLieGroup}}{ActionLieALgebroid}
Let $E \to N$ be a Lie algebroid. Then $E$ is locally an action Lie algebroid if and only if it admits locally a flat connection $\nabla$ with $R_\nabla^{\mathrm{bas}} = 0$. If there is such a local isomorphism, then it can be chosen in such a way that $\nabla$ describes the canonical flat connection.
\end{theorems}

\begin{remark}\label{remSimplyConnectedEqualsGlobal}
\leavevmode\newline
As clarification of the last sentence, under that isomorphism we have (locally) $E = N \times \mathfrak{g}$ for some Lie algebra $\mathfrak{g}$, and a basis of $\mathfrak{g}$, that is, a constant frame of $E$, will be parallel with respect to $\nabla$. Especially, the canonical flat connection of every action Lie algebroid has a vanishing basic curvature. Furthermore, over a simply connected base the isomorphism is global as we will see in the proof (because one can then construct a global parallel frame for $\nabla$; see the proofs in the references or in \cite[proof of Thm.~4.3.41]{MyThesis}).
\end{remark}

We now want to generalize Cor.~\ref{cor:FlatnessVonEichtrafos} by using Cor.~\ref{cor:RelationShipsOfCurvatures}, especially we need to understand the behaviour for scalar-valued functionals. For such functionals the infinitesimal gauge transformation is nothing else than the Lie derivative of some vector field in $\mathfrak{M}_E$, which we denoted by $\Psi_\varepsilon$.
Recall Remark \ref{NotASubalgebraXB}, we do in general not expect that $\Psi_\varepsilon \in \mathfrak{X}^E\bigl( \mathfrak{M}_E(M;N) \bigr)$ builds a subalgebra; however, since we restricted the set of those vector fields by defining $\delta_\varepsilon A$ in Prop.~\ref{prop:VariationOfA}, there may be hope for the structure of a subalgebra; this will be discussed now. 

\begin{theorems}{Bracket of gauge transformations a gauge transformation}{VektorfelderSindZumGlueckGeschlossen}
Let $M, N$ be smooth manifolds, $E \to N$ a Lie algebroid, $\nabla$ a connection on $E$ with $R^{\mathrm{bas}}_\nabla=0$. Furthermore let $\Psi_\varepsilon$ and $\Psi_\vartheta$ for $\varepsilon, \vartheta \in \mathcal{F}^0_E(M; {}^*E)$ be the unique elements of $\mathfrak{X}^E\bigl( \mathfrak{M}_E(M;N) \bigr)$ as given by Prop.~\ref{prop:VariationOfA}.\footnote{Recall that those $\Psi_\varepsilon$ are the vector fields describing the infinitesimal gauge transformation; see Def.~\ref{def:TotalInfGaugeTrafoYayy}.}

Then
\ba
\mleft[ \Psi_\varepsilon, \Psi_\vartheta \mright]
&=
- \Psi_{\llbracket\varepsilon, \vartheta \rrbracket}
\ea
for all $\varepsilon, \vartheta \in \mathcal{F}^0_E(M; {}^*E)$, where $\Psi_{\llbracket\varepsilon, \vartheta\rrbracket}$ is also the unique element of $\mathfrak{X}^E\bigl( \mathfrak{M}_E(M;N) \bigr)$ as given by Prop.~\ref{prop:VariationOfA}.
\end{theorems}

\begin{proof}
\leavevmode\newline
First recall that we have by Remark \ref{RemLeibnizeRegelaufProdukteWeshalbEConnectionNichtWichtigIst}
\bas
\delta_\varepsilon \omega
&=
\mathcal{L}_{\Psi_\varepsilon} \omega
\eas
for all $\omega \in \mathcal{F}^\bullet_E(M)$ and $\varepsilon \in \mathcal{F}^0_E(M; {}^*E)$. Therefore we want to use Cor.~\ref{cor:FlatnessVonEichtrafos}. As vector fields of $\mathfrak{M}_E(M;N)$, the action of $\mathcal{L}_{\Psi_\varepsilon}$ is uniquely given by its action on coordinates of $\mathfrak{M}_E(M;N)$, and these are essentially given by the components of the fields $(\Phi, A) \in \mathfrak{M}_E(M;N)$: Let $\mleft( x^i \mright)_i$ be local coordinate functions on $N$ and let $\mleft(e_a\mright)_a$ be a local frame of $E$, then coordinates of $\mathfrak{M}_E(M;N)$ are given by the functionals ${}^*\mleft(x^i\mright)$ and $\varpi_2^a$ because of
\bas
\mleft.{}^*\mleft(x^i\mright)\mright|_{(\Phi,A)}
&=
\Phi^i,
\\
\varpi_2^a(\Phi,A)
&=
A^a
\eas
for all $(\Phi, A) \in \mathfrak{M}_E(M;N)$. Recall the first calculation in the proof of Cor.~\ref{cor:RelationShipsOfCurvatures}, we get similarly
\bas
R_\delta(\varepsilon,\vartheta)\mleft( {}^*\mleft(x^i\mright) \mright) 
&= 
\varepsilon^a \vartheta^b ~ {}^*\underbrace{\mleft(
	\mathcal{L}_{\rho(e_a)} \mathcal{L}_{\rho(e_b)} x^i
	- \mathcal{L}_{\rho(e_b)} \mathcal{L}_{\rho(e_a)} x^i
	- \mathcal{L}_{\rho\mleft(\mleft[ e_a, e_b \mright]_E\mright)} x^i
\mright)}_{= \mleft( \mathcal{L}_{\mleft[ \rho(e_a), \rho(e_b) \mright]} - \mathcal{L}_{\rho\mleft(\mleft[ e_a, e_b \mright]_E\mright)} \mright) x^i = 0}
=
0
\eas
for all $\varepsilon, \vartheta \in \mathcal{F}^0_E(M; {}^*E)$, using that $\rho$ is a homomorphisma and Remark \ref{RemLeibnizeRegelaufProdukteWeshalbEConnectionNichtWichtigIst} such that $\delta_\varepsilon \mleft( {}^*\mleft( x^i \mright) \mright) = - \varepsilon^a ~ {}^*\mleft( \mathcal{L}_{\rho(e_a)} x^i \mright)$. By Cor.~\ref{cor:FlatnessVonEichtrafos} we also get
\bas
R_\delta (\varepsilon, \vartheta) \varpi_2^a
&=
0.
\eas
By $\delta_\varepsilon = \mathcal{L}_{\Psi_\varepsilon}$ on scalar-valued functionals we therefore get
\bas
\mleft(\mleft[ \mathcal{L}_{\Psi_\varepsilon}, \mathcal{L}_{\Psi_\vartheta} \mright]
+ \mathcal{L}_{\Psi_{\llbracket\varepsilon, \vartheta\rrbracket}}\mright)f
&=
0
\eas
for all $f \in C^\infty\bigl(\mathfrak{M}_E(M;N)\bigr)$,
which finishes the proof.
\end{proof}

\begin{remarks}{Curvature of $\delta$ on $\Phi$}{WasIstMitDemHiggsFeldBeiDerDeltaKruemmung}
Keeping the same situation and notation as in the previous proof, observe that we have
\bas
\delta_{{}^*\nu} \delta_{{}^*\mu} \Phi
&=
- \delta_{{}^*\nu} \bigl( {}^*(\rho(\mu)) \bigr)
=
{}^*\mleft( \nabla^{\mathrm{bas}}_\nu \bigl( \rho(\mu) \bigr) \mright)
=
{}^*\mleft( \rho\mleft( \nabla^{\mathrm{bas}}_\nu \mu \mright) \mright)
\eas
for all $\mu, \nu \in \Gamma(E)$, hence,\footnote{Recall Eq.~\eqref{EqLieKlammerAufPullBackSections}.}
\bas
\delta_{{}^*\nu} \delta_{{}^*\mu} \Phi
	- \delta_{{}^*\mu} \delta_{{}^*\nu} \Phi
	+ \delta_{{}^*\mleft( \mleft[ \nu, \mu \mright]_E \mright)} \Phi
&=
{}^*\mleft( \rho\mleft( 
	\nabla^{\mathrm{bas}}_\nu \mu 
	- \nabla^{\mathrm{bas}}_\mu \nu 
	- \mleft[ \nu, \mu \mright]_E
\mright) \mright)
=
{}^*\Bigl( \rho\bigl( 
	t_{\nabla^{\mathrm{bas}}}(\nu, \mu)
\bigr) \Bigr).
\eas
Therefore, if we want that this is zero, too, we would need that the torsion of the basic connection has values in the kernel of the anchor which is in general not the case. However, it is no harm that we do not have a zero value in general here. That is due to the fact that on one hand $\Phi$ just contributes via pull-backs in gauge theories, see \textit{e.g.}~\cite{CurvedYMH} and \cite{MyThesis}; on the other hand $\Phi$ is not vector-bundle valued and hence will not arise in any other form than as the map for the pullbacks in any Lagrangian or physical quantity. Even in the classical case, recall Eq.~\eqref{ActionAndRep}, a Lie algebra representation acting on $\Phi$ is just the evaluation of its induced action at $\Phi$.
\newline\newline
However, as we have seen in the proof, we got $R_\delta(\cdot,\cdot)\mleft( {}^*\mleft(x^i\mright) \mright) = 0$, and $\mleft.{}^*\mleft(x^i\mright)\mright|_{(\Phi,A)} = \Phi^i$ for all $(\Phi, A) \in \mathfrak{M}_E(M;N)$. That is, for the components of the Higgs field we have the desired behaviour, which is all we need.
\end{remarks}

Finally, we can generalize Cor.~\ref{cor:FlatnessVonEichtrafos}.

\begin{theorems}{Curvature of $\delta$ on arbitrary functionals}{AllgemEineGeileFormelFuerDieEichKruemmung}
Let $M, N$ be smooth manifolds, $E \to N$ a Lie algebroid, $\nabla$ a connection on $E$ with $R^{\mathrm{bas}}_\nabla=0$. Furthermore let $V\to N$ be a vector bundle, equipped with an $E$-connection ${}^E\nabla$ on $V$. Then
\ba
R_\delta(\varepsilon, \vartheta) L
&=
\mleft({}^*R_{{}^E\nabla} \mright)(\varepsilon, \vartheta)L
\ea
for all $L \in \mathcal{F}_E^k(M; {}^*V)$ ($k \in \mathbb{N}_0$) and $\varepsilon, \vartheta \in \mathcal{F}^0_E(M; {}^*E)$. In short, $R_\delta = {}^*R_{{}^E\nabla}$.
\end{theorems}

\begin{remark}
\leavevmode\newline
This also shows that $R_\delta$ is a tensor.
Moreover, as expected, for flat ${}^E\nabla$ we would get
\ba
R_\delta(\varepsilon, \vartheta) L
&=
0.
\ea
\end{remark}

\begin{proof}[Proof of Thm.~\ref{thm:AllgemEineGeileFormelFuerDieEichKruemmung}]
\leavevmode\newline
We want to use Cor.~\ref{cor:RelationShipsOfCurvatures}, so, for a given frame $\mleft( e_a \mright)_a$ we have
\bas
R_\delta(\varepsilon, \vartheta)L
&=
R_\delta(\varepsilon, \vartheta)L^a \otimes {}^*e_a
	+ \mleft({}^*R_{{}^E\nabla}\mright)(\varepsilon, \vartheta)L
\eas
for all $L \in \mathcal{F}^k_E(M; {}^*V)$ ($k \in \mathbb{N}_0$) and $\varepsilon, \vartheta \in \mathcal{F}^0_E(M; {}^*E)$. Hence, we just need to show that $R_\delta(\varepsilon, \vartheta)L^a = 0$. Again by Remark \ref{RemLeibnizeRegelaufProdukteWeshalbEConnectionNichtWichtigIst} we have $\delta_\varepsilon = \mathcal{L}_{\Psi_\varepsilon}$ on scalar-valued functionals, where $\Psi_\varepsilon$ still denotes vector fields as uniquely given by Prop.~\ref{prop:VariationOfA}. $\Psi_\varepsilon$ are elements of $\mathfrak{X}\bigl( \mathfrak{M}_E(M;N) \bigr)$, hence, 
\bas
(\underbrace{\delta_\varepsilon L^a}_{\mathclap{= \mathcal{L}_{\Psi_\varepsilon}L^a }})_p(Y_1, \dotsc, Y_k)
&=
\mathcal{L}_{\Psi_\varepsilon} \mleft(L^a_p(Y_1, \dotsc, Y_k)\mright)
\eas
for all $p \in M$ and $Y_1, \dotsc, Y_k \in \mathrm{T}_pM$. We know that $L^a\in\mathcal{F}^k_E(M)$, and therefore $L^a_p(Y_1, \dotsc, Y_k) \in C^\infty\bigl(\mathfrak{M}_E(M;N)\bigr)$, so, we just need to use Thm.~\ref{thm:VektorfelderSindZumGlueckGeschlossen} to get
\bas
\mleft(R_\delta(\varepsilon, \vartheta) L^a \mright)_p(Y_1, \dotsc, Y_k)
&=
\mleft(
	\mleft(\mleft[ \mathcal{L}_{\Psi_\varepsilon}, \mathcal{L}_{\Psi_\vartheta} \mright]
	+ \mathcal{L}_{\Psi_{\llbracket\varepsilon, \vartheta\rrbracket}}\mright) L^a
\mright)_p(Y_1, \dotsc, Y_k)
\\
&=
\mleft(\mleft[ \mathcal{L}_{\Psi_\varepsilon}, \mathcal{L}_{\Psi_\vartheta} \mright]
	+ \mathcal{L}_{\Psi_{\llbracket\varepsilon, \vartheta\rrbracket}}\mright)
\mleft(L^a_p(Y_1, \dotsc, Y_k)\mright)
\\
&\stackrel{\mathclap{ \text{Thm.~\ref{thm:VektorfelderSindZumGlueckGeschlossen}} }}{=}\quad~~
0,
\eas
which concludes the proof.
\end{proof} 

Let us conclude this section by showing that this finally implies that $\llbracket \cdot, \cdot \rrbracket$ is a Lie bracket. For this we need to use the first Bianchi identity of $E$-connections on $E$ itself with torsion, which is very similar to the one known for vector bundle connections with torsions,
\ba\label{eq:firstBianchi}
&R_{{}^E\nabla}(\mu, \nu) \eta + R_{{}^E\nabla}(\nu, \eta) \mu + R_{{}^E\nabla}(\eta, \mu) \nu 
\nonumber \\
&=
t_{{}^E\nabla}\mleft(t_{{}^E\nabla}(\mu, \nu), \eta\mright) + t_{{}^E\nabla}(t_{{}^E\nabla}(\nu, \eta), \mu) + t_{{}^E\nabla}(t_{{}^E\nabla}(\eta, \mu), \nu)
\nonumber \\
&\hspace{1cm}
+ \left({}^E\nabla_\mu t_{{}^E\nabla}\right)(\nu, \eta) 
+ \left({}^E\nabla_\nu t_{{}^E\nabla}\right)(\eta, \mu) + \left({}^E\nabla_\eta t_{{}^E\nabla}\right)(\mu, \nu),
\ea
for all $\mu, \nu, \eta \in \Gamma(E)$; the proof is as usual, using the Jacobi identity of $\mleft[ \cdot, \cdot \mright]_E$, but see \textit{e.g.}~\cite[first equation in Thm.~3.4.6]{MyThesis} for an explicit proof in that setting.

\begin{theorems}{Pre-bracket a Lie bracket}{PreKlammerEineSuperLieKlammer}
Let $M, N$ be smooth manifolds, $E \to N$ a Lie algebroid, $\nabla$ a connection on $E$ with $R^{\mathrm{bas}}_\nabla=0$. Then $\llbracket \cdot, \cdot \rrbracket$ is a Lie bracket.
\end{theorems}

\begin{proof}
\leavevmode\newline
By Prop.~\ref{prop:PropertiesOfThePreBracket} we already know antisymmetry and $\mathbb{R}$-bilinearity.
Thus, only the Jacobi identity is left to show, and the calculation is very similar to the calculation of proofs of the first Bianchi identity,
\bas
\mleft\llbracket \eta, \mleft\llbracket \vartheta, \varepsilon \mright\rrbracket \mright\rrbracket
&=
\mleft\llbracket \eta, \delta_\varepsilon \vartheta - \delta_\vartheta \varepsilon - \bigl( {}^*t_{\nabla^{\mathrm{bas}}} \bigr)\mleft( \vartheta, \varepsilon \mright) \mright\rrbracket
\\
&=
\underbrace{\delta_{\delta_\varepsilon \vartheta} \eta
	- \delta_{\delta_\vartheta \varepsilon} \eta
	- \delta_{\bigl( {}^*t_{\nabla^{\mathrm{bas}}} \bigr)\mleft( \vartheta, \varepsilon \mright)} \eta}
	_{\delta_{\llbracket\vartheta, \varepsilon\rrbracket} \eta}
\\
&\hspace{1cm}
	- \delta_\eta \delta_\varepsilon \vartheta
	+ \delta_\eta \delta_\vartheta \varepsilon
	+ \bigl( {}^*t_{\nabla^{\mathrm{bas}}} \bigr)\mleft( \eta, \mleft( \bigl( {}^*t_{\nabla^{\mathrm{bas}}} \bigr)\mleft( \vartheta, \varepsilon \mright) \mright) \mright)
\\
&\hspace{1cm}
	+ \delta_\eta \mleft( \bigl( {}^*t_{\nabla^{\mathrm{bas}}} \bigr)\mleft( \vartheta, \varepsilon \mright) \mright)
	- \bigl( {}^*t_{\nabla^{\mathrm{bas}}} \bigr)\mleft( \eta, \delta_\varepsilon \vartheta \mright)
	+ \bigl( {}^*t_{\nabla^{\mathrm{bas}}} \bigr)\mleft( \eta, \delta_\vartheta \varepsilon \mright)
\\
&=
\delta_\eta \delta_\vartheta \varepsilon
	- \delta_\eta \delta_\varepsilon \vartheta
	+ \delta_{\llbracket\vartheta, \varepsilon\rrbracket} \eta
\\
&\hspace{1cm}
	+ \delta_\eta \mleft( \bigl( {}^*t_{\nabla^{\mathrm{bas}}} \bigr)\mleft( \vartheta, \varepsilon \mright) \mright)
	- \bigl( {}^*t_{\nabla^{\mathrm{bas}}} \bigr)\mleft( \eta, \delta_\varepsilon \vartheta \mright)
	+ \underbrace{\bigl( {}^*t_{\nabla^{\mathrm{bas}}} \bigr)\mleft( \eta, \delta_\vartheta \varepsilon \mright)}_{\mathclap{= - \bigl( {}^*t_{\nabla^{\mathrm{bas}}} \bigr)\mleft( \delta_\vartheta \varepsilon, \eta \mright)}}
\\
&\hspace{1cm}
	+ \bigl( {}^*t_{\nabla^{\mathrm{bas}}} \bigr)\mleft( \eta, \mleft( \bigl( {}^*t_{\nabla^{\mathrm{bas}}} \bigr)\mleft( \vartheta, \varepsilon \mright) \mright) \mright)
\eas
for all $\varepsilon, \vartheta, \eta \in \mathcal{F}_E^0(M; {}^*E)$. Taking the cyclic sum, we collect the terms and get, using that $\nabla^{\mathrm{bas}}$ is used for the definition of $\delta$ on $E$-valued functionals,
\bas
&\mleft\llbracket \eta, \mleft\llbracket \vartheta, \varepsilon \mright\rrbracket \mright\rrbracket
	+ \mleft\llbracket \vartheta, \mleft\llbracket \varepsilon, \eta \mright\rrbracket \mright\rrbracket
	+ \mleft\llbracket \varepsilon, \mleft\llbracket \eta, \vartheta \mright\rrbracket \mright\rrbracket
\\
&=
\underbrace{R_{\delta}( \eta, \vartheta ) \varepsilon
	+ R_{\delta}( \varepsilon,\eta ) \vartheta
	+ R_{\delta}( \vartheta, \varepsilon) \eta}_{\stackrel{\text{Thm.~\ref{thm:AllgemEineGeileFormelFuerDieEichKruemmung}}}{=} ~ \mleft(^*R_{\nabla^{\mathrm{bas}}} \mright)(\eta, \vartheta) \varepsilon
		+ \mleft(^*R_{\nabla^{\mathrm{bas}}} \mright)(\varepsilon, \eta)\vartheta
		+ \mleft(^*R_{\nabla^{\mathrm{bas}}} \mright)(\vartheta, \varepsilon)\eta
	 }
\\
&\hspace{1cm}
	+ \bigl( {}^*t_{\nabla^{\mathrm{bas}}} \bigr)\mleft( \eta, \bigl( {}^*t_{\nabla^{\mathrm{bas}}} \bigr)\mleft( \vartheta, \varepsilon \mright) \mright) 
	+ \bigl( {}^*t_{\nabla^{\mathrm{bas}}} \bigr)\mleft( \varepsilon, \bigl( {}^*t_{\nabla^{\mathrm{bas}}} \bigr)\mleft( \eta, \vartheta \mright) \mright) 
\\
&\hspace{1cm}
	+ \bigl( {}^*t_{\nabla^{\mathrm{bas}}} \bigr)\mleft( \vartheta, \bigl( {}^*t_{\nabla^{\mathrm{bas}}} \bigr)\mleft( \varepsilon, \eta \mright) \mright) 
\\
&\hspace{1cm}
	+ \underbrace{\mleft( \delta_\eta \mleft({}^*t_{\nabla^{\mathrm{bas}}} \mright) \mright)}_{= - {}^*\mleft( \nabla^{\mathrm{bas}}_\eta t_{\nabla^{\mathrm{bas}}} \mright)} (\vartheta, \varepsilon)
	+ \mleft( \delta_\varepsilon \mleft({}^*t_{\nabla^{\mathrm{bas}}} \mright) \mright) (\eta, \vartheta)
	+ \mleft( \delta_\vartheta \mleft({}^*t_{\nabla^{\mathrm{bas}}} \mright) \mright) (\varepsilon, \eta)
\\
&=
\vartheta^a \varepsilon^b \eta^c ~ {}^*\biggl(
	\mleft(^*R_{\nabla^{\mathrm{bas}}} \mright)(e_c, e_a) e_b
		+ \mleft(^*R_{\nabla^{\mathrm{bas}}} \mright)(e_b, e_c)e_a
		+ \mleft(^*R_{\nabla^{\mathrm{bas}}} \mright)(e_a, e_b)e_c
\\
&\hspace{1cm}\hphantom{- \vartheta^a \varepsilon^b \eta^c ~ {}^*\biggl(}
		-t_{\nabla^{\mathrm{bas}}} \mleft( t_{\nabla^{\mathrm{bas}}}(e_a, e_b), e_c \mright)
		- t_{\nabla^{\mathrm{bas}}} \mleft( t_{\nabla^{\mathrm{bas}}}(e_b, e_c), e_a \mright)
		- t_{\nabla^{\mathrm{bas}}} \mleft( t_{\nabla^{\mathrm{bas}}}(e_c, e_a), e_b \mright)
\\
&\hspace{1cm}\hphantom{- \vartheta^a \varepsilon^b \eta^c ~ {}^*\biggl(}
	- \mleft( \nabla^{\mathrm{bas}}_{e_c} t_{\nabla^{\mathrm{bas}}} \mright) (e_a, e_b)
	- \mleft( \nabla^{\mathrm{bas}}_{e_a} t_{\nabla^{\mathrm{bas}}} \mright) (e_b, e_c)
	- \mleft( \nabla^{\mathrm{bas}}_{e_b} t_{\nabla^{\mathrm{bas}}} \mright) (e_c, e_a)
\biggr)
\\
&\stackrel{\mathclap{\text{Eq.~\eqref{eq:firstBianchi}}}}{=}~~~~~
0
\eas
for all $\varepsilon, \vartheta, \eta \in \mathcal{F}^0_E(M; {}^*E)$, where $\mleft( e_a \mright)_a$ is a local frame of $E$. Thence, the Jacobi identity follows.
\end{proof}

\begin{remark}\label{RemarkBracketIsVeryIndependent}
\leavevmode\newline
The proof is essentially based on the first Bianchi identity of curvatures. Hence, taking any other $E$-connection $\nabla^\prime$ on $E$ one could define the bracket $\llbracket \cdot, \cdot \rrbracket$ by using the torsion of $\nabla^\prime$ instead of $\nabla^{\mathrm{bas}}$, and then also define the infinitesimal gauge transformation $\delta$ with respect to $\nabla^\prime$ on $E$-valued form. By Thm.~\ref{thm:AllgemEineGeileFormelFuerDieEichKruemmung} we could not expect $R_\delta=0$ in general, but $\llbracket \cdot, \cdot \rrbracket$ should be nevertheless a Lie bracket due to the fact that the first Bianchi identity Eq.~\eqref{eq:firstBianchi} always holds, even without flatness, and that Thm.~\ref{thm:AllgemEineGeileFormelFuerDieEichKruemmung} provides the needed curvature terms for the Bianchi identity. The vanishing of the basic curvature was essential however, and by Remark \ref{PropsOfBasicConnection} the flatness of the basic curvature is a useful side effect but not needed for the Jacobi identity of $\llbracket \cdot, \cdot \rrbracket$. Furthermore, already Eq.~\eqref{EqDeltaInFrameKoord} points out that the definition of $\llbracket \cdot, \cdot \rrbracket$ is independent of the choice of $\nabla^\prime$ because $\delta_\varepsilon$ is just a Lie derivative on scalar-valued functionals, so that it is clear that it is always the same Lie bracket. By the very last statement of Prop.~\ref{prop:PropertiesOfThePreBracket}, we achieve a Lie bracket completely independent of connections, if the parameters are just functionals depending on the Higgs field $\Phi$.
\end{remark}

\section{Conclusion}

We have seen how one can formulate infinitesimal gauge transformations using pullbacks of Lie algebroid connections. We first studied this formulation in the classical formulation, and this discussion can also be seen as a motivation of some of the fundamental aspects and notions of theories like curved Yang-Mills-Higgs gauge theories as provided by Thomas Strobl and Alexei Kotov (see \textit{e.g.}~\cite{CurvedYMH} and the references therein). We then generalized these notions in a straightforward manner.

\begin{itemize}
	\item By Cor.~\ref{cor:CoolesCommutingDiagramForHiggsTrafosStuff} and Def.~\ref{def:VariationenOfAundPhi} we have learned that the infinitesimal gauge transformations $\delta_\varepsilon \Phi$ of the Higgs field can be understood as the condition for making a pullback of an $E$-connection on a vector bundle over the Higgs target manifold $N$; the foundation for this were Cor.~\ref{cor:VeryGeneralPullbackConnection} and Remark \ref{rem:CommutingDiagramOfPullbacks}. Hence, the definition of the infinitesimal gauge transformation of the Higgs field gives the existence of a suitable derivation describing infinitesimal gauge transformations by Thm.~\ref{thm:NewFormulaRecoversOldGaugeTrafoYay}; uniqueness however comes together with the infinitesimal gauge transformation of the field of gauge bosons:
	\item Regarding the infinitesimal gauge transformation $\delta_\varepsilon A$ of the field of gauge bosons, we learned by Prop.~\ref{prop:TangentSpaceOfSpaceOfFields} and Remark \ref{rem:TangentCommutingDiagram} that $\delta_\varepsilon A$ is not vertical in the sense of being an element of the vertical subbundle of $\mathrm{T}E$, but vertical in the sense of having values in the vector bundle $\mathrm{T}E \to \mathrm{T}N$ with an offset given by $\delta_\varepsilon \Phi$. Since $\delta_\varepsilon \Phi$ is a given information, we can apply a horizontal projection to $\delta_\varepsilon A$ without really loosing any information, technically leading to Prop.~\ref{prop:VariationOfA}, uniquely fixing $\Psi_\varepsilon = (\delta_\varepsilon \Phi, \delta_\varepsilon A)$. This construction is based on a specific motivation about how the infinitesimal gauge transformation of the minimal coupling should behave, given in Prop.~\ref{prop:InfinitesimalGaugeTrafoOfMinimalCoupleSmiley} and motivated by Cor.~\ref{cor:EichtrafovonDAPHIinClassicIstBabyEinfach}.
	\item Def.~\ref{def:TotalInfGaugeTrafoYayy} summarises all of that and provides the infinitesimal gauge transformation as a derivation lifting $\Psi_\varepsilon$; on scalar-valued functionals the infinitesimal gauge transformation is just the Lie derivative of $\Psi_\varepsilon$, while for $E$- and $\mathrm{T}N$-valued ones we use the basic connection to lift $\Psi_\varepsilon$.
	\item Finally, we discussed the algebra of such derivations, it was especially about whether it closes with respect to their commutator. Thm.~\ref{thm:VektorfelderSindZumGlueckGeschlossen} emphasizes that the vanishing of the basic curvature is essential for the closedness of the algebra subordinate to the vector fields $\Psi_\varepsilon$; while Thm.~\ref{thm:AllgemEineGeileFormelFuerDieEichKruemmung} shows that the closedness is up to the curvature of the used Lie algebroid connection behind the lift of $\Psi_\varepsilon$, as expected. The fundamental aspect needed for the proofs is provided by Thm.~\ref{thm:CurvatureOfBasicStuffIsEquivalentForGaugeTrafoCurvature}, providing the interpretation that the basic curvature measures whether or not the algebra of infinitesimal gauge transformations closes. 
	\item The parameters $\varepsilon$ are general functionals in this description, which is not completely avoidable, recall Remark \ref{ClassicalCommutatorRemark} (the part about the bookkeeping trick of the parameters) and the first bullet point of Remark \ref{RemarkUeberNablaRhoCurvatureForGauegTrafo} about the general unavoidable $\Phi$-dependency of $\varepsilon$ by construction. Hence, we needed to construct a Lie algebra for such parameters, essential for Thm.~\ref{thm:VektorfelderSindZumGlueckGeschlossen} and Def.~\ref{def:ErsteKruemmungsFormelFuerEichtrafos}. A candidate as Lie bracket is provided by Def.~\ref{def:PrebracketonPullbackLiealgebroid}, and Thm.~\ref{thm:PreKlammerEineSuperLieKlammer} accentuates that we have indeed a Lie bracket.
\end{itemize}

\textbf{Acknowledgements:} I want to thank Mark John David Hamilton, Anna Dall'Acqua, Alessandra Frabetti, Anton Alekseev and Maxim Efremov for their great help and support in making this paper. Also thanks to Jim Stasheff for his great remarks.

This paper started as part of my Ph.D.~at two universities (that type of Ph.D.~is called cotutelle), supervised by Anton Alekseev (Universit\'{e} de Gen\`eve) and Thomas Strobl (Universit\'{e} Claude Bernard Lyon 1):

This publication was produced within the scope of the NCCR SwissMAP which was funded by the Swiss National Science Foundation. I would like to thank the Swiss National Science Foundation for their financial support.

This work was supported by the LABEX MILYON (ANR-10-LABX-0070) of Universit\'{e} de Lyon, within the program "Investissements d'Avenir" (ANR-11-IDEX-0007) operated by the French National Research Agency (ANR), for which I am also grateful.

The paper was then finalised as part of my post-doc fellowship at the National Center for Theoretical Sciences (NCTS), which is why I also want to thank the NCTS.

%


\newpage



\renewcommand\refname{List of References}

\bibliography{Literatur}
\bibliographystyle{unsrt}


\end{document}